\documentclass[11pt]{article}
\usepackage[margin=4cm]{geometry}
\usepackage{titlepic}
\usepackage{graphicx}
\usepackage{float}
\usepackage{times}
\usepackage{url}
\usepackage{enumitem}
\usepackage[table]{xcolor}
\usepackage{csquotes}
\usepackage{multicol}
\usepackage{titlesec}
\usepackage[backend=biber, style=authoryear-ibid]{biblatex}
\bibliography{computepaper}
\usepackage[bitstream-charter]{mathdesign}
\usepackage[T1]{fontenc}
\usepackage{setspace}
\usepackage{multirow}
\usepackage{array}
\usepackage[font={normal,it},labelfont=bf]{caption}
\usepackage{subcaption}
\usepackage{booktabs}
\usepackage{longtable}
\usepackage{changepage}
\usepackage{textcomp}
\usepackage{hyperref}
\usepackage[nameinlink]{cleveref}
\usepackage[titletoc]{appendix}
\usepackage{afterpage}
\usepackage{microtype}

\usepackage{tocloft}

\renewcommand\thesection{\arabic{section}}

\usepackage{titlesec}
\titleformat{\section}
  {\huge}{\thesection}{2em}{}
\usepackage{titlesec}
\titleformat{\subsection}
  {\Large}{\thesubsection}{1em}{}

\renewcommand\thesection{\arabic{section}}
\renewcommand{\thesubsection}{\thesection.\Alph{subsection}}
\makeatletter
\def\blfootnote{\xdef\@thefnmark{}\@footnotetext}
\makeatother

\crefformat{section}{\textbf{#2Section~#1#3}}
\Crefformat{section}{\textbf{#2Section~#1#3}}
\Crefformat{appendix}{\textbf{#2Appendix~#1#3}}

\crefmultiformat{section}{\textbf{#2Sections~#1#3}}{ and \textbf{#2#1#3}}{, \textbf{#2#1#3}}{, and \textbf{#2#1#3}}
\Crefmultiformat{section}{\textbf{#2Sections~#1#3}}{ and \textbf{#2#1#3}}{, \textbf{#2#1#3}}{, and \textbf{#2#1#3}}

\crefmultiformat{appendix}{\textbf{#2Appendices~#1#3}}{ and \textbf{#2#1#3}}{, \textbf{#2#1#3}}{, and \textbf{#2#1#3}}
\Crefmultiformat{appendix}{\textbf{#2Appendices~#1#3}}{ and \textbf{#2#1#3}}{, \textbf{#2#1#3}}{, and \textbf{#2#1#3}}
\begin{document}

\begin{titlepage}

\newgeometry{margin=1.5cm}

\title{\Huge Computing Power and the \\ Governance of Artificial Intelligence \blfootnote{Each author contributed ideas and/or writing to the paper. However, being an author does not imply agreement with every claim made in the paper, nor does it represent an endorsement from any author's respective organization.} \blfootnote{$^{\ast}$ Denotes primary authors, who contributed most significantly to the direction and content of the paper. Both primary authors and other authors are listed in approximately descending order of contribution.} \blfootnote{$^{\dag}$ Indicates the corresponding authors: Girish Sastry (girish@openai.com), Lennart Heim (lennart.heim@governance.ai), and Haydn Belfield (hb492@cam.ac.uk). Figures can be accessed at \href{https://github.com/lheim/CPGAI-Figures}{\texttt{https://github.com/lheim/CPGAI-Figures}}.} }

\author{Girish Sastry,$^{\ast \dag 1}$ Lennart Heim,$^{\ast \dag 2}$ Haydn Belfield,$^{\ast \dag 3}$ \\ 
Markus Anderljung,$^{\ast 2}$  Miles Brundage,$^{\ast 1}$ Julian Hazell,$^{\ast 2,4}$ Cullen O’Keefe,$^{\ast 1,5}$ \\ Gillian K. Hadfield,$^{\ast 6,7}$ 
Richard Ngo,$^{1}$ Konstantin Pilz,$^{8}$ George Gor,$^{9}$\\ Emma Bluemke,$^{2}$
Sarah Shoker,$^{1}$ Janet Egan,$^{10}$ Robert F. Trager,$^{11}$ \\Shahar Avin,$^{12}$ 
Adrian Weller,$^{13}$ Yoshua Bengio,$^{14}$ Diane Coyle$^{15}$ 
\vspace{0.3in}\\
\small{$^1$OpenAI, $^2$Centre for the Governance of AI (GovAI),}\\
\small{$^3$Leverhulme Centre for the Future of Intelligence, Uni. of Cambridge,}\\ \small{$^4$Oxford Internet Institute, $^5$Institute for Law \& AI, $^6$University of Toronto}\\ \small{$^7$Vector Institute for AI, $^8$Georgetown University, $^9$ILINA Program, $^{10}$Harvard Kennedy School,}\\
\small{$^{11}$AI Governance Institute, Uni. of Oxford, $^{12}$Centre for the Study of Existential Risk, Uni. of Cambridge,}\\
\small{$^{13}$Uni. of Cambridge, $^{14}$Uni. of Montreal / Mila, $^{15}$Bennett Institute, Uni. of Cambridge}
}

\renewcommand*{\thefootnote}{\fnsymbol{footnote}}
\setcounter{footnote}{0}

\date{February 14, 2024}
\maketitle

\begin{abstract}\noindent \small Computing power, or "compute," is crucial for the development and deployment of artificial intelligence (AI) capabilities. As a result, governments and companies have started to leverage compute as a means to govern AI. For example, governments are investing in domestic compute capacity, controlling the flow of compute to competing countries, and subsidizing compute access to certain sectors. However, these efforts only scratch the surface of how compute can be used to govern AI development and deployment. Relative to other key inputs to AI (data and algorithms), AI-relevant compute is a particularly effective point of intervention: it is \textit{detectable}, \textit{excludable}, and \textit{quantifiable}, and is produced via an extremely \textit{concentrated supply chain}. These characteristics, alongside the singular importance of compute for cutting-edge AI models, suggest that governing compute can contribute to achieving common policy objectives, such as ensuring the safety and beneficial use of AI. More precisely, policymakers could use compute to facilitate regulatory \textit{visibility} of AI, \textit{allocate} resources to promote beneficial outcomes, and \textit{enforce} restrictions against irresponsible or malicious AI development and usage. However, while compute-based policies and technologies have the potential to assist in these areas, there is significant variation in their readiness for implementation. Some ideas are currently being piloted, while others are hindered by the need for fundamental research. Furthermore, naïve or poorly scoped approaches to compute governance carry significant risks in areas like privacy, economic impacts, and centralization of power. We end by suggesting guardrails to minimize these risks from compute governance.

\end{abstract}

\end{titlepage}

\restoregeometry

\renewcommand*{\thefootnote}{\arabic{footnote}}
\setcounter{footnote}{0}

\tableofcontents

\clearpage
\section{Introduction and Summary}\label{sec:introduction}

Artificial intelligence (AI) has made tremendous strides over the past decade, fueled in large part by a sharp exponential increase of computing power applied to the training of deep neural networks. This increased computing power (``compute'') has been a key enabler of the current wave of AI, including large language models and ``generative AI,'' for which general performance predictably improves as more compute is applied \autocite{weiEmergentAbilitiesLarge2022, ganguliPredictabilitySurpriseLarge2022, kaplanScalingLawsNeural2020}.

Increasingly powerful AI systems could profoundly shape society over the coming years; indeed, they are already affecting many areas of our lives, such as productivity, mobility, health, and education \autocite{pengImpactAIDeveloper2023}. The risks and benefits of AI raise questions about the governance of AI: what are the norms, institutions, and policies that can influence the trajectory of AI for the better \autocite{dafoeAIGovernanceResearch2018}? The central thesis of this paper is that governing AI \textit{compute} can play an important role in the governance of AI. Other inputs and outputs of AI development (data, algorithms, and trained models) are easily shareable, non-rivalrous intangible goods, making them inherently difficult to control; in contrast, AI computing hardware is tangible and produced using an extremely concentrated supply chain.

Policymakers are already making significant decisions about compute. Governments have invested heavily in the domestic production of compute, imposed export controls on sales of computing hardware to competing countries, and subsidized compute access to those outside of big technology companies \autocite{weinsteinExportControlsAI2023}. These early steps, however, do not exhaust the potential ways in which intervening on compute can be used to guide the development and deployment of AI.\footnote{While the author did not explore compute's role in AI governance in as much detail as we do, \textcite{hwangComputationalPowerSocial2018} was among the first to highlight its significance and outline some of its implications.}

Without prescribing specific policies, we argue that compute can be leveraged in many specific ways to enhance three key areas of governance. First, governance of compute can help increase regulatory \textit{visibility} into AI capabilities and use; second, it can steer AI progress by changing the \textit{allocation} of resources toward safe and beneficial uses of AI; third, it can enhance \textit{enforcement} of prohibitions against reckless or malicious development or use. Improvements in these three governance capacities can help achieve a range of policy objectives, like achieving public safety and ensuring equitable access to AI capabilities.

However, just as compute alone does not determine AI capabilities, governance of compute is not the whole story of AI governance. For example, approaches beyond compute governance are likely needed to address small-scale uses of compute that could pose major risks, like specialized AI applied to military use.\footnote{One emerging approach is to move towards measurements of the AI system’s capabilities directly \autocite{shevlaneModelEvaluationExtreme2023, anthropicAnthropicResponsibleScaling2023, openaiPreparedness2023}. See \textcite{maasAdvancedAIGovernance2023} for other levers of AI governance.}

Moreover, if not implemented carefully, compute governance can pose risks to privacy and other critical values. Since compute governance is still in its infancy, policymakers have limited experience in managing its unintended consequences. To mitigate these risks, we recommend implementing key safeguards, such as focusing on governance of industrial-scale compute and incorporating privacy-preserving practices and technology.\footnote{We discuss these risks and guardrails in \hyperref[sec:risks]{\Cref{sec:risks}}.}

This paper discusses a range of policy options and considerations available to different governing entities with decision-making authority. We use the term ``policymaker'' generically to refer to (ideally legitimate) authorities that can implement changes to norms, policies, processes, laws, and specific behaviors. This does not just include governments, and is meant to be an expansive definition. For example, national security policymakers, decision-makers at AI companies, lawmakers, standard-setting bodies, and international coalitions of governments are all included. Throughout this paper, we will specify which policymakers are most relevant to particular discussions.\footnote{This choice is mainly to balance abstraction and precision. We hope that this paper will also be useful to anyone interested in AI governance, including civil society and advocacy organizations.}

\begin{figure}[ht]
    \centerline{\includegraphics[width=1.4\linewidth]{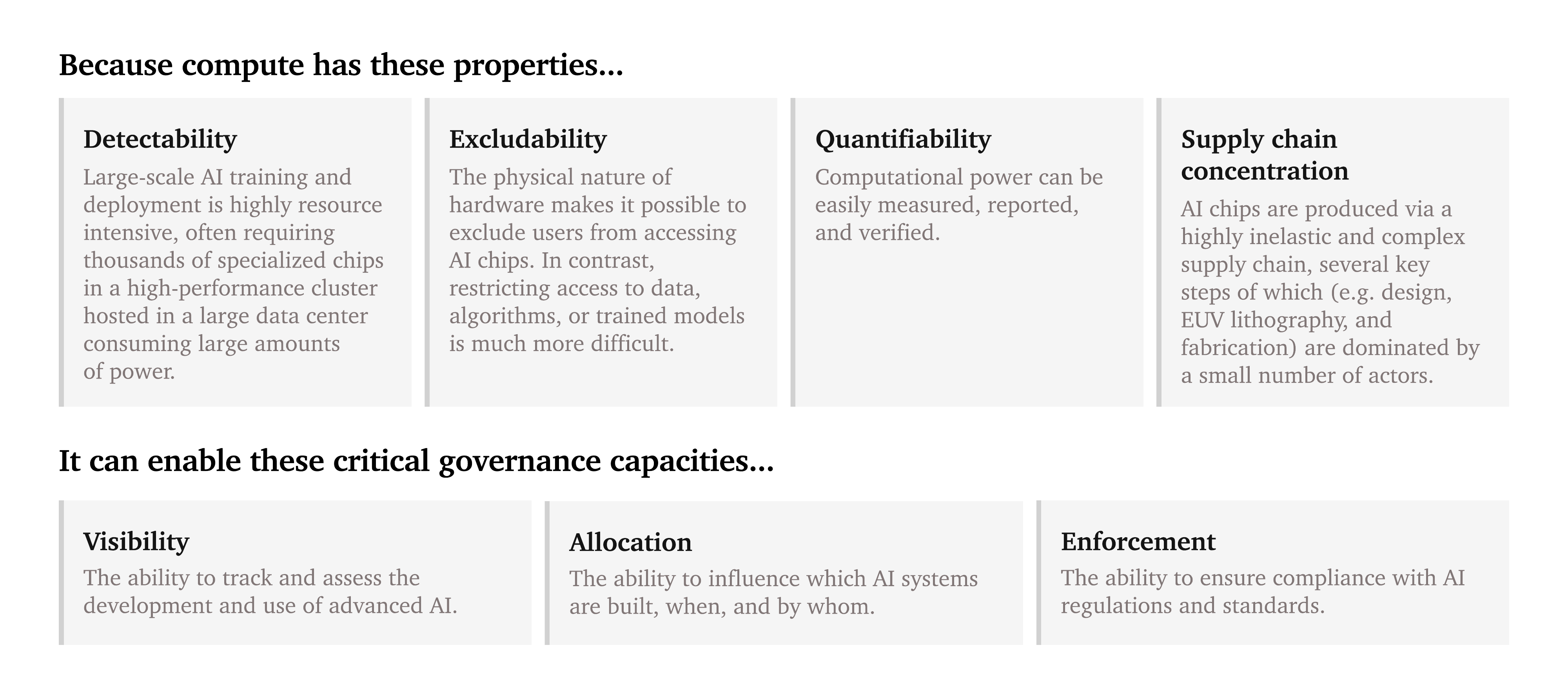}}
    \caption{\textbf{Summary of the core concepts in the report.} Compute is attractive for policymaking because of four properties. These properties can be leveraged to design and implement policies that enable three critical capacities for the governance of AI.}
    \label{fig:summary}
\end{figure}

The remainder of the paper is structured as follows.

In \Cref{sec:overview}, ``\textbf{Overview of AI Capabilities, AI Governance, and Compute},'' we provide basic context on several topics that serve as foundations for later sections. We discuss human capital, data, algorithms, and compute as the key inputs of AI development. We then characterize the steps of the AI lifecycle (consisting of design, training, enhancement, and deployment)---each of which presents a possible point of intervention (and has a unique compute footprint). We go on to discuss the impacts AI could have on society to motivate the importance of its responsible governance. To contextualize later sections, we then review ongoing efforts in governing compute.

In \Cref{sec:policymaking}, ``\textbf{Why Compute Governance Is Attractive for Policymaking},'' we explain the features of compute that make it an attractive tool for AI governance. This stems from compute’s singular importance to frontier models, and several properties of compute that augment its efficacy as a governance strategy:
\begin{itemize}
    \item[] \hyperref[ssec:detectability]{\textit{Detectability:}} Large-scale AI development and deployment is highly resource-intensive, often requiring thousands of specialized chips in a high-performance cluster hosted in a large data center consuming large amounts of power.
    \item[] \hyperref[ssec:excludability]{\textit{Excludability:}} The physical nature of hardware makes it possible to exclude users from accessing AI chips.\footnote{We use ``AI chips'' in this paper to refer to data center-grade, high-end chips targeted at AI use cases.} In contrast, restricting access to data, algorithms, or trained models is much more difficult.
    \item[] \hyperref[ssec:quantifiability]{\textit{Quantifiability:}} Computational power can be easily measured, reported, and verified.
    \item[] \hyperref[ssec:supplychain]{\textit{Supply chain concentration:}} AI chips are produced via a highly inelastic and complex supply chain, several key steps of which (e.g., design, EUV lithography, and fabrication) are dominated by a small number of actors.
\end{itemize}

Readers already convinced of compute's importance and special properties, but who wonder how compute governance might be extended beyond existing efforts, may consider jumping to \Cref{sec:enhance}, ``\textbf{Compute Can Enhance Three AI Governance Capacities},'' where we explore how compute can be used to enhance key governance capacities: (a) increasing the \textit{visibility} of AI development through monitoring compute, (b) changing the \textit{allocation} of compute to enable beneficial development, and (c) using compute for \textit{enforcement} of norms and regulations around AI.

We provide several illustrative policy mechanisms for visibility, allocation, and enforcement. The authors vary significantly in their views of which of these, if any, would be desirable. As important as \textit{whether} these mechanisms are adopted is the question of \textit{how} they are designed, implemented, and updated: subtle details of design and implementation could determine whether a compute governance policy is beneficial or detrimental on balance. To emphasize this point, we also note how these mechanisms could cause bad outcomes if designed or implemented poorly.

The illustrative mechanisms we explore are:
\begin{itemize}
    \item[A] \hyperref[ssec:visibility]{Visibility}
    \begin{enumerate}
        \item Using public information about compute quantities to estimate actors’ AI capabilities (now and in the future)
        \item Required reporting of training compute usage from cloud providers and AI developers
        \item International AI chip registry
        \item Privacy-preserving workload monitoring
    \end{enumerate}
    \item[B] \hyperref[ssec:allocation]{Allocation}
    \begin{enumerate}
        \item Differentially advancing beneficial AI development
        \item Redistributing AI development and deployment across and within countries
        \item Changing the overall pace of AI progress
        \item Collaborating on a joint AI megaproject
    \end{enumerate}
    \item[C] \hyperref[ssec:enforcement]{Enforcement}
    \begin{enumerate}
        \item Enforcing “compute caps” via physical limits on chip-to-chip networking
        \item Hardware-based remote enforcement
        \item Preventing risky training runs via multiparty control
        \item Digital norm enforcement
    \end{enumerate}
\end{itemize}

In \Cref{sec:risks}, ``\textbf{Risks of Compute Governance and Possible Mitigations},'' we synthesize our previous discussion of the possible limitations of compute governance. We emphasize the following (non-exhaustive) risks from compute governance:
\begin{itemize}
    \item[A] \hyperref[ssec:unintended-consequences]{Unintended Consequences}
    \begin{enumerate}
        \item Threats to personal privacy
        \item Opportunities for leakage of sensitive strategic and commercial information
        \item Risks from centralization and concentration of power
    \end{enumerate}
    \item[B] \hyperref[ssec:issues-feasibility-efficacy]{Issues of Feasibility and Efficacy}
    \begin{enumerate}
        \item Algorithmic and hardware progress
        \item Low-compute narrow models with dangerous capabilities
        \item Incentives for diversion, evasion, circumvention, and decoupling
    \end{enumerate}
\end{itemize}

Given those potential downsides, we suggest some \hyperref[ssec:guardrails]{guardrails} for compute governance:
\begin{enumerate}
    \item Exclude small-scale AI compute and non-AI compute from governance
    \item Research and implement privacy-preserving practices and technologies
    \item Only use compute-based controls for risks where ex ante controls are justified
    \item Periodically revisit controlled computing technologies
    \item Implement all controls with substantive and procedural safeguards
\end{enumerate}

We also provide two appendices: \Cref{sec:appA}: ``\textbf{The Compute-Uranium Analogy}'', and \Cref{sec:appB}: ``\textbf{Research Directions}''.

\section[Overview of AI Capabilities, AI Governance, and Compute]{Overview of AI Capabilities, \\AI Governance, and Compute}\label{sec:overview}

In this section, we provide an overview of key empirical context for the arguments and ideas in the following sections.

The section proceeds as follows: First, we describe how AI capabilities are created, and the role of compute in that process. Second, we define AI governance and describe key themes and trends in this area. Finally, we give four current examples of compute being leveraged for AI governance purposes.

\subsection{Creating AI Capabilities}\label{ssec:creating_capabilities}

Artificial intelligence (AI) refers to the science and engineering of building digital systems capable of performing tasks commonly thought to require intelligence, with this behavior often being learned rather than directly programmed.\footnote{Adapted from \textcite{brundageTrustworthyAIDevelopment2020}. When learning is involved, this subset of AI is often referred to as “machine learning” (ML). We focus on ML in this paper, given the strong empirical performance of ML-based systems compared to others.} The three key technical inputs to producing AI capabilities are data, algorithms, and compute, also referred to as the ``AI triad'' \autocite{buchananAITriadWhat2020}.\footnote{We focus on the subset of AI referred to as deep learning specifically here, rather than all of AI, given its disproportionate role in current high-profile deployments and policy discussions. Compared to some other AI techniques such as classical planning, deep learning is disproportionately compute-intensive, which admittedly biases our analysis toward the conclusion that compute is important, though we think this focus is justified by the predominance of deep learning.} People provide the necessary technical and scientific expertise (``talent,'' or human capital) to orchestrate the AI triad in order to produce a trained model.

\begin{figure}
    \centerline{\includegraphics[width=1.4\linewidth]{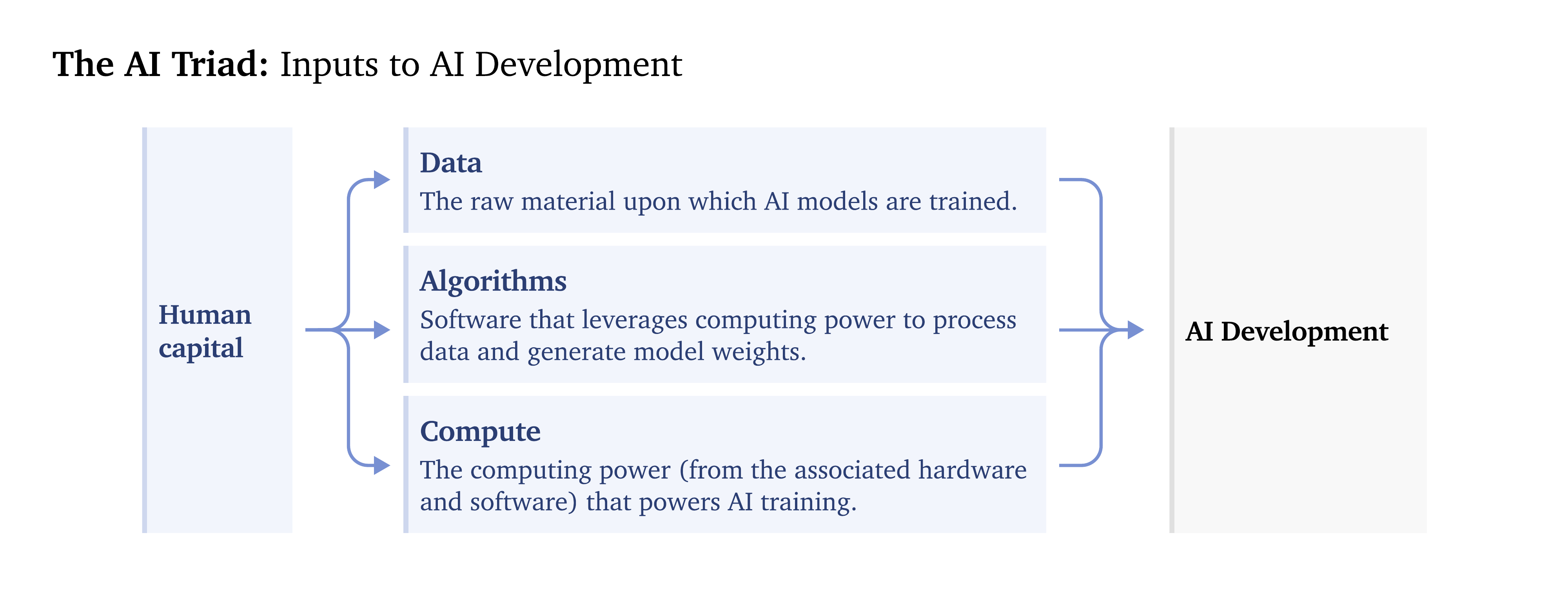}}
    \caption{\textbf{The AI Triad.} The three key technical inputs to AI are data, algorithms, and compute. Human capital is required for all inputs.}
    \label{fig:ai_triad}
\end{figure}

Data, algorithms, compute, and human capital each play pivotal roles in the development and deployment of AI. \textit{Data} is the raw material that is processed by compute; put differently, compute is the ``engine'' fueled by large amounts of data.\footnote{The volume of data used to train cutting-edge AI systems has grown dramatically over the last decade \autocite{villalobosTrendsTrainingDataset2022}.} There is a growing industry focused on producing this data,\footnote{For example, there is a growing industry focused on ``data labeling''---paying humans to perform tasks so that AI systems can be trained on that data, or to grade AI systems on their current performance. Data labeling is estimated to be a \$5 billion dollar market in 2023, much of it outsourced to developing countries due to lower wages \autocite{kshetriDataLabelingArtificial2021}.} and significant investment in new ways of generating valuable training data with less human involvement.\footnote{For example, see \textcite{baiConstitutionalAIHarmlessness2022}.} \textit{Algorithms} dictate the operations that are performed on data to produce AI capabilities.\footnote{Better algorithms essentially improve capabilities without increasing the required investment \autocite{pilzIncreasedComputeEfficiency2023}. Algorithmic breakthroughs such as the Transformer architecture significantly increased the efficiency with which compute and data are converted into capable models \autocite{vaswaniAttentionAllYou2017, hernandezMeasuringAlgorithmicEfficiency2020, erdilAlgorithmicProgressComputer2022, hoffmannTrainingComputeoptimalLarge2022}.} Algorithms encompass the source code that defines everything from the architecture of AI models to the specific methodologies employed in the training. \textit{Computing power} (and the associated hardware and software), is used to execute algorithms, and serves as the ``substrate'' for the information processing involved in AI. Finally, \textit{human capital} is important to produce data, algorithms, and compute and to operate the training process itself.\footnote{For example, the Transformer architecture \autocite{vaswaniAttentionAllYou2017} was invented using similar amounts of data and compute to what was available previously. Human capital is also used to train AI systems: humans essentially ``teach'' machine learning models by demonstrating how to do a task or providing feedback.}

Compute has played a particularly prominent role in recent AI progress. The advent of the deep learning era around 2010–2012 can be attributed to the initial use of GPUs (Graphics Processing Units---specialized chips originally developed for graphics rendering) for training AI systems \autocite{krizhevskyImageNetClassificationDeep2017, amodeiAICompute2018, sevillaComputeTrendsThree2022}. This enabled AI systems to grow significantly in size, providing the ``deep'' in ``deep learning.'' AI chips provide significant efficiency and performance boosts to AI systems \autocite{khanAIChipsWhat2020}. Development of frontier AI systems has become increasingly synonymous with large compute budgets, access to large computing clusters,\footnote{We use the word ``cluster'' to refer to any amount of compute that can be viewed as a single system (even if each computing element is geographically distributed). In the context of AI, these are typically geographically concentrated in large data centers, to reduce inefficiencies from communication cost.} and the proficiency to leverage them effectively \autocite{besirogluComputeDivideMachine2024}. However, it is important to note that not all AI applications require vast amounts of compute; specialized AI systems have displayed impressive abilities in some domains, even by using much less compute than frontier systems.\footnote{For more discussion on this point, see \Cref{sec:risks}.}

Most current progress in AI leverages a technology called artificial neural networks. After a neural network model is trained, it becomes capable of executing different tasks, such as writing computer code, generating images, or acting and responding to sensory input. These trained models are then often deployed as a general-purpose system, such as a chatbot, or as a sub-component of some other product or service.

\begin{figure}
    \centerline{\includegraphics[width=1.4\linewidth]{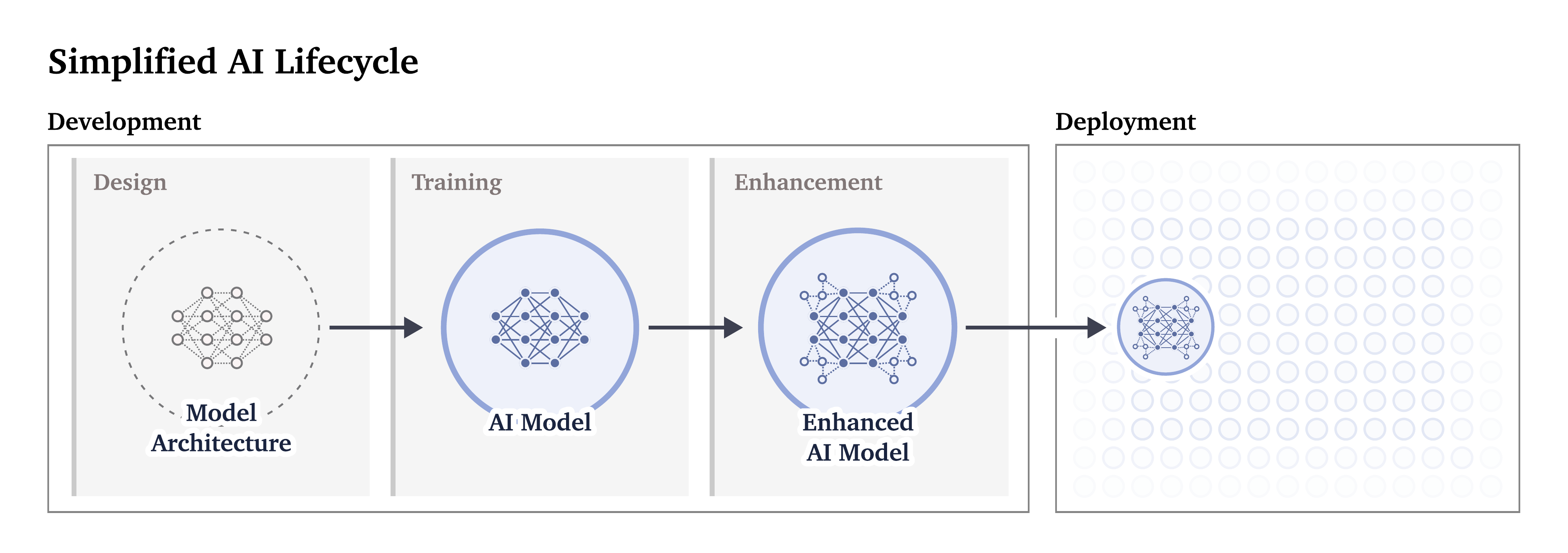}}
    \caption{\textbf{A Simplified AI lifecycle.} In the compute-intensive Development stage, the model is designed, trained, and enhanced. The model is then put to use in the Deployment Stage. Many copies of the model can be run during Deployment.}
    \label{fig:simp_lifecycle}
\end{figure}

A simplified model of the AI lifecycle consists of two main phases: the \textit{development} phase and the \textit{deployment} phase (\Cref{fig:simp_lifecycle}). In the development phase, AI systems are trained and optimized, whereas in the deployment phase, these systems are put toward solving a variety of tasks, based on the knowledge and skills they learned during training \autocite{oecdBlueprintBuildingNational2023}.

In the development phase, AI systems are designed, trained, and enhanced. Design involves determining the general characteristics of an AI model (e.g., how many layers the neural network will have), the dataset that will be used, and how to train the model (e.g., how many times to ``look at'' each data point). Training is a process that involves learning from vast amounts of data, often sourced from the internet (e.g., public domain websites or images). Training is also the most compute-intensive part of AI development, i.e., performing a large number of computational operations (often measured as ``floating point operations'').\footnote{``Floating point operations'' are used when a high-degree of precision is required to represent numbers in a computer, and are common for tasks that require large-scale mathematical calculations. However, recent progress in AI has raised the possibility of using lower-precision representations of numbers (and ``integer'' representations), which increases the processing speed of each operation. Presently, most AI training predominantly uses floating point numbers, but this could change in the future \autocite{ghaffariIntegerArithmeticEnough2022}.} The compute required for training is determined by factors such as the system’s architecture, the size of the architecture (i.e., the number of trainable ``weights''), the volume and quality of data presented to the system, the number of times this data is reused, and the training algorithm. Other ``enhancements'' like fine-tuning and reinforcement learning from human feedback are also effective at increasing the usefulness and capabilities of an AI system. Enhancement typically requires much less compute than pre-training.

A trained model can then be distributed and deployed for various applications, marking the beginning of the deployment phase. In this phase, the model performs ``inferences'' by processing inputs and making predictions (e.g., about which word would come next in a sentence, or what the answer to a question is). The inference compute needed for deployment is essentially a product of the architecture and parameter size of the model and the number of instances of the model being deployed. Although there are many methods to make inference more efficient,\footnote{Such methods include pruning, distillation, fine-tuning, and others \autocite{millerTechniquesMakeLarge2023, menghaniEfficientDeepLearning2023}.} it is reasonable to say that, all things being equal, larger and more performant AI systems require higher compute budgets for a single inference. Often, trained models are deployed as part of a larger AI system, which includes non-machine-learning components (like user interfaces and access controls).

Both large-scale training and inference processes require centralized, high-performance computing systems optimized for AI workloads housed within data centers. Due to the immense scale of current model deployment, the majority of all AI compute is now used for inference, even though a single training run requires far more compute than a single inference. For example, widely used AI applications such as internet search, voice recognition, and language translation all require large-scale compute infrastructure to serve billions of users; running these applications at scale requires many billions of inferences from AI models.\footnote{For example, AWS estimated that 90\% of its workload is inference \autocite{pattersonCarbonEmissionsLarge2021}. We discuss other reasons this is likely true in \Cref{ssec:importance_compute_frontier}.}

\subsection{AI Governance}\label{ssec:ai_governance}

``AI governance'' refers to the study or practice of local and global governance systems---including norms, policies, laws, processes, and institutions---that govern or should govern AI research, development, deployment, and use \autocite{huaAiAntitrustReconciling2021}.\footnote{AI governance involves the establishment of regulations, standards, best practices, and decision-making processes by governments and society to ensure the development and use of AI are beneficial and align with societal well-being \autocite{dafoeAIGovernanceResearch2018}. See also \textcite{maasConceptsAdvancedAI2023}, \textcite{oheigeartaighOvercomingBarriersCrosscultural2020}, \textcite{dafoeAIGovernanceOverview2023}, and \textcite{schuettThreeLinesDefense2023}.}

As AI systems gain increased capability across a wide range of domains, they have the potential for incredibly beneficial applications in health care, energy, entertainment, and many other business and public services \autocite{abramoffAutonomousArtificialIntelligence2023, segerDemocratisingAIMultiple2023}. The use of AI systems is widely expected to have a positive impact on productivity and living standards \autocite{brynjolfssonGenerativeAIWork2023, czarnitzkiArtificialIntelligenceFirmlevel2023, bailyMachinesMindCase2023}, but the realized benefits will depend on the regulatory and governance structures adopted. AI could also pose risks that are more extreme in nature \autocite{shevlaneModelEvaluationExtreme2023}. These include highly effective and widespread surveillance to oppress populations \autocite{petersonGeopoliticalImplicationsAI}, large-scale influence operations \autocite{goldsteinGenerativeLanguageModels2023}, biological weapons \autocite{moutonOperationalRisksAI2023}, threats to international stability \autocite{imbrieAISafetySecurity2019, horowitzAIInternationalStability2021, shokerConfidencebuildingMeasuresArtificial2023}, and the potential for AI to deliberately cause harm due to misalignment \autocite{ngoAlignmentProblemDeep2023}. Mismanagement of such risks could lead to human disempowerment or even extinction \autocite{caisStatementAIRisk, russellHumanCompatible2019}. 

Compute governance---the topic of this paper---is one tool for AI governance. Other tools for AI governance include, for example, model performance standards on tests or evaluations and rules establishing requirements about the training data, technical methods, and personnel used to produce AI \autocite{shevlaneModelEvaluationExtreme2023}.

\subsection{Compute Governance Today}\label{ssec:governance_today}

Governments around the world are already targeting compute. This is mostly in the context of geopolitical efforts to ensure that their countries are able to thrive in the unfolding AI revolution and to prevent confirmed or suspected misuses from adversaries.\footnote{However, these efforts are unequally distributed: it is mostly a handful of countries, concentrated in the Global North, that are engaging in compute governance. We discuss these equity issues further in \Cref{ssec:allocation}.} We point to this not to suggest that what is being done is wise or effective. But these cases demonstrate that compute governance is not a purely theoretical idea: it is already happening today. Here we discuss four examples: investing in domestic compute capacity, subsidizing compute access to those outside big technology companies, imposing export controls on competing countries, and setting compute-based reporting thresholds. We also discuss some emerging concerns with the role of compute in AI governance. These actions---and the concerns raised in response---emphasize the need for a holistic theory and appraisal of compute governance, which this paper aims to provide.

\textbf{\textit{Investment in domestic compute capacity}} \\\\
Compute is a key resource for modern economies and societies, so the amount of compute possessed by different states is a key topic of interest to those states \autocite{oecdBlueprintBuildingNational2023, oecdExpertGroupCompute}. Access to compute is arguably comparable in economic and societal importance to access to the internet and the infrastructure of undersea cables that support it, and perhaps even to energy infrastructure. Much as they have with those other resources, many governments have become increasingly interested in the vulnerabilities that compute dependence may create. Access to compute provided by foreign-located and/or foreign-owned data centers may be vulnerable to espionage, sabotage, price hikes, political interference, or geopolitical interventions \autocite{belfieldGreatBritishCloud2023, hogarthAINationalism2018, chanderDataNationalism2015}.

Affecting the distribution of compute between countries is becoming a key point of intervention by governments.\footnote{For example, the OECD AI Compute and Climate group whose mission is to promote compute access \autocite{oecdExpertGroupCompute}.} The EU and the U.S. have both provided \$50 billion in subsidies to semiconductor manufacturing in their respective CHIPS Acts \autocite{browneEuropeApprovesIts2023, shepardsonSenateApproves522022}. In the U.S., Europe, and China there is significant government interest in acquiring sovereign cloud computing centers \autocite{chanderDataNationalism2015, pilzComputeScaleBroad2023}. There has been extensive discussion of both compute and AI ``sovereign capability'' in the U.K., France, and Germany \autocite{belfieldGreatBritishCloud2023}.\footnote{``Sovereign'' capability can refer to a capability either located in a particular country, or located and owned by a company or other group within a particular country. See the distinction between ``own, collaborate, and access'' \autocite{ukcabinetofficeGlobalBritainCompetitive2021}. For compute sovereignty, see \textcite{alephalphaAlephAlphaRaises2023, lawranceXavierNielInvests2023, wanatEuAIAct2023}.}

The U.S., China, and Russia have long-standing supercomputing programs including, for example, the U.S. Department of Energy’s Advanced Scientific Computing Research program. Projects for civilian use include Japan’s consistent investment in supercomputing, including the nearly \$1 billion Fugaku, and Australia’s National Research Infrastructure (though these focus more on scientific computing rather than AI). Governments are investing in publicly funded and owned national compute infrastructure specifically for AI in the U.S. (the NAIRR), the U.K. (AIRR), and the EU (EuroHPC) \autocite{nairrtfStrengtheningDemocratizingArtificial2023, ukdepartmentforscienceinnovation&technologyFrontierAICapabilities2023, eurohpcEuroHPC}.

\textbf{\textit{Subsidizing compute access}}\\\\
Currently, most AI compute is concentrated in the hands of private industry \autocite{besirogluComputeDivideMachine2024, verdegemDismantlingAICapitalism2022}. Because the distribution of compute between AI developers affects markets and outcomes for consumers and citizens, there may be good reasons to support increased use of AI computing infrastructure by other sectors, including academia, civil society, and governments.

Training large AI models and delivering access to them at scale requires access to large amounts of compute. Without that, building this class of models is out of reach: there are experiments one simply cannot run, and products (and services) one cannot build. Some companies (like Meta, Google, and Amazon) are of a sufficient scale that they own their own compute, but most AI developers rely on accessing cloud compute from infrastructure-as-a-service (IaaS) companies. This market (outside of China) is dominated by three companies, termed ``hyperscalers'': Amazon (through Amazon Web Services), Microsoft (through Azure), and Google (through Google Cloud Platform). Today, most major developers of large models are either subsidiaries of the hyperscalers, or have entered into ``compute partnerships'' with them. This includes Anthropic, Cohere, Google DeepMind, Hugging Face, OpenAI, Stability AI, and many others \autocite{benaichStateAIReport2022, anthropicAnthropicPartnersGoogle, huggingfaceHuggingFaceAWS2023}.

The compute available to academics has not grown at anywhere near the rate available to the public sector \autocite{ahmedDedemocratizationAIDeep2020, besirogluComputeDivideMachine2024}. The compute disparity between industry and other AI developers such as academics is one reason that many AI and computer science professors have gone to work full-time or part-time in industry \autocite{zwetslootAIFacultyShortages2022}. This may have concerning effects such as fewer professors available to train the next generation of PhD graduates, and less research focused on non-commercial public goods or verifying companies’ claims.

Given these considerations, changing the distribution of compute between AI developers is considered a key point of intervention by some policymakers. Compute access through the U.S. National AI Research Resource (NAIRR) and U.K. AI Research Resource (AIRR) is explicitly intended to address the imbalances discussed above \autocite{nairrtfStrengtheningDemocratizingArtificial2023}. We say more about what additional steps might be taken with compute subsidies in \Cref{sec:policymaking}.

Policymakers face a choice between public and private provision of compute access. Compute credits for existing big cloud providers are easier to immediately administer, as they do not require establishing new institutions and they leverage private clouds’ existing expertise. However, they can reinforce the power of the largest cloud providers. While this can benefit the countries in which these cloud providers are based---providing them greater control and influence---it may increase vulnerabilities for other countries. This choice is therefore especially stark for countries that are not the U.S. or China.

\textbf{\textit{Imposing export controls}}\\\\
Over the past several years, some countries have imposed export controls on semiconductors and semiconductor manufacturing equipment, to slow the technological advancement of their geopolitical adversaries (and especially their military capabilities) by denying them access to the most advanced forms of compute \autocite{fedasiukSiliconTwist2022}. For example, the October 7, 2022, U.S. chip export restrictions \autocite{usbisImplementationAdditionalExport2022} prohibited the sale of the chips most relevant to AI to Chinese organizations, and enforced stringent controls on advanced semiconductor manufacturing equipment and software essential for creating cutting-edge chips to impede China's ability to independently produce competitive (AI) chips \autocite{allenChokingChinaAccess2022}. The U.S. updated these restrictions on October 17, 2023 \autocite{bureauofindustryandsecurityExportAdministrationRegulations2023}.

The scope of this ``small yard, high fence'' approach \autocite{thewhitehouseRemarksNationalSecurity2022} is particularly focused on the AI chips used in data centers and excludes consumer devices, such as gaming chips. By focusing on these specific characteristics, U.S. export restrictions intend to regulate AI data center compute to prevent misuse by foreign actors while avoiding unnecessarily impeding other uses of computing hardware (such as gaming).\footnote{For example, the rule includes exceptions to the export controls for consumer-grade chips \autocite{usbisImplementationAdditionalExport2022}.} However, the increasing power of consumer chips could enable them to be used for purposes that the controls aimed to prevent.\footnote{We discuss some of these drawbacks in more detail in \Cref{sec:risks}.}

Restricting compute access for specific actors might be a key method for utilizing compute to avert harm and encourage adherence to certain norms. However, this approach comes with drawbacks, such as exacerbating geopolitical tensions and intensifying economic incentives for domestic compute producers, curbing potentially advantageous applications in the affected regions, and centralizing power among nations and organizations with compute access. In \Cref{sec:enhance}, we examine these drawbacks of broad technology denial and advocate for further research into more refined alternatives to these strategies.

\textbf{\textit{Compute-based reporting in the Executive Order}} \\\\
The Biden-Harris Administration's Executive Order 14110 issued on October 30, 2023, ``Ensuring the Safe, Secure, and Trustworthy Development and Use of Artificial Intelligence,'' introduces a range of AI governance measures. Significantly, Section 4 of this order leverages computational power as a criterion for classifying AI systems that warrant additional scrutiny due to potential safety and security concerns \autocite{thewhitehouseExecutiveOrderSafe2023}.

Previously, only AI companies knew the specifics of their frontier training runs, including the details of their models and the measures taken to ensure their security. The U.S. government typically became aware of new advanced models only after their public announcement, often leaving uncertainties about the associated risks. The new executive order mandates U.S. AI companies to proactively notify the government about any ongoing or planned activities concerning the training, development, or production of frontier models. It also requires these companies to share the results of red-team safety tests, and instructs the new AI Safety Institute within the National Institute of Standards and Technology (NIST) to develop evaluation standards. These requirements apply to foundation models trained with more than $10^{26}$ operations, or $10^{23}$ operations for models trained using primarily biological sequence data. This threshold is designed to capture future developments in AI. At the time of writing, no publicly known AI model meets the $10^{26}$ operations  threshold~\autocite{epochAITrends2023}, whereas one model appears to meet the biological sequence data threshold~\autocite{maugBiologicalSequenceModels2024}.

\begin{figure}[!ht]
    \centerline{\includegraphics[width=1.4\linewidth]{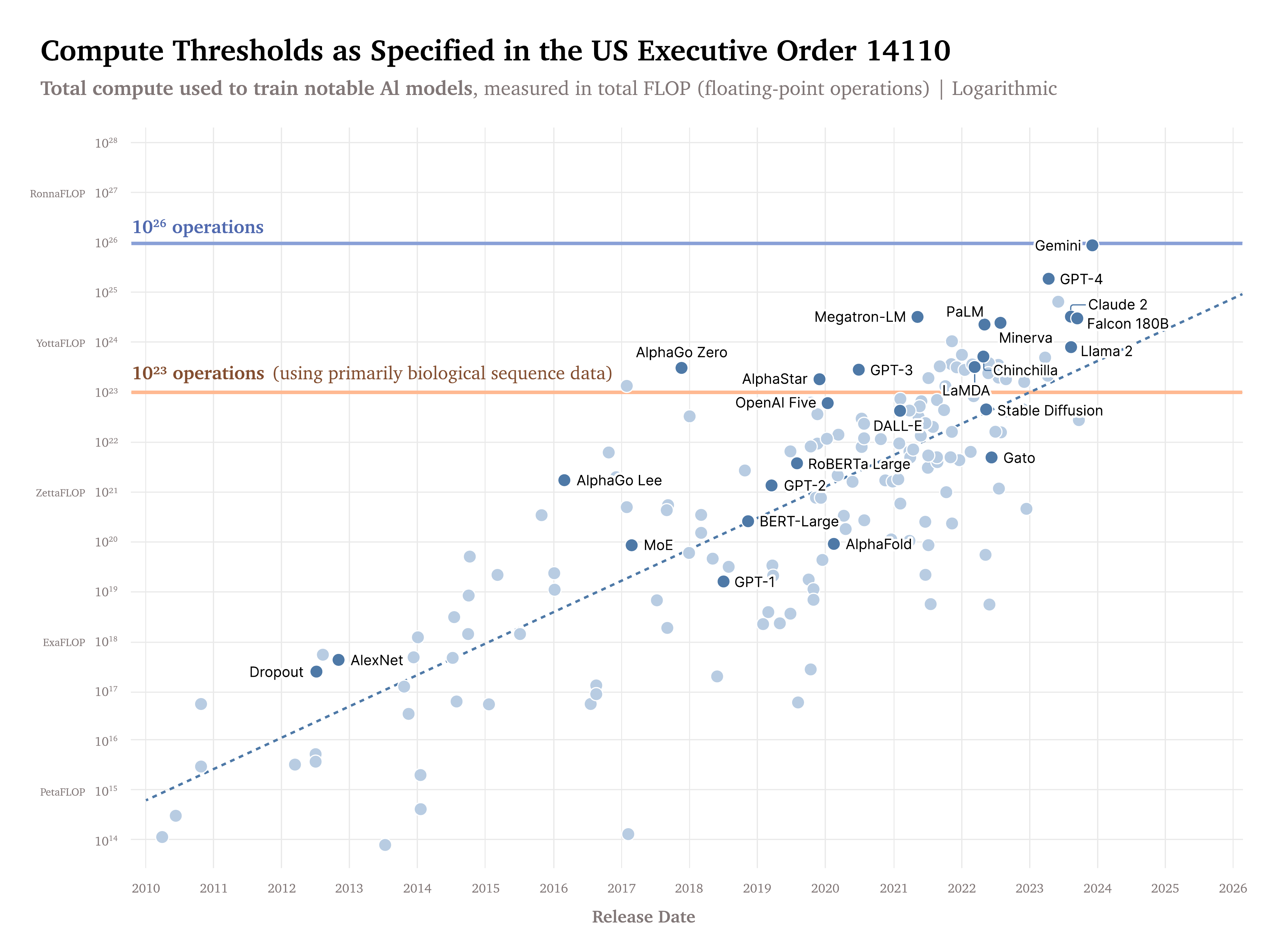}}
    \caption{\textbf{Training compute used for notable ML models has been doubling every six months since the emergence of the Deep Learning Era.} Executive Order 14110 introduced a notification requirement for models trained with more than $10^{26}$ operations (and $10^{23}$ operations if trained on using primarily biological sequence data).}
    \label{fig:train_compute}
\end{figure}

Moreover, Executive Order 14110 includes reporting requirements for large compute clusters that could potentially be used in such training runs.\footnote{The computing cluster needs to meet an aggregated computing performance of more than 1020 operations per second, a transitive connection of more than than 100 Gbit/s, and be housed in a single data center. The requirements include reporting ``acquisition, development, or possession, including the existence and location of these clusters and the amount of total computing power available in each cluster'' \autocite{thewhitehouseExecutiveOrderSafe2023}.} This rule also encompasses compute provided as a service (e.g., cloud computing), if a foreign entity accesses compute resources above the mentioned training compute threshold and if trained on a cluster that meets the previous definition. This Know-Your-Customer provision had already been proposed to patch a potential loophole of the previously mentioned October 7 2022 U.S. chip export controls \autocite{eganOversightFrontierAI2023, whittlestoneResponseUKFuture2023, fistChineseFirmsAre2023, heimAccessingControlledAI2023}. We discuss extensions and related options in \Cref{ssec:visibility}.

\newpage
\textbf{\textit{Emerging concerns with compute governance}} \\\\ 
Concerns about compute governance have grown alongside these new compute governance efforts and proposals. This further emphasizes the need for greater understanding of the role of compute in AI development and a balanced appraisal of the promises and perils of AI governance.

Responses to export controls on semiconductors have been mixed. A number of commentators have noted risks to the economic interests of the United States and its allies \autocite{fengCostsChinaSemiconductor2022}, who generally benefit greatly from trade with export control targets like China. Compute manufacturers are among the most critical, often emphasizing their dependence on China for supplies for the same chips subject to recent export controls \autocite{ting-fangASMLSaysDecoupling2023, goswamiNvidiaCEOChipmakers2023}. China has indeed imposed retaliatory export controls on raw materials needed for chipmaking \autocite{heChinaJustStopped2023}. There are also reports of China amassing chip-making equipment and materials ahead of anticipated controls \autocite{panTechWarChinese2023}. Others worry that the U.S. imposed the export controls too early, and that keeping China reliant on supply chains dominated by democracies would have been more prudent \autocite{scharreDecouplingWastesLeverage2023}. The recent advances in Chinese chipmaking capacity, such as the fabrication of a 7nm chip\footnote{The meaning and significance of the 7nm designation are explained in \Cref{ssec:feasibility_governance}.} for Huawei phones \autocite{liuHowHuaweiSurprised2023} have increased concerns about the controls accelerating China’s progress towards AI chip supply chain independence and thus diminishing U.S. capacity to control access to compute. However, others point out 
that China was already working towards such independence long before the October 2022 export controls \autocite{allenChinaNewStrategy2023}. There are also serious doubts about whether the export controls are being effectively targeted and enforced \autocite{patelChinaAISemiconductors2023}. 

Other specific compute governance proposals have attracted similar controversy. For example, one prominent idea for regulating frontier AI systems is to require a license to access a large amount of AI compute or use large amounts of AI compute for specific purposes \autocite{anderljungFrontierAIRegulation2023, smithDevelopingDeployingAI2023}. This idea is intended to enable a more anticipatory approach to governing the development of the highest-risk AI systems. A number of objections have been raised to this cluster of ideas, including the possibility of licensing creating barriers to competition, centralization of power, or opportunities for regulatory capture \autocite{thiererExistentialRisksGlobal2023, howardAISafetyAge2023}. More prosaically, barriers to trade in compute and AI could slow growth in one of the most promising economic sectors, which has historically benefited enormously from low barriers to entry, competition, and trade \autocite{fengCostsChinaSemiconductor2022, thiererPermissionlessInnovationContinuing2014}.

Numerous proposals remain untested in real-world scenarios, and the manner of their implementation could significantly impact their effectiveness. For instance, if strategically and commercially vital compute information is disclosed to regulators (as stipulated in the executive order), it may become a prime target for espionage. Consequently, the diligence and security applied to managing this information could play a crucial role.

Using training compute-based thresholds as the sole foundation for policy has also prompted concern. One reason is that training compute usage is only a high-level proxy for a model’s capabilities; it alone does not provide a comprehensive assessment. As the science of AI risk assessment advances, higher-fidelity measurements of AI capabilities could become possible. In turn, these capability measurements can enable better-targeted policies \autocite{shevlaneModelEvaluationExtreme2023, openaiPreparedness2023}. Other issues include, for example, the necessity of changing the compute thresholds over time as algorithmic and hardware progress occur \autocite{pilzIncreasedComputeEfficiency2023}, and the possibility of unforeseen low-compute enhancements that drastically change an AI system’s capabilities \autocite{bommasaniDrawingLinesTiers2023}.

We encourage readers to keep these possible risks and limitations of compute governance in mind when evaluating compute governance proposals. We do our best to acknowledge them when they apply, and also discuss recurring genres of risks and limitations in \Cref{ssec:limitations}. In \Cref{ssec:guardrails}, we discuss guardrails that could be included in compute governance proposals to reduce their risks. These concerns also highlight the need to be thoughtful and flexible in compute governance design and implementation: poor execution of compute governance carries serious risks that could destroy much of the promise compute governance holds.

\section{Why Compute Governance Is Attractive for Policymaking}\label{sec:policymaking}

In this section, we note two reasons why compute is an appealing lever for AI governance. First, compute plays a crucial role in developing and deploying cutting-edge AI systems. All else equal, the amount of compute used is one of the most reliable indicators of the potential impact of a system, during both development and deployment. AI systems consistently develop more sophisticated capabilities as more computing power is used to train them.\footnote{As predicted by ``scaling laws,'' described in more detail later in this section.} Because of this, the amount of compute used to train frontier systems has rapidly increased over the last decade, and now often costs tens of millions of dollars \autocite{cottierTrendsDollarTraining2023}. After training, the impacts of a model correlate with how widely it is deployed;\footnote{This is not a simple linear relationship: some inferences will be significantly more impactful than others.} some frontier AI systems are deployed to millions of users, which also requires a large amount of compute.\footnote{While widespread deployment requires a large amount of compute in total, it does not necessarily require a large amount of centrally owned compute---for example, after a model’s weights are released publicly, it can be downloaded and run independently by many individuals.} Therefore, identifying and regulating the use of large amounts of compute has the potential to significantly influence the impacts of AI.

\begin{figure}[!ht]
    \centerline{\includegraphics[width=1.4\linewidth]{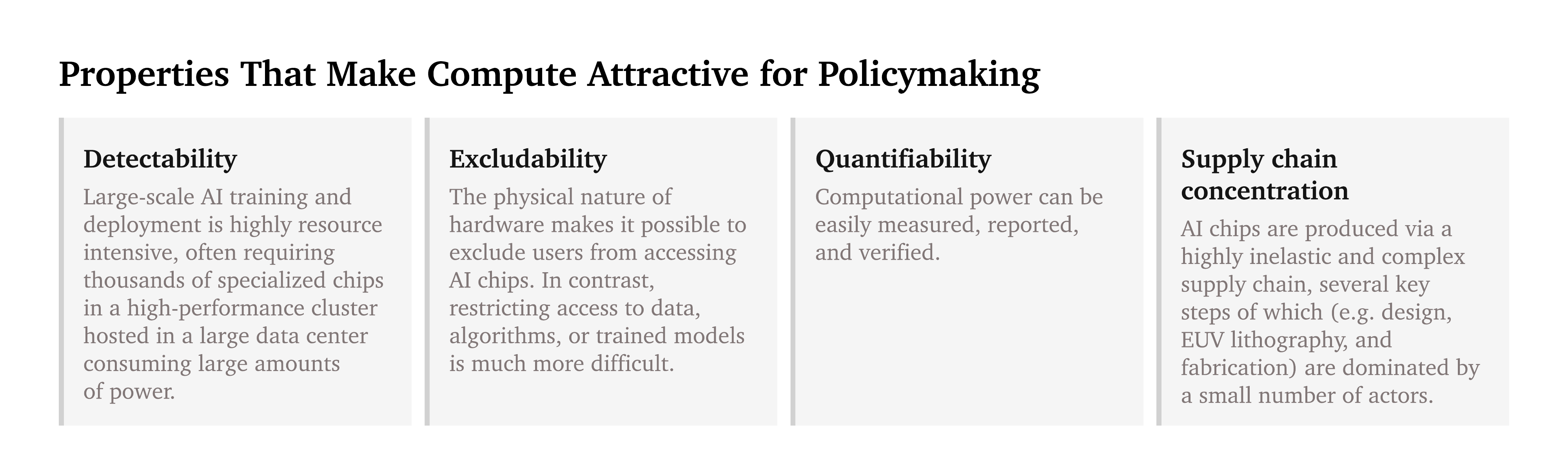}}
    \caption{\textbf{The feasibility of compute governance} is underpinned by four properties: detectability, quantifiability, excludability, and supply chain concentration.}
    \label{fig:feasibility}
\end{figure}

\begin{figure}[!ht]
    \centerline{\includegraphics[width=1.4\linewidth]{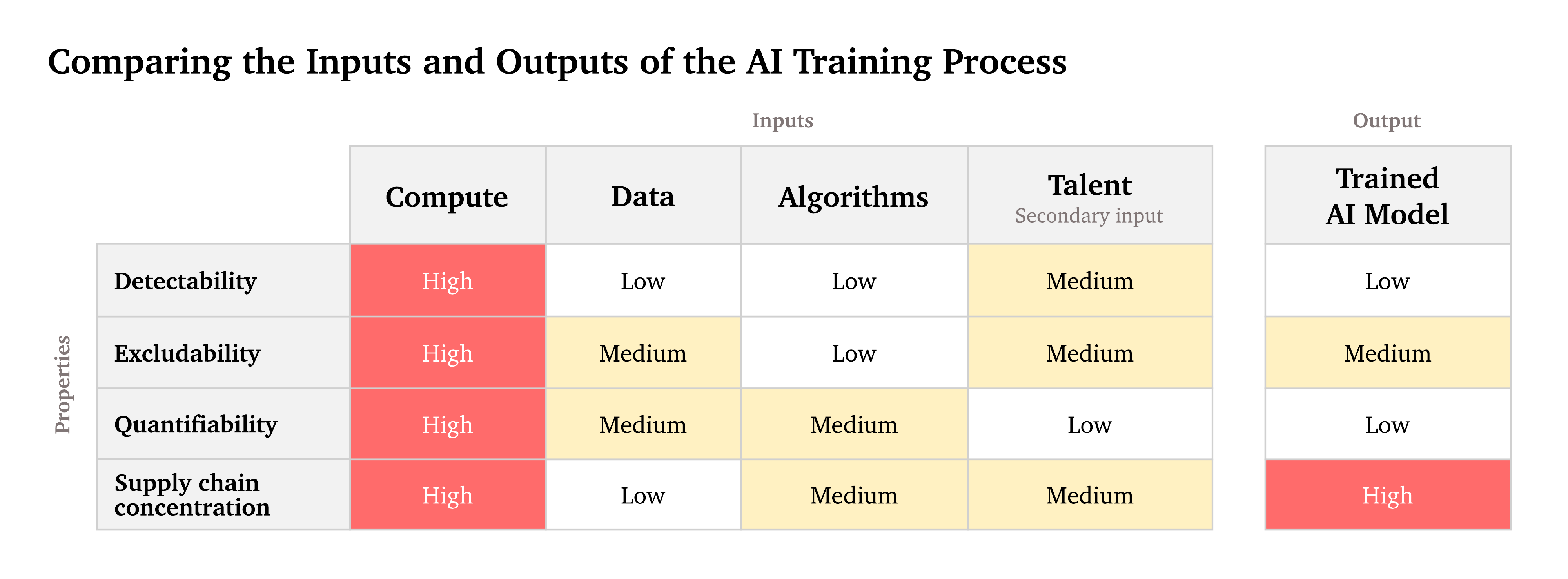}}
    \caption{\textbf{Comparing the four properties of the key inputs and outputs of the AI training process.} Compute scores highly on all four properties, suggesting that compute governance may be feasible, and perhaps more effective than governance of other inputs or outputs.}
    \label{fig:input_output}
\end{figure}

Second, governing compute is technologically \textit{feasible}: it seems possible for society to monitor and restrict the computational resources used to develop and deploy AI, should it choose to do so. This is a consequence of four features of compute that other inputs to AI progress don’t share: \textbf{detectability, excludability, quantifiability,} and \textbf{supply chain concentration} (\Cref{fig:feasibility}). Computing hardware is a rivalrous physical good that can be identified, counted, and tracked; this is made easier by the fact that the supply chains used to produce it have several key bottlenecks. By contrast, many other inputs and outputs (including training data, algorithms, and trained models) are easily shareable, non-rivalrous intangible goods. Additionally, computing hardware can be quantified in relatively objective ways (e.g., technical features like operations per second, communication bandwidth, and memory), allowing quantification of the overall compute used to develop an AI system. Almost all other inputs (in particular, human capital) are much harder to quantify. For a summary of our comparison, see \Cref{fig:input_output}. 

The rest of this section defends these two main claims, which provide a foundation for our investigation of possible approaches to compute governance in later sections. These two claims also suggest an analogy between compute and uranium in the context of nuclear governance; \Cref{sec:appA} explores this analogy further.

\subsection{The Importance of Compute for Frontier Models}\label{ssec:importance_compute_frontier}

Compute is a particularly key input for frontier models, which frequently introduce new AI capabilities. Compute constitutes a large fraction of the costs of frontier AI labs, due to the enormous amounts used \autocite{knightOpenAICEOSays2023, ukdepartmentforscienceinnovation&technologyFrontierAICapabilities2023, competitionandmarketsauthorityAIFoundationModels2023}. The compute used to train notable machine learning systems has doubled roughly every six months on average, growing by a factor of 350 million over the last 13 years (\Cref{fig:compute_train_a}) \autocite{epochAITrends2023}.\footnote{According to Epoch’s data, the doubling time between 2010 and March 2023 was 5.6 months. They define ``notable machine learning systems'' as follows: ``All models in our dataset are mainly chosen from papers that meet a series of necessary criteria (has an explicit learning component, showcases experimental results, and advances the state-of-the-art) and at least one notability criterion (>1000 citations, historical importance, important SotA advance, or deployed in a notable context). For new models (from 2020 onward), it is harder to assess these criteria, so we fall back to a subjective selection.'' \autocite{sevillaComputeTrendsThree2022}}. This increase cannot be explained by the increasing price-performance ratio of GPUs, which has followed a slower pace, doubling roughly every two to two and a half years \autocite{hobbhahnTrendsMachineLearning2023, hobbhahnTrendsGPUPriceperformance2022}. Instead, the six-month doubling pace seems to be sustained by the expensive use of ever-larger compute clusters with more chips, enabled by increased investment \autocite{cottierTrendsDollarTraining2023}. One consequence of the high demand for compute is scarcity: even companies with multibillion-dollar budgets must wait months or years to have large compute orders fulfilled.

\begin{figure}[!ht]

    \begin{subfigure}{\textwidth}
        \centerline{\includegraphics[width=1.4\linewidth]{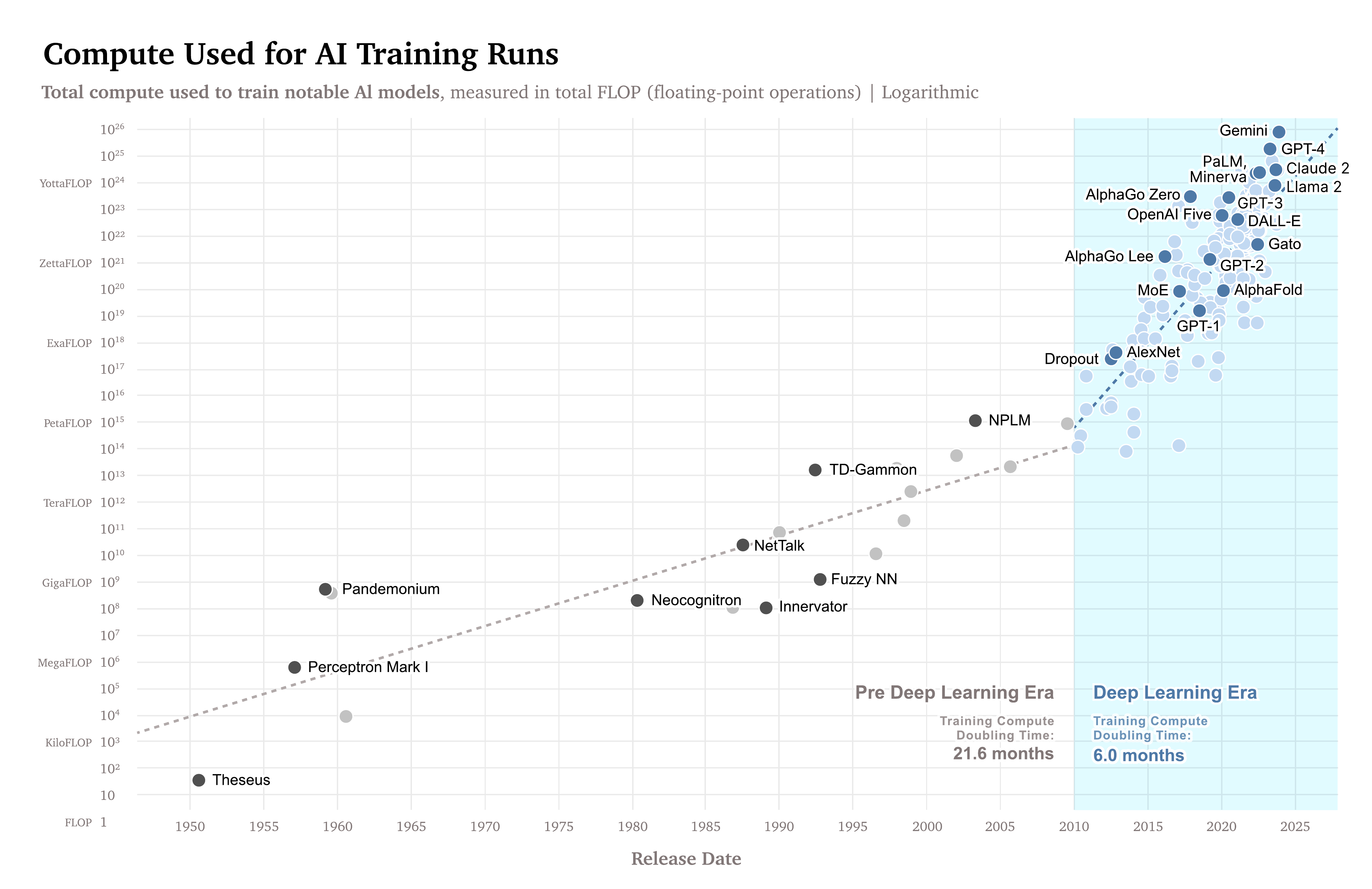}}
        \caption{\textbf{Pre-2010 Trend.} Compute usage for training AI systems before 2010 doubled every 1.8 months. This tracks Moore’s Law-esque improvements in compute price-performance (doubling every two years).}
        \label{fig:compute_train_a}
    \end{subfigure}

    \caption{\textbf{The importance of compute AI in a historical context.} (Data from \textcite{epochAITrends2023, sevillaComputeTrendsThree2022}.)}

\end{figure}

\begin{figure}[!ht]
\ContinuedFloat
    \begin{subfigure}{\textwidth}
        \centerline{\includegraphics[width=1.4\linewidth]{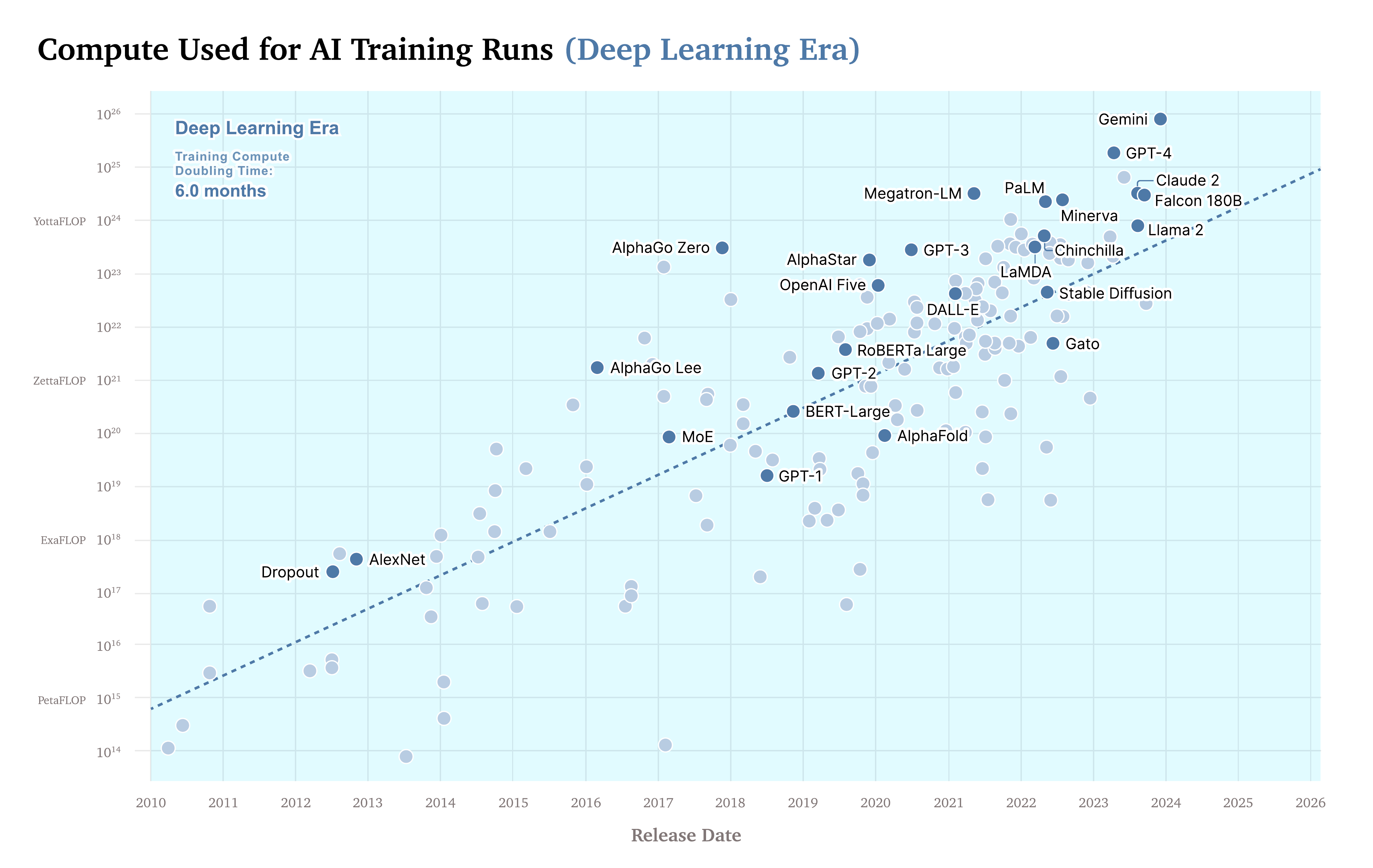}}
        \caption{\textbf{Post-2010 Exponential Growth.} Since 2010, the amount of compute used to train the largest AI models has been growing rapidly, with a doubling time of approximately six months. This shift signifies that the most general and capable models of today tend to be trained with the most compute.}
    \label{fig:compute_train_b}    
    \end{subfigure}
    
    \caption{\textbf{The importance of compute AI in a historical context.} (Data from \textcite{epochAITrends2023, sevillaComputeTrendsThree2022}.)}    \label{fig:compute_train}

\end{figure}

AI developers are not using massive amounts of compute for frivolous reasons: investments in compute have reliably delivered capability improvements \autocite{owenHowPredictableLanguage2024}. In his influential essay ``The Bitter Lesson,'' \autocite{suttonBitterLesson2019} AI pioneer Rich Sutton observed that, historically, AI researchers tried to hand-design knowledge into their systems. This approach led to short-term progress. 
Sutton argued that, since the 1950s and more evidently since 2010, breakthroughs in AI have more often come from an alternative approach that relies on scaling compute with simple algorithms that can effectively use this increased compute. This approach relies on machine learning to ``figure out'' the knowledge that humans had previously been ``hard-coding'' into machines. Furthermore, with more available compute, researchers can also run more experiments to validate algorithmic ideas.

\begin{figure}[!ht]
    \centerline{\includegraphics[width=1.4\linewidth]{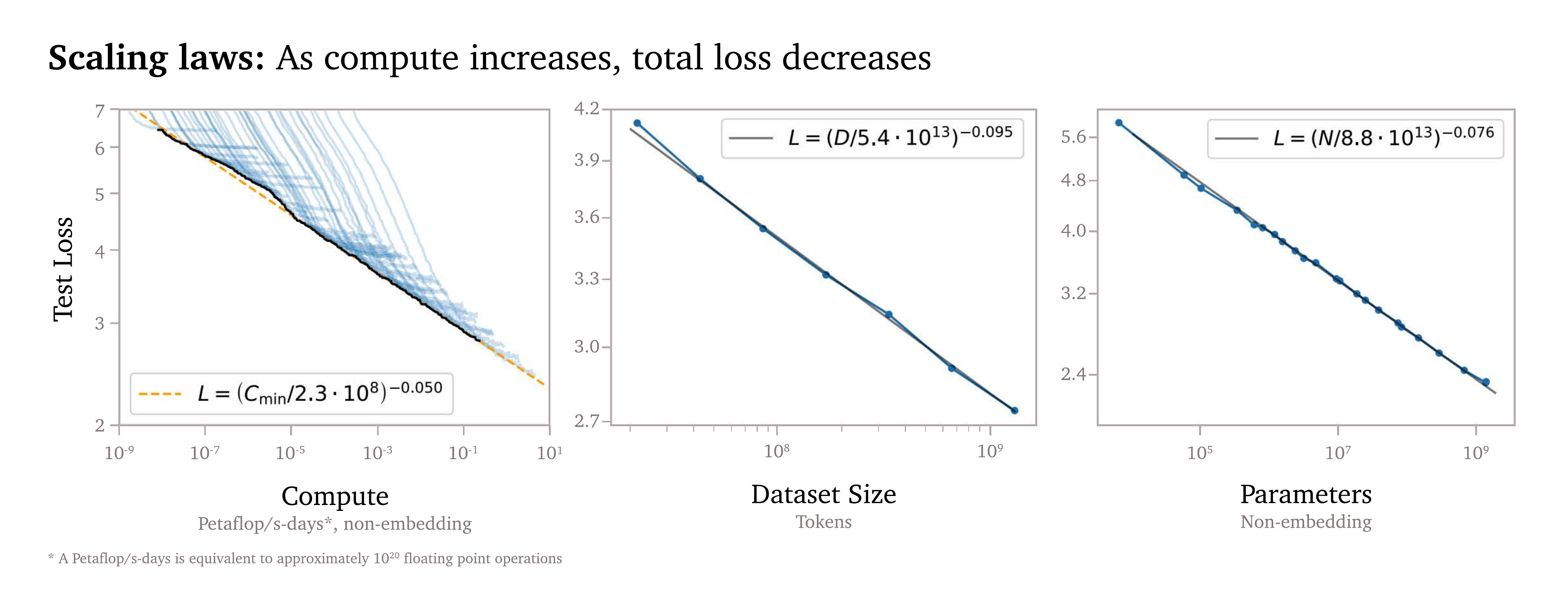}}
    \caption{\textbf{Scaling laws indicate that a fundamental measure of performance decreases as compute, dataset size, and parameters increase.} Reproduced from \textcite{kaplanScalingLawsNeural2020}. Note that subsequent research by \textcite{hoffmannTrainingComputeoptimalLarge2022} has found that the scaling laws in question are differently shaped, though this did not change the general conclusion that there are strong returns to scale.}
    \label{fig:scaling}
\end{figure}

In addition to these anecdotal and qualitative observations of compute-intensive frontier systems, the relationship has been investigated quantitatively through the study of ``scaling laws,'' which describe how the performance of a particular AI model scales with respect to the model’s inputs for a given architecture and algorithm (\Cref{fig:scaling}). The relationship between AI performance and model size, data, and training compute has tended to follow a power law, with fundamental measures of performance\footnote{For language systems, performance is typically measured as the cross-entropy loss on the next-word prediction task.} continuing to improve smoothly as these variables increase. These laws have been instrumental in understanding and predicting performance improvements \autocite{villalobosScalingLawsLiterature2023, kaplanScalingLawsNeural2020, hoffmannTrainingComputeoptimalLarge2022}. However, while scaling laws predict system performance on training objectives, they are not always reliable predictors of performance improvements on individual downstream task performance, which can be sudden and unexpected \autocite{ganguliPredictabilitySurpriseLarge2022, weiEmergentAbilitiesLarge2022}.\footnote{These results have been called into question, noting that the suddenness is partly a result of how performance being assessed with discontinuous measures, such as getting a math question exactly right, without giving points for getting close to the right answer \autocite{schaefferAreEmergentAbilities2023}. However, others have responded that performance on discontinuous measures is crucial for real-world impact and that continuous ``surrogate measures'' meant to predict performance on discontinuous measures are difficult to identify ahead of time \autocite{weiCommonArgumentsRegarding2023}.} Not only are scaling laws a way of quantifying Sutton’s ``Bitter Lesson'', but they also show the importance of algorithmic innovations: better neural network architectures and training algorithms exhibit steeper scaling laws.

Compute is essential not only for training AI models, but also for deploying and operating them. Just as operating expenses outpace initial fixed costs for many large-scale projects, the majority of available AI compute resources are used for operating AI models rather than training them.\footnote{For example, Google estimated that 15\% of its global energy use went toward machine learning workloads, of which 60\% was for inference in 2019, 2020, and 2021 \autocite{pattersonCarbonFootprintMachine2022}. NVIDIA estimated 80\% to 90\%, and AWS estimated that 90\% of its workload is inference \autocite{pattersonCarbonEmissionsLarge2021}. The computational needs for running a single copy of a trained model (inference) are significantly lower than that needed for training it---perhaps only a few dozen chips. However, the majority of computational power for AI systems may not necessarily be used for training runs. Countless everyday actions, such as chatbot interactions (e.g., ChatGPT), Google searches, or inquiries to virtual personal assistants like Siri or Alexa, generate outputs from a model via inference. As consumer AI usage increases, the share of compute used for inference may increase even further.}\label{fn:inference} Frontier AI models are so large that they cannot be efficiently operated at large-scale with household amounts of typical consumer hardware. Instead, for models in high demand, inference requires thousands of AI chips housed in specialized data centers to adequately serve the needs of thousands of users \autocite{pilzComputeScaleBroad2023}. The wider the deployment of AI systems (which requires more compute), the more impact they will likely have (both beneficial and harmful).\footnote{However, there are many caveats to this correlation. The impact could vary significantly based on the application domain and other factors. Some inferences, or even certain users, could pose considerably higher risks than others. Hence, the relationship between deployment compute and the impact of AI systems is not as clear-cut as that observed in the context of training compute and AI capabilities.}

The recent rise of large language models also helps illustrate compute’s centrality to creating and governing frontier AI models. Computing hardware has been the key factor in affecting who is able to build cutting-edge large language models \autocite{bommasaniOpportunitiesRisksFoundation2022, ganguliPredictabilitySurpriseLarge2022, tamkinUnderstandingCapabilitiesLimitations2021}. Google and OpenAI were early investors in large-scale AI training runs, and consequently played a significant role in the early development of language model research and norm development \autocite{devlinBERTPretrainingDeep2019, shevlaneOffensedefenseBalanceScientific2020, segerDemocratisingAIMultiple2023}. Compute has thus become the de facto ``currency'' of access to large language models; many AI companies charge for outputs on a per-token basis, which aims to account for the compute used for inference.\footnote{Tokens from larger models are typically more expensive than tokens from smaller models, reflecting their higher cost to produce and higher quality. However, there are numerous techniques by which more tokens from smaller models can be used to match the performance of fewer tokens from larger models---for example, running many copies of a large language model (LLM) in parallel to generate many candidate options and then choosing between them can improve performance \autocite{jonesScalingScalingLaws2021, villalobosTradingComputeTraining2023}.} Access to compute also influenced the speed with which capabilities diffused throughout the broader AI research ecosystem: the first actors to replicate GPT-3 were relatively ``compute-rich'' actors or had received large grants from such actors \autocite{cottierReplicationEmulationGPT32022}.

\subsection{The Feasibility of Compute Governance}\label{ssec:feasibility_governance}

Several properties of AI compute suggest it can serve as an effective governance instrument. We focus on four: \textit{detectability, excludability, quantifiability,} and \textit{supply chain concentration}.

\subsubsection{Detectability}\label{ssec:detectability}

\begin{figure}[!ht]
    \centering
    \begin{subfigure}{.5\linewidth}
      \centering
      \includegraphics[width=.9\linewidth]{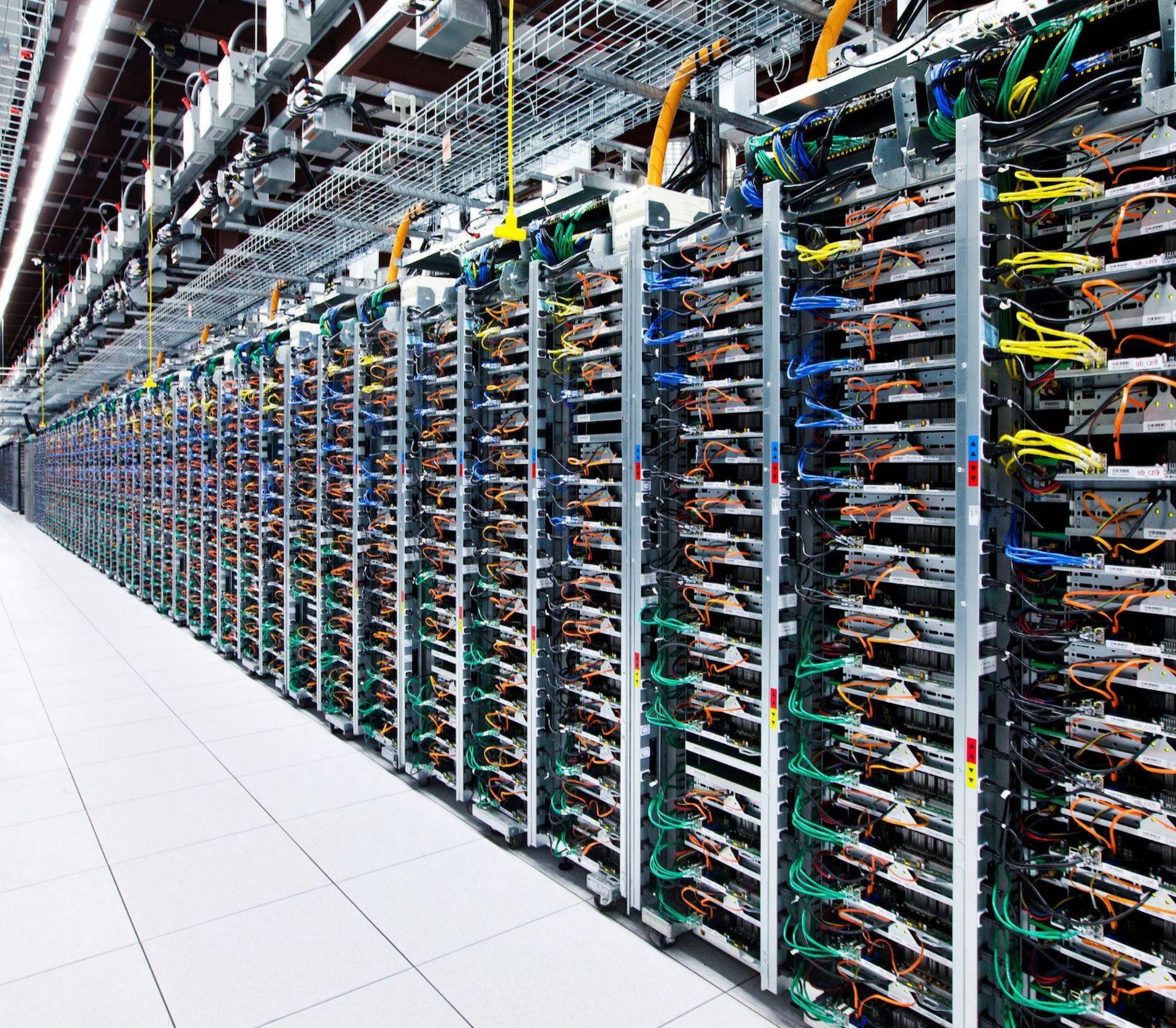}
    \end{subfigure}%
    \begin{subfigure}{.5\linewidth}
      \centering
      \includegraphics[width=.9\linewidth]{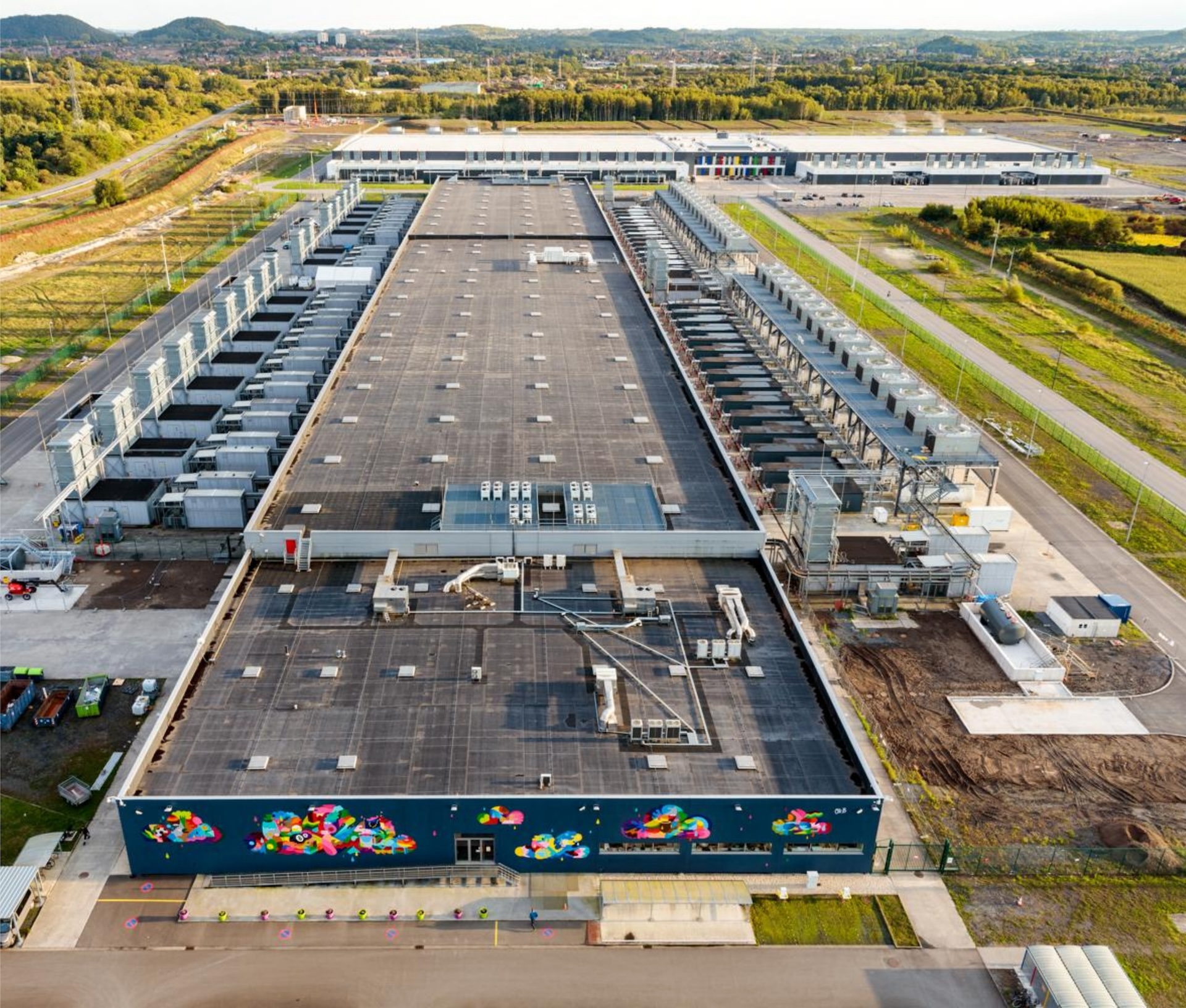}
    \end{subfigure}
    \caption{\textbf{Internal and external views of a data center} (from \textcite{googleDataCentersPhotogallery}.}
    \label{fig:data_centers}
\end{figure}

The physicality and resource intensity of AI supercomputers makes them highly detectable and thereby governable.\footnote{This detectability might be undermined should decentralized training, spread across many data centers and/or using lower-quality compute, become more viable. We discuss this more in \Cref{ssec:limitations}.} AI supercomputers consist of tens of thousands of AI chips connected with high-bandwidth networking equipment and consume up to dozens of megawatts of power---equivalent to tens of thousands of U.S. households.\footnote{For example, AWS recently announced an AI supercomputer consisting of 20,000 H100 chips \autocite{nvidiaAWSNVIDIACollaborate2023}. Given 10.2 kW of power consumption per 8-chip DGX system \autocite{nvidiaNVIDIADGXH100}, this cluster would consume more than 25 MW, even before accounting for networking, storage, and cooling.} They are hosted in large data centers---industrial facilities spanning the equivalent of up to several football fields---that require large-scale cooling and power infrastructure ( \Cref{fig:data_centers}) \autocite{pilzComputeScaleBroad2023}. The construction of such a facility costs up to several billion dollars and involves a complex permitting and power allocation process.\footnote{\textcite{bachHowSmallCity2023} describes a large Microsoft data center in Iowa of the type used to train GPT-4. Like most data centers of hyperscalers, it likely had a power capacity of above 100 MW \autocite{pilzComputeScaleBroad2023}. \textcite{pilzComputeScaleBroad2023} estimate that only around 140 data centers of this size class existed in 2023.} The visibility of supercomputer use has also been used to quantify the climate impact of modern AI systems \autocite{oecdMeasuringEnvironmentalImpacts2022, pattersonCarbonFootprintMachine2022, pattersonCarbonEmissionsLarge2021, hendersonSystematicReportingEnergy2020}.

However, there are also challenges to detecting AI training runs by tracking data centers. While most data centers are likely easy to identify on geospatial imagery, some may be concealed underground\footnote{For instance, see \textcite{steersTop10Underground2022} for a compilation of underground data centers. However, this has yet to be demonstrated for AI supercomputers.} or hidden within other industrial facilities. Furthermore, even successfully detecting AI data centers is not sufficient for identifying AI models hosted on those data centers. This would require the data center owners to monitor and report information about how their computers are used----which would raise privacy concerns----and to distinguish AI workloads from the non-AI workloads also hosted by the majority of data centers.

\subsubsection{Excludability}\label{ssec:excludability}

Compute has a high degree of excludability and rivalry, key attributes of a private good (as distinct from a public good \autocite{samuelsonPureTheoryPublic1954}). Unauthorized users can be easily excluded from accessing AI chips. Someone wishing to use AI chips---i.e., run desired computations on them---must either possess the chips themselves, or (more commonly) rent the right to use the chips from a cloud compute provider that is in possession of the chips themselves \autocite{pilzComputeScaleBroad2023}. In both scenarios, the entity in possession of a chip generally maintains the ability to prevent others from using it.\footnote{Users send instructions to the chip (i.e., directions about which computations to run) via physical networking infrastructure. The person in possession of the chip will naturally have the right and ability to determine and configure the networking connected to that chip, and therefore control the process by which users can send instructions to the chip to make use of it. Crudely, the person in possession of the chip could exclude others by simply disconnecting it from the networking or power supply. Of course, more nuanced, computational methods of control at the network access level are generally used.} While hackers can theoretically gain access to and exploit an actor’s compute, they can easily be expelled once their intrusion is detected.\footnote{Intrusion detection could be achieved by monitoring energy and compute usage.} Therefore, compute can be allocated or withheld from actors or particular use cases.

The excludability and rivalry of compute can perhaps best be understood in contrast to the two other elements of the AI triad: data and algorithms \autocite{buchananAITriadWhat2020}. Both are intangible. Data and algorithms can be kept private prior to publication, but once published it is difficult to control their use \autocite{arrowEconomicsInformationExposition1996}, and they become ``digital public goods'' with low excludability and rivalry \autocite{gruenBuildingPublicGoods2017}. This has been referred to as the ``copy problem'' \autocite{traskPrivacyTradeoffsStructured2020}. Once a paper has been downloaded from the website hosting it, it can be copied and reshared virtually costlessly, even if we remove the original copy from the original host website. By contrast, computing hardware has a finite throughput: if one actor is using some computing power, another actor cannot use that same computing power at the same time.

To prevent the unsanctioned copying of data or ideas, society primarily relies on institutional tools (e.g., intellectual property rights, contracts, criminal law). However, these policies are far from perfectly effective, especially across jurisdictional borders. The ability to \textit{reliably} exclude people from accessing these informational goods is much weaker than for physical goods, as evidenced by, for example, the history of nuclear technology, discussed further in \Cref{sec:appA}.\footnote{In particular, one worry is that rules excluding persons from informational goods in AI will disadvantage law-abiding and/or domestic actors, while law-breaking or foreign actors may be undeterred by laws intended to constrain access to information.} The U.S. government’s unsuccessful attempts in the 1970s to restrict access to the RSA encryption algorithm serve as an apt example of the challenges inherent in governing algorithms.\footnote{Discussed in \textcite{fischerAIPolicyLevers2021}, and \Cref{sec:appA}.} The risk of cybertheft of organizational internal assets increases the difficulty of regulating algorithms. Thus, the control and tracking of AI capabilities by monitoring where certain AI algorithms are used or whether some actor is using a particular algorithm becomes a complex task.\footnote{Nonetheless, we expect that frontier AI organizations will become more reserved about their employed algorithms than they have been in the past. Compare GPT-2 \autocite{radfordLanguageModelsAre2019} with GPT-4 \autocite{openaiGPT4TechnicalReport2023}. This will influence the diffusion of AI algorithms into the research community.}

\subsubsection{Quantifiability}\label{ssec:quantifiability}

The computing power attainable from hardware is also easily quantified. It is generally easier to regulate behavior when it is quantifiable---when we can more precisely measure some activity, it is easier to identify it and promote, limit, or deter it.

Computational resources can be quantified by the quantity and quality of their chips.\footnote{This section discusses quantifiability in terms of computational infrastructure, focusing on metrics related to hardware capabilities such as computational performance. Compute can also be used to quantify AI systems, specifically through the amount of training compute they’ve utilized.
These two forms of quantification serve distinct, yet occasionally intersecting, regulatory purposes. While this second type---quantifying AI systems based on training compute---is a standalone criterion that can be applied to subject these systems to particular regulations (that do not leverage compute), the first type concentrates on the hardware’s capabilities and is important for the governance of compute. Moreover, these two metrics can be employed in a complementary manner. Knowing the specifications of an AI compute cluster allows one to determine whether a particular cluster is capable of training a system with given compute requirements. Additionally, the hardware can be leveraged to verify adherence to these thresholds. While there are measures to quantify training compute, these are not yet fully standardized \autocite{sevillaEstimatingTrainingCompute2022, brundageTrustworthyAIDevelopment2020}.} Most prosaically, chips can be counted. Chips also possess measurable specifications---such as computational performance (in operations/s),\footnote{As previously stated, this paper primarily discusses the metric of ``operations per second'' when evaluating the computational performance of AI chips. This differs from the more commonly cited ``floating point operations per second.'' The focus on ``operations per second'' is intended to provide a more holistic measurement, especially in the context of recent advancements in AI training that increasingly utilize lower precision training methods.} chip-to-chip communication bandwidth, memory capacity,\footnote{The memory is not always part of the chip. However, in the case of cutting-edge chips that leverage high-bandwidth memory, they are part of the same chip packaging.} and memory bandwidth---that indicate quality. Further, training and deploying advanced AI models typically involves tens of thousands of advanced AI chips, requiring significant amounts of ancillary infrastructure---such as high-speed networking and cooling and energy infrastructure---housed in large-scale data centers. This infrastructure can be used to estimate actors’ computational resources, as well as to verify and set clear thresholds on access.\footnote{We expand on these sorts of mechanisms in \Cref{sec:enhance}.}

The quantifiability of compute contrasts strongly with another input to AI progress: human capital or ``talent.'' Individuals are not as transparent as compute or data \autocite{belfieldComputeAntitrustRegulatory2022}. The governance of talent is rightly limited by civil liberties like privacy, and freedom of association and thought (outside of specific and sometimes contentious cases, such as subsidizing research and education and granting or denying student and work visas).\footnote{Note that talent can itself be a means of governance even if it’s not the best target of governance. AI researchers and engineers can choose to limit access to their talent. There have been cases of workplace activism within the AI community, deterring their employers from working on certain projects. A notable instance is the protest by Google employees against Project Maven, leading Google to retract from the defense contract \autocite{belfieldActivismAICommunity2020}. Moreover, a 2019 survey indicated a considerable number of AI researchers would likely participate in workplace activism if asked to work on projects they object to \autocite{zhangEthicsGovernanceArtificial2021}.} Quantifying and comparing talent directly is difficult, making it a less useful indicator of AI capabilities. For example, while some metrics can be predictive of the high productivity of some scientists over others (e.g., h-index, or citation counts), such measures have significant limitations (e.g., they are field dependent or favor older researchers).\footnote{See \textcite{clancyAcademicCitationsMeasure2022} for a discussion of the benefits and limitations of citation counts as a scientific metric.}

\subsubsection{Supply Chain Concentration}\label{ssec:supplychain}

\begin{figure}[!ht]
    \centerline{\includegraphics[width=1.4\linewidth]{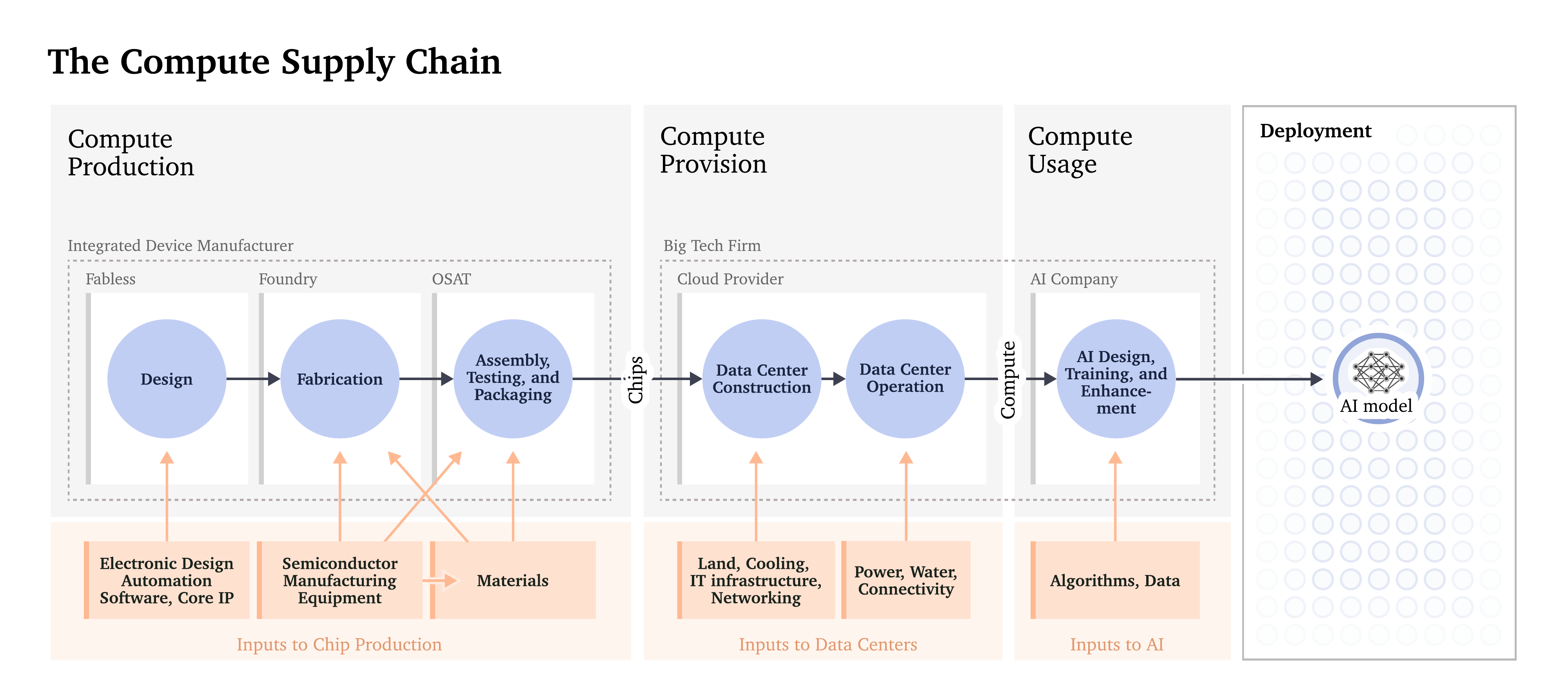}}
    \caption{\textbf{An overview of the AI compute supply chain.} First, chips are produced through a process of design, fabrication, and testing. They are then distributed and accumulated in data centers. Compute users---such as AI developers---can then train and run AI systems from these AI supercomputers.}
    \label{fig:compute_supply_chain}
\end{figure}

\begin{figure}[!ht]
    \centerline{\includegraphics[width=1.4\linewidth]{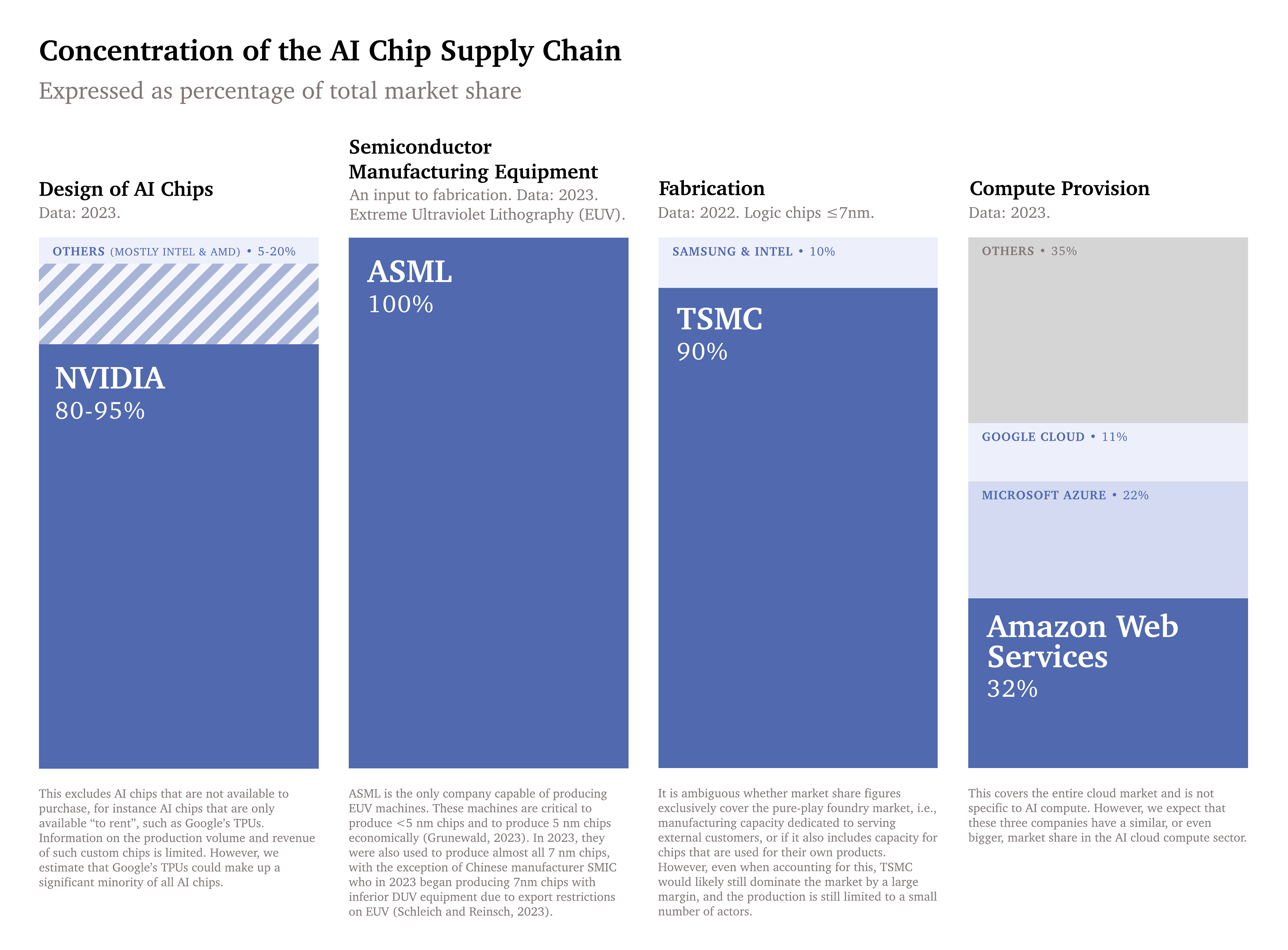}}
    \caption{\textbf{The supply chain for AI chips is highly concentrated.} Several critical steps---including AI chip design and production---have fewer than three suppliers. Even AI development at the frontier consists of only tens of organizations. These facts enhance the governability of the compute supply chain–and how difficult it is to compete at the cutting edge of chip production \autocite{nellisNoBigCustomers2023, morganDatacenterGPUGravy2024, dasSemiconductorMonopolyHow2023, tarasovASMLOnlyCompany2022, grunewaldIntroductionAIChip2023, schleichContextualizingNationalSecurity2023, hwangIndustryWatchSemiconductor2022, richterInfographicAmazonMaintains2023, ofcomCloudServicesMarket2023}. }
    \label{fig:supply_concentration}
\end{figure}

A key factor enhancing the detectability, excludability, and quantifiability of compute is the concentration of the global supply chain for high-end ($\le$ 7 nm)\footnote{In chip manufacturing, each generation of chipmaking technology has a designated ``process node'' or ``technology node,'' measured in nanometers, with smaller nodes being more advanced. Historically, this nomenclature referred to the minimum size of actual features on a chip (smaller features meaning more features could be packed on a chip) \autocite{semiconductorindustryassociationOverallRoadmapTechnology2003}. However, this nomenclature no longer actually corresponds to the physical feature sizes \autocite{mooreBetterWayMeasure2020}, though smaller node sizes continue to correspond to more advanced chipmaking capabilities.} chips. The large majority of the world’s most advanced AI chips are manufactured by a single company (TSMC),\footnote{In 2020, TSMC dominated approximately 90\% of the pure-play foundry market (i.e., manufacturing capacity dedicated to serving external customers) for technology nodes of 10 nm and below, with Samsung accounting for the remaining 10\% \autocite{hwangIndustryWatchSemiconductor2022, hilleTSMCHowTaiwanese2021}. When considering all technology nodes, TSMC and Samsung together constituted 74.3\% of the market share in the pure-play foundry category \autocite{chiaoTop10Foundries2023}.} which is critically reliant on extreme ultraviolet (EUV) lithography machines, also only manufactured by a single company (ASML) \autocite{patelGapsNewChina2023, tarasovASMLOnlyCompany2022}.\footnote{However, some argue that substitutes for EUV could potentially be used to produce high-end chips, though at a significant efficiency penalty \autocite{patelGapsNewChina2023}.} Several other links in the supply chain are also dominated by a few providers, including data center GPU design (where NVIDIA has a market share of over 90\% \autocite{ukcmaNVIDIAArmReport2021, nellisNoBigCustomers2023}), and cloud compute services (dominated by a few large providers \autocite{ofcomCloudServicesMarket2023}). The supply chain is also inelastic, as the entry barriers are high and supply is difficult to increase quickly \autocite{ofcomCloudServicesMarket2023}.\footnote{Competition authorities have been exploring the possibility of increasing competition in these markets \autocite{ukcmaNVIDIAArmMerger2022, ofcomCloudServicesMarket2023, ftcFTCSuesBlock2021}.} This is especially evident for EUV lithography machines, which took multiple decades and billions of dollars in investments to develop \autocite{millerChipWar2022}. These empirical factors regarding the supply chain could change over time, potentially affecting governability, but examples like the U.S. export controls on semiconductor manufacturing equipment (discussed below) illustrate the existing potential for governance today. \Cref{fig:compute_supply_chain} illustrates the compute supply chain, whereas \Cref{fig:supply_concentration} focuses on its concentration. 

It is not feasible to regulate every instance of AI model deployment, nor is it desirable (as discussed in \cref{ssec:limitations}). Today, however, a significant fraction of frontier AI model-related development and deployment compute could be regulated and governed because it is hosted in a relatively small number of data centers housing large numbers of AI chips \autocite{pilzComputeScaleBroad2023}.\footnote{While we expect a large amount of edge compute used for inference (such as inference-optimized chips in smartphones), we do not expect them to be suited for training or executing the most powerful AI models, which require high-bandwidth interconnected compute including high network connectivity to serve users. We also don’t expect that a single actor can control most inference edge compute given the strong decentralized nature of these devices.}

\subsection{Regulating Development Versus Regulating Deployment}\label{ssec:development_vs_depoyment}

The arguments above outline why compute governance is a promising approach to governing the development of AI. However, they don’t establish that it’s \textit{necessary} to reliably prevent major harms from AI. It could be that other approaches to AI governance could achieve similar outcomes---in particular via focusing on the \textit{deployment} of AI systems. In most other sectors, regulation focuses on restricting harmful use of products, e.g., by restricting the sale of products that fail to meet specifications, or by holding manufacturers liable for harms caused by their products.

We expect that regulation of AI deployment will be a part of any frontier AI regulatory regime. However, we argue that without regulations on the development of AI, regulation on AI deployment would not be adequate to protect against the most severe risks from AI, due to (at least) two key shortcomings \autocite{heimCasePreemptiveAuthorizations2023, anderljungFrontierAIRegulation2023, koltAlgorithmicBlackSwans2023, mathenyHereSimpleWay2023}.

First, it will be very difficult to identify all relevant deployments of any given model with high reliability. Individual copies of a model can be run using a relatively small amount of compute, making it extremely difficult to detect which computers they’re being run on. Copies can also easily be distributed to many different actors---for example, via sharing the weights online. Even models whose weights aren’t released publicly, such as GPT-4, could be stolen via hacking or insider espionage, then deployed by the attackers \autocite{jeffreyInformationSecurityConsiderations2022, cottierUnderstandingDiffusionLarge2022, nevoSecuringArtificialIntelligence2023, anderljungFrontierAIRegulation2023}. Those attackers may be criminal enterprises or state adversaries, who are difficult to monitor and who would be less constrained by legal penalties placed on them \autocite{anderljungProtectingSocietyAI2023, anderljungFrontierAIRegulation2023}.

Second, some models may pose risks that are disproportionate to the scale or sensitivity of the tasks for which they’re deployed. Regulators could aim to detect only particularly sensitive deployments of models, like models that are given access to critical infrastructure; or they could target particularly wide deployments of models. But if the effects of a model’s actions ripple beyond its immediate deployment environment, then they still may pose large-scale risks. For example, in the context of biosecurity, if a model is used to design novel pathogens, those designs could easily be shared very widely \autocite{berkeCanQuiteDevelop2023}. Similarly, models highly capable at understanding computer systems might be used to design highly-sophisticated computer viruses that proliferate across the internet \autocite{ukdepartmentforscienceinnovation&technologyFrontierAICapabilities2023}. Future models may also develop the capability to autonomously pursue unintended goals \autocite{berglundTakenOutContext2023, shevlaneModelEvaluationExtreme2023, ngoAlignmentProblemDeep2023, cotraSpecificCountermeasuresEasiest2022}. These capabilities might allow a model to spread itself like a computer worm, hacking and spreading through networks and causing severe disruption \autocite{carlsmithPowerseekingAIExistential2022}. Risks like these could arise even when models are deployed internally within AI companies, without any external deployments; indeed, they may be more severe in those cases, since if security precautions are not taken, it could be easier for internally-deployed models to access private code and data (including their own weights).

Regulations on the deployment of frontier models must therefore be supplemented by regulation of the development of those models \autocite{anderljungFrontierAIRegulation2023}. One method of detecting and monitoring development would involve tracking the inputs necessary for this process; for the reasons given above, compute is likely the most feasible such input.\footnote{Governance of inputs to a technology is already done in cases where the consequences of misuse or accident are severe. For instance, the Chemical Weapons Convention regulates the production, use, and stockpiling of specific chemicals (and precursors thereof) that can be used to create chemical weapons \autocite{opcwChemicalWeaponsConvention2023}. For similar reasons, access to and sale of nuclear materials is regulated. Misusable AI systems, by analogy, can exploit vast attack surfaces, result in extreme and widespread harms, and be difficult or impossible to reverse thereafter \autocite{anderljungProtectingSocietyAI2023}.} An ``upstream'' approach can provide more assurance than governance focused solely on AI systems and applications themselves. It also allows us to ensure that sufficient beneficial and defensive applications of AI are produced, by steering inputs toward such applications \autocite{koltAlgorithmicBlackSwans2023}, as discussed further in \Cref{ssec:allocation}.

\section[Compute Can Enhance Three AI Governance Capacities]{Compute Can Enhance Three AI \\Governance Capacities}\label{sec:enhance}

The arguments in \Cref{sec:policymaking} give us reasons to further explore governing AI via compute.

In this section, we argue that compute can be used to improve society’s capacity to govern AI in at least three key ways:\footnote{One can view each area as a governance ``capacity'' that contributes to effective governance. This is analogous to the concept of state capacity \autocite{lucianaStateStateCapacity2013}.} increasing the \textit{visibility} of AI to policymakers, \textit{allocating} AI capabilities, and enhancing \textit{enforcement} of norms and laws. We provide illustrative examples of how these capacities can be used for AI governance.\footnote{We are not aware of a standard such taxonomy, though there is related work. Here we use a bespoke taxonomy, which we arrived at via trial and error in organizing several compute-related policy mechanisms. We were particularly inspired by Elinor Ostrom’s work on commons management in emphasizing visibility and enforcement \autocite{ostromGoverningCommonsEvolution2015}, and by the idea of differential technological development \autocite{sandbrinkDifferentialTechnologyDevelopment2022} in emphasizing the importance of allocation. 
We also recognize that these categories overlap and interact with each other. For example, withholding compute from an actor that violates norms could be seen as using the allocation capacity to enhance enforcement. Similarly, the visibility capacity can help regulators detect whether allocation goals are being achieved, or where possible enforcement might be warranted.
}

\textit{Visibility} refers to the ability to understand how actors use, develop, and deploy AI, and which actors are most relevant to frontier AI model development and (to a lesser extent\footnote{This is because, while inference costs across all users are generally many multiples of training costs, an individual user may be able to perform large amounts of inference using much less compute than required for training (see \cref{ssec:creating_capabilities}). Furthermore, it is often possible to compress, distill, or otherwise optimize large models so that they can run on a wider variety of hardware than would be suitable for training \autocite{appleDeployingTransformersApple2022, cuencaFasterStableDiffusion2023}. Computationally cheap post-training interventions can also meaningfully change model behavior, including by making it less safe \autocite{gadeBadLlamaCheaplyRemoving2023}. Thus, compute governance will be less effective at governing small deployments, especially when model weights are readily available (e.g., due to model release) \autocite{anderljungFrontierAIRegulation2023, segerDemocratisingAIMultiple2023}. Nevertheless, compute governance can still play an important role in detecting which individual actors have and/or use the largest inference capacities, which may correlate with various risks and opportunities, as discussed in \Cref{ssec:feasibility_governance}.}) deployment. This visibility is crucial: it allows policymakers to anticipate problems, make more accurate decisions, track key outcomes within a country, and negotiate and implement agreements between countries---e.g., new international institutions for governing AI \autocite{hoInternationalInstitutionsAdvanced2023}, treaties, or more informal confidence-building measures \autocite{shokerConfidencebuildingMeasuresArtificial2023}.

\textit{Allocation} refers to the ability to direct and influence the trajectory of AI development by changing the distribution of AI capabilities among different actors and projects. For example, a government may want to steer AI development (e.g., to correct for market failures) toward beneficial and defensive uses, disincentivizing harmful and malicious ones, increasing the fraction of public interest-oriented AI development, or expanding access to AI capabilities.

\textit{Enforcement} refers to the ability to respond to violations of norms or laws related to AI, such as  reckless development and deployment that violates established safety requirements, or deliberately malicious uses of the technology. In the context of AI governance broadly, enforcement can occur through mechanisms like the legal system, informal social norms, industry self-regulation, or other procedures.

In each area, taking compute seriously can open up new policy options. To illustrate this, we discuss several policy ideas in each of the three categories. These ideas are brief and exploratory; more analysis will be needed to gain confidence that they are feasible and desirable, and to understand how they might interact with each other. Here we focus primarily on what is possible; we revisit the question of desirability in \Cref{sec:risks}.

\begin{figure}[ht]
    \centerline{\includegraphics[width=1.4\linewidth]{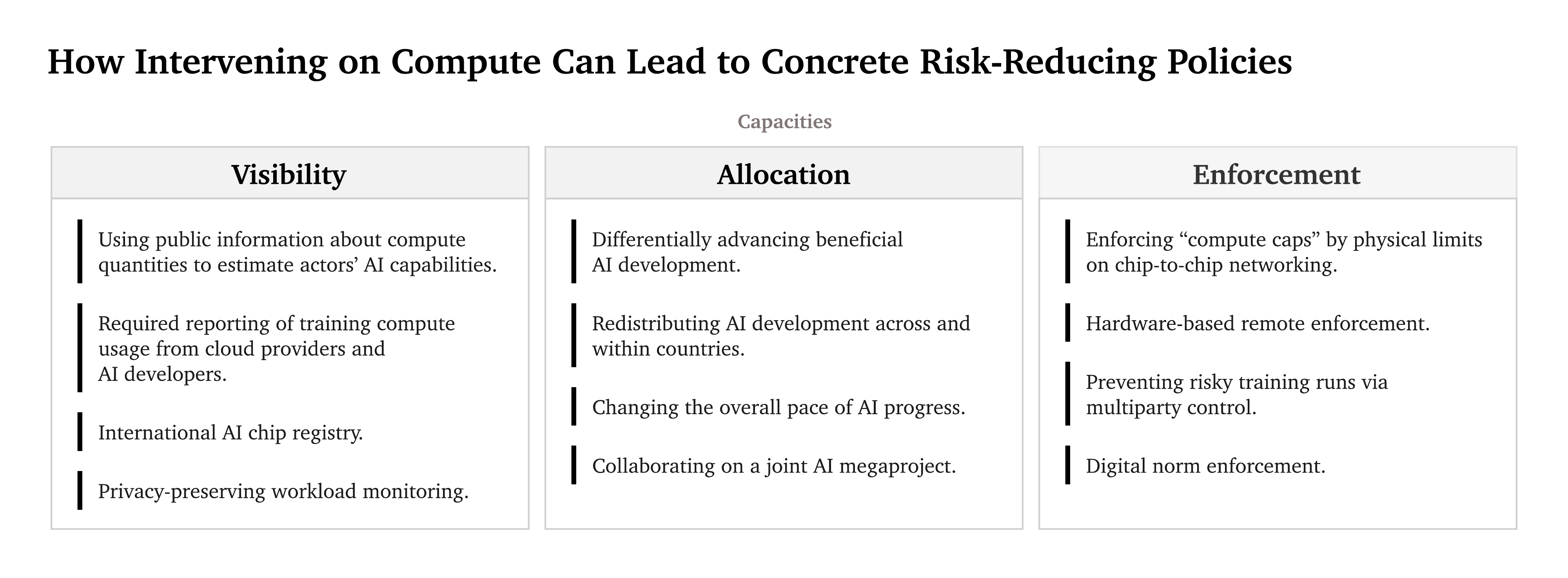}}
    \caption{Examples of how intervening on compute can lead to concrete risk-reducing policies in the areas of visibility, allocation, and enforcement.}
    \label{fig:intervening}
\end{figure}

\subsection{Visibility}\label{ssec:visibility}

If effective and proactive governance of advanced AI is to be achieved, policymakers must have a means of reliably identifying actors developing and deploying advanced AI systems. They must also be able to measure the properties of those systems themselves. Suppose that there is a law mandating safety measures for training frontier AI systems \autocite{anderljungFrontierAIRegulation2023}. If a firm violates that law, then the application of legal penalties is only possible if the legal system knows that the violation occurred. Similarly, to forecast AI advancements, policymakers need insight into the trajectory of AI capabilities---akin to how the Intergovernmental Panel on Climate Change (IPCC) forecasts climate scenarios.

On the global stage, visibility is also crucial. Successful international agreements, like arms control and nonproliferation treaties, often depend on transparent signaling and verification of compliance, a process laden with social and technical intricacies \autocite{gallagherPoliticsVerification1999}.\footnote{Analogously, this transparent signaling problem is also a key struggle with governing lethal autonomous weapons systems (LAWS): how do you verify autonomy if, from the outside, it behaves and appears identical to a non-autonomous system \autocite{horowitzIntroductionAutonomyWeapon2015}?} The more effectively a state can convey this information, the more feasible these agreements become.\footnote{Including, for example, selectively conveying relevant information without necessarily sharing other, possibly sensitive, information. The question of whether information sharing is beneficial is heavily contextual (see \textcite{emery-xuUncertaintyInformationRisk2023} on knife-edge results).} Compute governance can offer policymakers additional tools to enhance regulatory visibility across these different contexts.

In this section, we explore four policy mechanisms that leverage compute to increase regulatory visibility:
\begin{enumerate}
    \item Using public information about compute quantities to estimate actors’ AI capabilities (now and in the future)
    \item Required reporting of large-scale training compute usage from cloud providers and AI developers
    \item International AI chip registry
    \item Privacy-preserving workload monitoring
\end{enumerate}

All attempts to create greater visibility will face common risks, especially if they rely on nonpublic information. In particular, we highlight the risk that visibility efforts will violate individuals’ privacy or threaten the security of strategically sensitive information. We discuss these risks further in \Cref{ssec:limitations}, and possible approaches for mitigating them in \Cref{ssec:guardrails}.

\textbf{\textit{Using public information about compute quantities to estimate actors’ AI capabilities (now and in the future)}} \\\\
Governments who wish to identify the set of actors that could build the most capable general-purpose AI systems can first look to existing reporting and open source intelligence about compute. Building cutting-edge models requires enormous supercomputers that house large numbers of specialized chips. Because of this, while publications from AI companies often do not reveal the exact amount of compute used in a particular advance, it is usually possible to get a rough sense of which actors are compute-rich.\footnote{For example, researchers outside the AI industry have made estimates of the compute usage of notable AI systems \autocite{epochAITrends2023}.} For example, debates on the EU AI Act have noted that it is mainly U.S. rather than EU companies that are compute-rich \autocite{futureoflifeinstituteEmergingNoneuropeanMonopolies2022}. A similar dynamic has been observed in academia and industry, described as the ``compute divide'' \autocite{ahmedDedemocratizationAIDeep2020, besirogluComputeDivideMachine2024}.

While leveraging open source intelligence to identify frontier AI developers can help reduce uncertainty to an extent, this coarse-grained method is insufficient on its own for high degrees of visibility. A method with more visibility for the U.S. was introduced in October 2023 by Executive Order 14110 that now requires companies that ``acquire, develop, or possess a potential large-scale computing cluster'' to report ``the existence and location of these clusters and the amount of total computing power available in each cluster'' to the U.S. government \autocite{thewhitehouseExecutiveOrderSafe2023}.

On the global level, information about compute infrastructure can also be used to estimate different states’ AI capabilities. Given the size and energy requirements of data centers required to house AI supercomputers, geospatial intelligence\footnote{These methods can also be supplanted with classified capabilities to attain visibility over significant data center construction unaccounted for in open source reporting.} could also be used to evaluate countries’ potential AI capabilities and their compliance with future international agreements.\footnote{However, note that the majority of large data centers host general-purpose hardware rather than AI supercomputers. Geospatial intelligence may thus be insufficient to verify the use or non-use of AI. Additionally, motivated actors could conceivably implement countermeasures to evade detection, such as by hiding data centers underground, though doing so would likely significantly increase cost.}

Because compute is an important indicator of novel and general-purpose AI capabilities, policymakers can also leverage compute information to improve foresight and forecasts about which actors will be relevant and what AI capabilities might exist in coming years. This can then help them anticipate and preempt future risks. One can make initial forecasts about future progress by leveraging scaling laws; trends in compute growth, allocation, and efficiency; and trends in algorithmic progress and growth of AI talent \autocite{besirogluProjectingComputeTrends2022}. One example of this methodology attempts to estimate the number of operations required to train an AI model that is capable of more cost-effectively performing most human-level intellectual labor \autocite{cotraDraftReportAI2020, davidsonWhatComputecentricFramework2023, barnettDirectApproach2023}.
\newpage
{\textbf{\textit{Required reporting of large-scale training compute usage from cloud providers and AI developers}}\label{bit:report_training}} \\\\
Knowing the geographic location and ownership of large concentrations of AI chips can only tell a regulator so much about the usage of that compute. Most data centers outside of China that can train large AI models are concentrated in the hands of a few large cloud providers---primarily Amazon Web Services (AWS), Microsoft Azure, and Google Cloud \autocite{pilzComputeScaleBroad2023, belfieldComputeAntitrustRegulatory2022}. Yet the use of these data centers is largely rented out to paying customers. Most AI development occurs on rented chips accessed remotely ``in the cloud''. Requiring compute providers to institute ``Know Your Customer'' (KYC) requirements and report large compute usage to regulators\footnote{This KYC practice is required in Executive Order 14110 for foreign users of cloud compute~\autocite{thewhitehouseExecutiveOrderSafe2023}.} can complement knowledge of the total quantities and ownership of compute \autocite{eganOversightFrontierAI2023}. Accurate compute usage data can also help to evaluate the environmental impact of energy-intensive AI training and deployment processes. Reporting practices could assist in balancing these environmental costs against the broader benefits, guiding more sustainable AI development \autocite{oecdMeasuringEnvironmentalImpacts2022, hendersonSystematicReportingEnergy2020, luccioniEstimatingCarbonFootprint2022, pattersonCarbonFootprintMachine2022}.

Along with other mechanisms, required reporting can also serve as a foundation for post-incident liability and incident response (see \Cref{ssec:allocation} below). If model outputs can be attributed to a model, then regulators could work with compute providers to immediately shut down the offending system and identify who was responsible for deploying the model. Governance practices similar to this are common. For example, the hosts of malicious websites, such as ones where illegal drugs are sold, often remain anonymous, and the best available governance intervention is to shut down the servers hosting these websites. Access and close contact with the host---similar to the role of the compute providers we are discussing here---can help with prompt action. Strong procedural guardrails will also be needed to ensure that states use incident response powers in the public interest.

Policymakers have recently discussed reporting requirements for AI developers as well. For example, Executive Order 14110 uses training compute thresholds to trigger additional scrutiny on a potentially risky training run.\footnote{It places three broad requirements on AI companies: to notify the government before a frontier training run, to report large data centers and large foreign cloud computing jobs, and to share the results of safety tests.} If reporting mechanisms could eventually be made trustworthy (e.g., with strong information security and accurate information) and paired with other mechanisms such as external auditing, then a regulator could gain assurance that no excessively risky frontier AI systems are being developed. As a risk-reducing policy, reporting compute usage critically relies on compute usage as a proxy for risk. But, as we discuss in \Cref{ssec:importance_compute_frontier}, compute usage is a good high-level proxy for risk for general-purpose frontier AI systems, but not necessarily for some narrow AI capabilities. We discuss more limitations of compute thresholds in \Cref{sec:risks}.

We note that this information would likely be both strategically and commercially sensitive. A regulator aggregating such information would have great insight into the state of frontier capabilities and their attendant commercial and national security opportunities and risks. This information would be a uniquely attractive target for commercial and sovereign espionage. Even the migration of individual staff across boundaries between the regulator and competing firms, or across national boundaries, could have substantial competitive and security consequences. This is especially true to the extent that reported information might provide insight into how to advance capabilities that firms or countries might be able to rediscover more quickly than they could independently develop. Thus, required reporting could inadvertently undermine the very objectives it aims to achieve through regulatory mechanisms.

{\textbf{\textit{International AI chip registry}}\label{bit:chip_registry}} \\\\
Another option to increase visibility would be to track the flow and stock of new cutting-edge AI chips destined for AI supercomputers. Policymakers could require chip producers, sellers, and resellers to report transfers of AI chips. These transfers could be registered in a ledger, which could then be audited to detect and assign liability for diversion \autocite{fistPreventingAIChip2023, shavitWhatDoesIt2023}. Because of the concentrated supply chain previously discussed, this has the potential to provide policymakers with precise information on the amount of compute possessed by various actors, enabling governance plans that require knowledge of compute flow.\footnote{See \textcite{fistChineseFirmsAre2023} for a discussion of this idea in the context of current U.S. export controls and \textcite{thadaniMappingSemiconductorSupply2023} for an overview of the supply chain of semiconductors across geographies by sales.}

Implementing an international AI chip registry could involve cooperation from players in the AI chip value chain. Semiconductor fabs, assembly and test firms, and end users (especially cloud providers) could track these chips to ensure a chain of custody and a secure supply chain without diversion or smuggling. A physical unique identifier could be added to each AI chip during production. How exactly to cost-effectively add a unique identifier while retaining chip integrity is an open question, but there exist a variety of ideas worthy of exploration.\footnote{For example, one preliminary idea is that of a physically unclonable function \autocite{maesPhysicallyUnclonableFunctions2013}, which is a method of uniquely fingerprinting a physical device. This can help provide resistance against tampering attempts. Less costly mechanisms could include procedures used in export control compliance, such as end-user checks to verify that chips have not been diverted from their last reported user \autocite{shavitWhatDoesIt2023, kurlandEnduseMonitoringOverview2017}.} Given the enormous difficulty of manufacturing AI chips, it would also be difficult for someone to build a fab to manufacture untraced ``ghost chips'' anywhere near the state of the art.\footnote{This would be analogous to the problem of ``ghost guns'': privately manufactured firearms that lack a serial number and are therefore less traceable \autocite{thrushGhostGunsFirearm2021}. One might still worry that unscrupulous fabs might not properly register all of their output. While in practice it seems unlikely that fabs would underreport their output to manufacture ``ghost chips,'' there are mechanisms to detect such underreporting. For example, one could install in-line instrumentation on manufacturing equipment or scrutinize procurement activities for undeclared purchases of chip manufacturing materials \autocite{bakerNuclearArmsControl2023}.}

An international effort to track AI chips would be a significant expansion of governments’ visibility into computational activities. Before committing to such an effort, it is well worth worrying about how such an effort could be misused. For example, what privacy interests could such tracking infringe upon? How could corrupt or oppressive policymakers misuse this information? To what extent could small-scale consumers be exempted, and scrutiny focused only on large operators? Before establishing such a registry, these questions would have to be answered and weighed against possible benefits.\footnote{We list some of these limitations in \Cref{sec:risks}, and particular research directions in \Cref{sec:appB}.}

On the other hand, governments already track the cross-border movements of people and many economic transactions.\footnote{For example, ``Each person engaged in a trade or business who, in the course of that trade or business, receives more than \$10,000 in cash in one transaction or in two or more related transactions, must file [IRS] Form 8300'' \autocite{irsInstructionsForm83002023}.} It may be possible to limit chip tracking requirements to specific chips (or volumes of chips) where individuals’ privacy interests are less present, while still retaining the possible benefits of chip-tracking. See \Cref{sec:risks} for more discussion of the risks and possible mitigations.

{\textbf{\textit{Privacy-preserving workload monitoring}}\label{bit:workload_monitoring}} \\\\
If regulators can understand where large-scale compute is located and who is using it, is it possible to understand what the compute is being used for? In principle, this information is encoded in the workloads that are run by AI supercomputers. In practice, these workloads are not always legible: a chip, for example, only sees a sequence of extremely low-level instructions. Furthermore, these workloads are very important to their users, and may contain private or sensitive information. Therefore, naive approaches to workload monitoring could not only be impractical but also potentially disastrous, posing serious risks to privacy and human rights.

However, there may be methods that offer noninvasive insights into what compute is being used for. Data center operators naturally possess information about the volume of compute used by their customers,\footnote{For example, customers are often billed by the chip-hour, so cloud providers need to track that information for accurate billing.} which can rule out the development of some systems.\footnote{Such as frontier models or other high-compute systems.} Other insights could be derived from both individual AI chip data and aggregated metrics from the entire AI compute cluster. For example, the training and inference phases have different computational signatures, and observations about the computing cluster and the network communication patterns could help to distinguish between them.\footnote{For example, clusters used for inference require constant internet traffic to serve customers, whereas clusters used for training typically access training data hosted locally~\autocite{heimAccessingControlledAI2023}.} 

Other technical changes could provide greater privacy-preserving transparency into AI workloads. Cryptographic mechanisms on AI chips could allow AI developers to securely log their workloads, which they could subsequently present to inspectors to attest their workloads \autocite{shavitWhatDoesIt2023}. Such logging could be made more difficult to spoof by adding cryptographic mechanisms on chips \autocite{sommerhalderHardwareSecurityModule2023, sabtTrustedExecutionEnvironment2015}. Additionally, techniques like ``proof-of-learning'' \autocite{jiaProofoflearningDefinitionsPractice2021} could allow developers to precisely account for the quantity of compute used in a training run. Regulators could then require developers to link these proofs, reflecting the amount of compute used, with the specific data center where the work was carried out. Such a process would allow regulators to more accurately monitor and verify the usage of a data center's compute resources. It also provides a clearer distinction between the compute resources used for training purposes and those that were not.

Privacy-preserving workload monitoring is an example of using privacy-preserving practices and technologies as a part of compute governance. In the future, these practices and technologies could equip regulators with visibility and oversight capabilities while also preserving the strategic and commercial interests of AI developers. As an analogy, privacy-preserving practices and technologies have been an important part of nuclear weapons agreements \autocite{negusPrivacypreservingComputationNuclear2021}. Further technical and policy research in this area for AI could be extremely valuable.\footnote{We discuss this category in \Cref{ssec:guardrails}.}

Transparency into AI workloads could have important implications at an international level. If large compute investments are made without sufficient transparency about how that compute is used, fear and suspicion could drive growing investments by competing countries. A historical example of a similar dynamic is the ``missile gap'' controversy of the Cold War, where erroneous estimates of Soviet missile capabilities resulted in dangerous political pressure to strengthen the U.S.’s missile program in response \autocite{lickliderMissileGapControversy1970, belfieldWhyPolicyMakers2022}. Increasing the transparency and verifiability of compute usage can significantly alleviate information asymmetries and competitive race dynamics \autocite{shavitWhatDoesIt2023, eganOversightFrontierAI2023}, though in certain specific cases it could instead increase race dynamics \autocite{emery-xuUncertaintyInformationRisk2023}.\footnote{Such techniques are somewhat analogous to ``information barriers'' in the domain of nuclear verification, where one might provide enough information to confirm that a warhead has the properties claimed, but without revealing further information. Moreover, clever combinations of compute tracking, APIs, inspections, researcher interviews, and other means could help navigate the transparency-security trade-off often found in arms control contexts \autocite{coeWhyArmsControl2020}.} While many of the technical mechanisms to enable such verifiable information-sharing are nascent, greater research and investment could help increase visibility into AI capabilities, development, and deployment, and thus make strong international agreements on AI viable.\footnote{For example, Kissinger \& Allison \autocite{kissingerPathAIArms2023}, argue that AI is digital (and therefore extremely hard to control in an arms control context): ``Second, AI is digital. Nuclear weapons were difficult to produce, requiring a complex infrastructure to accomplish everything from enriching uranium to designing nuclear weapons. The products were physical objects and thus countable. Where it was feasible to verify what the adversary was doing, constraints emerged. AI represents a distinctly different challenge. Its major evolutions occur in the minds of human beings. Its applicability evolves in laboratories, and its deployment is difficult to observe. Nuclear weapons are tangible; the essence of artificial intelligence is conceptual.''}

Poorly scoped or insecure AI workload monitoring proposals could, however, threaten personal privacy or the security of commercially sensitive information. We discuss these risks further in \Cref{ssec:limitations}, and possible approaches for mitigating them in \Cref{ssec:guardrails}.

\subsection{Allocation}\label{ssec:allocation}

Policymakers have preferences over how AI is developed and deployed. They must then decide how to advance these preferences. If policymakers can identify actors that are more or less likely to use AI in preferred or dispreferred ways, they could promote preferred uses of AI and decelerate dispreferred uses by changing the allocation of compute among actors and projects. We call this method of steering AI progress ``\textit{allocation}.''\footnote{The unique features of compute mentioned in \Cref{ssec:importance_compute_frontier} and \Cref{ssec:feasibility_governance} make allocation via compute more feasible than allocation by data or algorithms. However, there is intense economic debate about the merits of advancing allocative goals through cash transfers versus in-kind transfers \autocite{gentiliniCashFoodTransfers2007, gentiliniWhyDoesInkind2023}, with many economists believing there are good theoretical reasons to favor cash transfers over non-cash methods of redistribution \autocite{kaplowWhyLegalSystem1994}.} Perhaps the paradigmatic examples of allocation today are major government investments in domestic AI supercomputing capacity (\Cref{ssec:governance_today}) and the allocation of government-owned supercomputers to users, as in the NAIRR (\Cref{ssec:ai_governance}). We identify several existing and proposed examples of steering AI progress via allocation:
\begin{enumerate}
    \item Differentially advancing beneficial AI development
    \item Redistributing AI development and deployment across and within countries
    \item Changing the overall pace of AI progress
    \item Collaborating on a joint AI megaproject
\end{enumerate}

\textbf{\textit{Differentially advancing beneficial AI development}}\\\\
As a general-purpose class of technologies, AI can be applied for both socially beneficial and socially detrimental purposes \autocite{brundageMaliciousUseArtificial2018}.\footnote{This is true of both AI technologies as a class (i.e., some particular AI systems are overall beneficial while others are overall detrimental), and many individual AI systems (i.e., the same individual AI systems can be used for both beneficial and detrimental purposes). The question of \textit{who} receives these benefits (or harms) is also critical, as they are unequally distributed, and political actors may wish to affect these distributions. A full review of the beneficial and detrimental applications of AI is beyond the scope of this paper.} Policymakers seeking to maximize social welfare may therefore wish to intentionally increase the amount of resources available to beneficial forms of AI research and development---for example, applications to climate, agriculture, energy, public health, or education.\footnote{To be sure, default market incentives will often create enormous surpluses for consumers and third parties. However, (1) these market activities may carry negative externalities that should be minimized, mitigated, or internalized, and (2) market activities will fail to adequately value public goods and some other types of goods. Intentional efforts to correct these market failures may therefore be warranted, and subsidization of compute for the provision of goods undersupplied by the market is one way to accomplish this.} Compute is one such resource, and one that is especially critical to frontier AI models (as discussed in \Cref{ssec:importance_compute_frontier}).\footnote{There is, of course, a long history in computer science of making data (for example, \textcite{usgovDataGov}) and algorithms freely available for use by a wide variety of actors, such as through open licensing frameworks \autocite{osiOpenSourceDefinition2006, definitionOpenDefinition}. This has undoubtedly enabled a large number of beneficial applications in AI and other forms of computing. There is also a more recent trend of creating, curating, and/or publishing datasets specifically to study and address important social issues \autocite{sefalaConstructingVisualDataset2021, microsoftAIEarthData} and software licenses that specifically disallow unethical uses of the licensed technology \autocite{oesWhatWeBelieve}. Because compute is rivalrous, open access to compute (if even a coherent concept) would not be an optimal way to ensure that beneficial AI projects receive adequate computing support. At present, nonprofit and academic projects often struggle to secure enough computing resources when bidding against well-resourced actors for the limited supply of compute \autocite{nairrNationalArtificialIntelligence2023}.\label{fn:free_data}}

Initiatives to increase compute access to pro-social actors are already underway. This includes governmental,\footnote{See \Cref{fn:free_data}, \textcite{ukgovernmentBristolSetHost2023, eurohpcEuroHPC}.} nonprofit \autocite{hofvarpnirstudiosBuildingComputeAI}, and corporate social responsibility \autocite{ortizGoogleCloudResearch2021} efforts to increase compute access to actors who cannot afford it at market rates in the volume they require for development and deployment purposes.

While broad efforts to increase nonprofit actors’ access to compute are laudable,\footnote{However, if not accompanied by proper oversight, these efforts could carry the same risks as AI development in general. While we think that the overwhelming majority of nonprofit and academic actors are likely to use subsidized compute access to prioritize the provision of public goods and socially beneficial technologies, experience in other domains has shown that poorly overseen scientific funding for nonprofit actors can subsidize unjustifiably risky or unethical research \autocite{esveltManipulatingVirusesRisking2021}. Subsidized compute for less-resourced actors must therefore still be subject to oversight and other forms of governance.} more targeted interventions may be even more effective if the goal is to incentivize the development of particular technologies. One way to achieve this is through ``differential technological development,'' a principle that calls for relevant actors to intervene in the types of technologies developed and their relative timing \autocite{sandbrinkDifferentialTechnologyDevelopment2022}. A core idea of differential technological development is that risks from new technologies can be lessened by prioritizing the development of risk-reducing technologies. Policymakers can use compute allocation to accelerate the development of technologies that reduce societal risks, including those from AI.\footnote{Building safe AI systems might necessitate such targeted investments. For example, some have theorized a ``safety tax,'' wherein producing safe AI is much more expensive than producing AI prone to accidents \autocite{leikeDistinguishingThreeAlignment2022}.} Reallocation of compute (e.g., via subsidies) may also be help to incentivize safe development \autocite{jensenIndustrialPolicyAdvanced2023}. However, we note that increased allocations of compute also require human capital that can effectively make use of that compute \autocite{musserMainResourceHuman2023}.

\begin{figure}[!ht]
    \centerline{\includegraphics[width=1.3\linewidth]{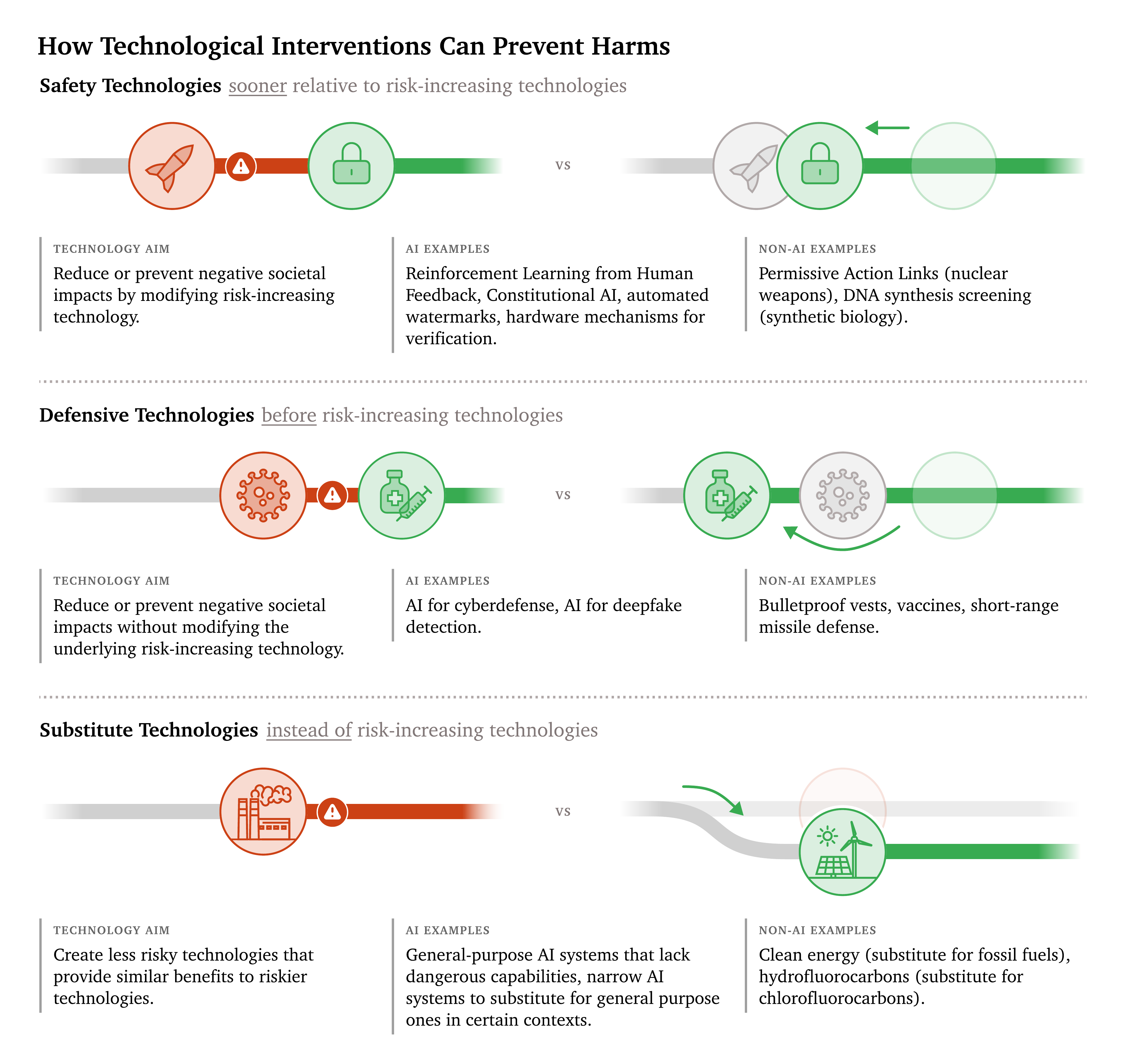}}
    \caption{\textbf{How differential technological development can reduce negative societal impacts.} Developing safety and defensive technologies sooner than riskier technologies and choosing to develop substitute technologies to risk-increasing technologies can reduce negative social impact. Adapted from \textcite{sandbrinkDifferentialTechnologyDevelopment2022}.}
    \label{fig:reduce_negative}
\end{figure}

If defensive AI applications---like AI systems for cyberdefense---are feasible \autocite{garfinkelHowDoesOffensedefense2019, segerOpensourcingHighlyCapable2023} and are built, policymakers could distribute access to such defensive technologies widely. For example, this could occur through liberal provision of subsidized inference capacity for defensive uses of AI or even a requirement that defense-dominant technologies developed using subsidized compute be open-sourced \autocite{howardAISafetyAge2023}.\footnote{See \textcite{segerOpensourcingHighlyCapable2023} for discussion of the benefits of and alternatives to open-sourcing as a strategy to deliver public benefits of large models.}

Differential technological development is of particular importance to compute governance due to the limitations imposed by algorithmic and hardware progress, discussed in \Cref{ssec:limitations}. These long-term trends imply that the cost of achieving any given level of AI capabilities will fall over time, making it more feasible to use less specialized compute \autocite{pilzIncreasedComputeEfficiency2023}.

So, over time, ``ungovernable'' compute\footnote{More precisely, compute that should not be subjected to significant compute governance measures as doing so would prove ineffective or impose unwarranted collateral damage.} will be capable of achieving greater capabilities than it is today \autocite{pilzIncreasedComputeEfficiency2023}. If some of those capabilities pose large risks, one way that society may be able to defend itself is by differentially allocating more governable, more powerful compute toward applications that can defend against risks from ungovernable compute \autocite{pilzIncreasedComputeEfficiency2023}. Examples could be cybersecurity and biosecurity applications, to defend against cyber and biological threats created or amplified by ungovernable compute.\footnote{This doesn’t necessarily mean that the set of chips subjected to any particular compute governance measure should expand over time. As discussed in \Cref{ssec:guardrails}, we propose limiting many of the compute governance mechanisms discussed here to AI chips---a small, distinct, difficult-to-produce, and expensive subset of all chips---and their surrounding infrastructure. Applying these mechanisms to all computer chips would be much more difficult (because, among other things, their supply chain is far less concentrated) and impose too large of a burden on privacy, centralization of power, and economic growth. For more, see \Cref{ssec:limitations}.}

\textbf{\textit{Redistributing access to AI development and deployment across and within countries}}\\\\
If AI becomes one of the most economically and strategically important technologies of the 21st century \autocite{danielsNationalPowerAI2021}, the geographic distribution of access to compute, and therefore the ability to develop and deploy AI without hindrance and oversight from other states, may influence the global distribution of power and prosperity. The fact that the AI compute supply chain is highly concentrated (\Cref{ssec:feasibility_governance}) means that a handful of countries have the ability to determine which countries can receive compute, should they wish to do so.

Two modes of geographical compute redistribution are worth considering. The first, ``negative redistribution,'' makes it harder for specific countries, such as geopolitical rivals or countries that fail to implement AI regulations \autocite{usbisImplementationAdditionalExport2022, bureauofindustryandsecurityExportControlsSemiconductor2022, haeckDutchCozyUS2023, nellisUSCurbsAI2023, khanMaintainingChinaDependence2020, daviesUSStrugglesMobilise2022, millerChipWar2022}, to acquire compute.\footnote{Another consideration that could motivate negative redistribution is that, independent of which states have compute, proliferation of computing capacity to a large number of states could make international coordination around responsible uses of compute more difficult \autocite{askellRoleCooperationResponsible2019, armstrongRacingPrecipiceModel2016}.} Instances of this are already underway through U.S. export controls, as we detail in \Cref{ssec:creating_capabilities}. The second, ``positive redistribution,'' ensures that specific countries have access to compute, thereby promoting other policy goals, such as global equity and sustainable development. This positive redistribution---particularly for the purpose of advancing global equity---is the focus of this subsection.\footnote{We note that these two goals may sometimes be in tension. For example, all else being equal, policymakers might reasonably prefer that AI compute remain in countries with strong state capacity, so as to prevent its misuse or diversion. However, state capacity is highly correlated with development \autocite{dinceccoStateCapacityHistorical2022, jenkinsOxfordHandbookHistorical2022, acemogluStateCapacityEconomic2015, gelosoStateCapacityEconomic2020}. Thus, any allocative efforts that prevent states with low state capacity from receiving compute will disproportionately deny compute to less developed states, further entrenching computational inequities \autocite{verdegemDismantlingAICapitalism2022, knightOneFathersAI2018, haoArtificialIntelligenceCreating2022, birhaneAlgorithmicColonizationAfrica2020, muggahArtificialIntelligenceWill2023}. None of this is to deny that there are states in the Global South with state capacity to administer adequate compute or other AI regulations \autocite{kenyaKenyaDataProtection2019, thomsonreutersfoundationAIGovernanceAfrica2023}. This disparity also highlights the importance of supporting efforts to improve state capacity in the Global South \autocite{okoloAIGlobalSouth2023}, so that trade-offs between these two goals are lessened.}

The disparity in AI development between the Global North and South has widened \autocite{yuAIDivideGlobal2023}. Compute, like many global resources, is unequally distributed between countries \autocite{boakyeStateComputeAccess2023, halopeHowCountriesCan2022}. A handful of countries, concentrated in the Global North, host the vast majority of AI data centers from major cloud compute providers, themselves headquartered in the Global North \autocite{googleTPURegionsZones, googleGPURegionsZones, amazonAWSGovCloudUS, microsoftAzureProductsRegion}.\footnote{Of course, customers can and do lease cloud compute capacity across borders.} Unequal access to compute and other key resources hinders the Global South’s ability to capitalize on the opportunities presented by this class of technologies \autocite{boakyeReapingRewardsNext2022, gulArtificialIntelligenceFrontier2019, chanLimitsGlobalInclusion2021}.\footnote{Other factors also contribute to this disparity, like a lack of trained AI experts, digital illiteracy, weaker governance frameworks from governments, infrastructural barriers, and the lack of sufficient indigenous datasets \autocite{okoloResponsibleAIAfrica2023, cipitStateAIAfrica2023}.} Increasing the Global South’s access to compute may therefore be an important method for decreasing global inequality and supporting domestic AI capacity therein \autocite{segerDemocratisingAIMultiple2023}.\footnote{Democratization of AI development through increased compute access in Global South countries would enhance visibility and thus bolster international coordination of AI governance \autocite{adanCaseIncludingGlobal2023}.}

How might we increase the Global South’s access to compute? Construction of large data centers requires specialized knowledge and enabling infrastructure, such as large-scale electrical generation and transmission and water delivery, that may not be immediately available in many Global South countries. Accordingly, simply reserving AI chips for delivery to the Global South is likely to be suboptimal for now, though possibly appropriate in some cases. Longer-term capacity-building programs, combining technical and financial assistance, could increase domestic capacity to build and operate AI compute infrastructure in the Global South.\footnote{Detailed recommendations for implementing a capacity-building program are beyond the authors’ own capacities, and in any case beyond the scope of this paper. However, we note that global capacity-building programs have been highlighted as one of the United Nations’ Sustainable Development Goals \autocite{unGoal17Revitalize}, and have been used in other fields such as civilian nuclear power \autocite{ghanaAnnouncesNewSupport2023, iaeaCapacityBuildingOperating} and seabed mining \autocite{isaCapacitydevelopmentTrainingTechnical2022}.}

Multistate collaborations (either entirely in the Global South, or bridging the South---North divide) to construct large-scale AI compute for use by the Global South, possibly modeled on or borrowing from public compute projects in the Global North (discussed in \Cref{ssec:governance_today}), could also spread risk and reap benefits from scale. However, even if these proposals were begun today, they would likely take years to make a dent in global computational inequality. Accordingly, a nearer-term measure might simply be to reserve some fraction of existing computing capacity for AI development or deployment in the Global South at subsidized costs.\footnote{As with any means-testing program, there would be nontrivial issues regarding how to determine whether these resources were reaching their intended beneficiaries, and how to prevent others from receiving the benefits not intended for them.}

The above ideas could also be applied to the allocation of compute within countries. The AI industry–and a few organizations within that industry–possesses a disproportionate amount of AI compute relative to academia, startups, or community-based AI efforts. It may become important to reduce this compute gap, especially when attending to risks from concentration of power in the hands of a few actors. This redistribution is one goal of ideas like the NAIRR \autocite{besirogluComputeDivideMachine2024}.

\textbf{\textit{Changing the overall pace of AI progress}} \\\\
Given compute’s importance to frontier AI development (as described in \Cref{sec:policymaking}), it is a powerful lever for influencing the pace of the field of AI as a whole (as opposed to only some aspects of AI development). Accelerating the pace of AI development aims to reap the benefits of more innovation \autocite{jonesScalingScalingLaws2021}. However, some have argued that slowing or pausing certain AI development and deployment is warranted \autocite{futureoflifeinstitutePauseGiantAI2023}.

Some compute governance interventions have likely already accelerated AI progress. Government support in different countries for semiconductor manufacturing capacity has made it easier for companies to lower costs, invest in research and development, and scale up production \autocite{chenFacilitatorsNationalInnovation2013}. Some further speed-ups are likely possible (e.g., via increased tax support for semiconductor manufacturers or direct government purchases of compute). This might be justified by the innovation and economic growth that could result.\footnote{The magnitude and even the moral value of the impact could vary depending on how these impacts are distributed across the world.} However, given the already rapid pace of developments and the growing amount of private sector investment in compute, it may become increasingly difficult for governments to take such an active role in speeding up AI progress.

Meanwhile, slowing down (or more radically pausing) AI development has received attention in recent years \autocite{futureoflifeinstitutePauseGiantAI2023}. In light of the extremely high opportunity costs of doing so, discussants have offered several justifications. One justification is security: if leading AI developers are not secure enough to defend against theft or misuse by opportunistic terrorist groups or ill-intentioned states, then slowing, pausing, or even destroying software that is vulnerable to theft might be warranted. Slowing AI development might also be warranted if the general rate of AI progress outstrips the progress in safety and security measures, or if society is not sufficiently prepared to integrate AI \autocite{danaherPhilosophicalDisquisitionsArtificial2023}.

One approach to restrict the pace of AI development (even in the absence of multilateral regulation\footnote{Note that decelerating unilaterally may be ineffective in a competitive environment \autocite{armstrongRacingPrecipiceModel2016}: a unilateral pause by a particular company would not necessarily be matched by others, nor would a unilateral pause by companies in the Global North necessarily be mirrored by companies located elsewhere. Pace-setting regulation that binds all actors would be one way of solving this sort of coordination problem. However, there is reason to doubt the feasibility of such regulation, especially if it needs to span multiple, rival geopolitical blocs \autocite{thiererExistentialRisksGlobal2023}.}) would be to modulate the quantity of inputs available. Of the major inputs to AI progress, compute is perhaps the easiest to verifiably modulate, for the reasons given in \Cref{sec:policymaking}. One simple method of modulating compute availability, therefore, could be to limit, by regulation, the amount of AI compute that can be produced every year. This would set a theoretical upper bound on the amount of compute that could be dedicated to AI progress at any given time, and also slow down the rate at which compute usage grows (thereby possibly allowing safety progress to ``catch up'').

A crude ``compute quota'' like this would have a number of drawbacks. Any attempt to limit output of AI chips will likely raise compute prices, thus harming consumers \autocite{gellhornIntroductionAntitrustEconomics1975}, especially those already struggling to afford compute. Depending on exactly how much supply is limited, the quota could diminish chipmakers’ profits, causing persistent political opposition to the quota by powerful firms.\footnote{The political economy of regulatory compute quotas depends on the current pricing dynamics in AI chips. Under certain assumptions, some specified supply restrictions would be profit-maximizing for producers (see, e.g.,  \textcite{gellhornIntroductionAntitrustEconomics1975}). Accordingly, producers often favor public policies that restrict output levels as a form of rent-seeking. However, there is no guarantee that the optimal level of compute outputs for modulating AI progress would be the profit-maximizing level for producers. Furthermore, as noted above, AI chipmaking is already a concentrated market. This suggests that producers already enjoy substantial price-setting power, and accordingly that additional artificial supply constraints imposed by regulation are likely to diminish, not increase, profits. Thus, despite the theoretical possibility that a compute quota would increase producer profits and therefore garner their support, in practice this seems unlikely.} Artificial supply constraints are also likely to lead to decreased investment in chipmaking capacity, which is both contrary to the revealed preferences of many governments, and may be an issue if higher compute production becomes desirable in the future.\footnote{This is because building out additional fabrication capacity takes years and many billions of dollars. If fabs are already producing at capacity, lifting the quota may not yield additional chip output for several years.} A quota, on its own, would not add to the government’s ability to select who gets chips, other than by possibly pricing certain actors out of the market.

An alternative possible means of using compute supply restrictions to modulate the pace of AI progress could be a government-operated ``compute reserve.'' This could first involve government authorities\footnote{Either a single government, or a consortium.} acquiring most or all cutting-edge AI chips produced by leading chip manufacturers. Government acquisition of chips would likely not be via expropriation, but rather via direct purchases at the fair market value of the AI chips.\footnote{One example pathway is via use of the U.S. Department of Defense’s authority under the Defense Production Act, which allows the government to place orders for standard products and require that these orders be served ``first in line.''} This would also maintain incentives to build out new fabs, and thereby create the option to more easily increase the flow of compute in the future.

After chip acquisition, the reserve authorities could then resell those chips or lease cloud capacity on them, controlling the flow of compute in order to control the rate of progress. Before the chips are ``released'' to the market for use in higher-risk projects, the reserve operator could possibly recoup some costs by allowing the chips to be used for non-AI purposes (e.g., graphics rendering), less risky AI projects, or the joint megaproject discussed in the next subsection.\footnote{See \Cref{ssec:visibility} for discussion of compute use monitoring, which may enable the reserve operator to ensure that chips ``in reserve'' are being used for non-accelerative purposes.} The reserve operator could also choose to block unvetted actors from buying or leasing large numbers of AI chips.

A compute reserve might be administered by and for multiple countries, speeding up or slowing down the flow of compute into the global economy.\footnote{A compute reserve might review evidence on a regular basis---say, every six months---in order to determine the effectiveness of risk mitigations undertaken in the prior time period, and decide on new compute influxes accordingly. To illustrate with one concrete example, the reserve might conclude in its first release decision that AI progress was proceeding somewhat too quickly for society to adapt, and that, e.g., all purchases made in the prior six months would be fulfilled up to 50\% of their size, with the remaining 50\% of each order instead purchased by the reserve and retained by the reserve for at least the next six months. Or it might conclude that current and near-term levels of compute---plus some additional margin of algorithmic and data-driven progress---would not pose significant societal risks, but the generation after that might. The reserve might thus share specific safety and societal resilience metrics for labs and governments to focus on in order to demonstrate at the next review that release should continue. One benefit of a compute reserve is that, unlike a petroleum reserve, the reserve administrator could still theoretically allow the chips in reserve to be used for non-acceleratory purposes (thereby recouping costs) during ``braking'' periods, while maintaining the ability to allow acceleratory uses later.} The presence of analogous institutions in other domains that attempt to control the supply of key resources (such as the Strategic Petroleum Reserve and OPEC with oil, or central banks with the money supply) points to the feasibility of a compute reserve, though the goal of a compute reserve might instead be to balance innovation and growth with safety and security.\footnote{Modern independent central banks are designed to be as free as possible from partisan or private commercial interests, though they do focus on their respective countries or regions. Note that the compute reserve would not function to ``roll back'' AI progress but would instead slow it, since it would (1) continue to allow increases in total compute, (2) not affect access to existing compute, and (3) not affect the other AI progress inputs, i.e., algorithmic progress and data (except indirectly). It could, once instituted, however, eventually modulate the speed of progress up or down.} Participation in the compute reserve could be incentivized by being the main or sole route through which advanced AI chips can be accessed.

The compute reserve also has significant downsides. It may give excessive power to member states, or the individual policymakers implementing the reserve. By increasing the demand for chips, it would also likely increase the cost of compute.\footnote{However, it is possible the compute operator would become such a large purchaser of compute that it would be able to negotiate for lower prices, if legally allowed to do so.} It would require large up-front capital costs from governments to acquire the chips. Due to hardware progress, the chips held in reserve would become less valuable over time, meaning that acquisition cost of the chips could be wasted if the reserve operator could not recoup costs through ``in-reserve'' usage.

More broadly, the power to modulate the overall pace of progress in an entire technical field is a sweeping one, and one that society rarely entrusts to policymakers.\footnote{But, see \textcite{maasPathsUntakenHistory2022}.} It could doubtless be misused in many ways. Whether---and under what conditions---it would be wise to entrust policymakers with such a power remains an important open question.

\textbf{\textit{Collaborating on a joint AI megaproject}}\\\\
The term ``CERN for AI'' is sometimes used to refer to the idea of an international scientific megaproject focused on AI \autocite{kempAdviceHighlevelPanel2019, kerryGlobalAiCooperation2022, zhangEnhancingInternationalCooperation2022, stixFoundationsFutureInstitution2022, hogarthWeMustSlow2023, hammondWeNeedManhattan2023, hausenloyProposalInternationalCoordination2023}. This term is inspired by previous international scientific megaprojects, like the European Organization for Nuclear Research (CERN), the International Space Station (ISS), and the International Thermonuclear Experimental Reactor (ITER). All three of these projects are notable as collaborations that include cooperation among geopolitical rivals.\footnote{CERN and the USSR had various scientific cooperation agreements since 1967, and Russia had observer status from 1991 to 2022 \autocite{cernRussianFederationObserver2023}. The ISS involves cooperation between NASA and Roscosmos, among other space agencies. ITER is funded by seven member parties: the United States, China, Russia, the European Union, India, Japan, and South Korea.}

These international scientific megaprojects (CERN, ISS, and ITER) all have high fixed capital costs that are beyond the budgets of individual universities (and even some countries). Countries pooled funding to build the capital-intensive, expensive, specialized, shared infrastructure for scientific experiments in the public interest. Similarly, the capital-intensiveness of compute (particularly that required for frontier AI models over the coming decade), suggests that an analogous ``CERN for AI'' could share the cost of building and operating a large compute cluster (and possibly next-generation fabs).

A truly international CERN for AI could offer an important alternative to large-scale corporate or  national projects. Corporate projects face significant legitimacy problems after a certain stage of development because they involve a private actor making large-scale decisions that could affect humanity as a whole \autocite{bengioAICatastrophicRisk2023}. An international project faces fewer legitimacy concerns, especially if its membership consists of a representative set of democratically accountable actors from all regions of the world. Another potential benefit of consolidating significant fractions of frontier AI development in an international institution is that it may be relatively efficient to ensure the safety and security of that development in such a scenario, as opposed to more decentralized development.\footnote{Note that we are not saying this argument definitely holds. One counterargument would be that parallel safety and security ``bets'' will lead to faster innovation, some of which will then be shared across different institutions. Our point is not to suggest that this or any other lever should certainly be implemented but to give some intuition for why one might consider it and how compute might enable it.} This could be true if there are large upfront safety and security risks for each additional frontier AI developer. Should any of the participant countries express significant concerns about the safety or security implications of the next phase, the project could temporarily halt to address those concerns prior to moving forward.

A CERN for AI could have different technical objectives. Most generally, it could simply provide computing resources for any large research project in AI. A ``CERN for Frontier AI'' could focus on training frontier models, with the objective of doing so safely and for broadly shared societal benefit. A ``CERN for AI for Good'' could focus on public goods, e.g., AI applied toward clean energy, medical research or achieving sustainable development goals \autocite{vinuesaRoleArtificialIntelligence2020}. Finally a ``CERN for AI Safety,'' could focus on a particular public good: improving our understanding of and ability to control the behavior of AI systems. Because this project would be publicly funded and organized, it would have a different set of incentives than private sector projects. This would therefore change the competitive dynamics of the market and could incentivize private AI companies to compete on a range of dimensions \autocite{coylePromisePerilGenerative2023, mazzucatoMissionEconomy2021}.

A CERN for AI could consider several strategies to disseminating access to its model.\footnote{We note here that benefit-sharing is a key and contentious topic in many international institutions, agreements, and discussions. We do not have space for a full discussion here. One argument for benefit sharing is that all people in the world share some level of risk from AI development, so should also share the benefits. Another is that humanity as a whole has created the ``data commons'' used for pre-training, so deserves to share the benefits. However, many international agreements can be seen as a ``deal.'' For example, in the Convention on Biological Diversity, the Nagoya Protocol on Access to Benefit Sharing can crudely be viewed as a ``payment for genetic resources'' deal. The nature of such potential deals requires further research.} One method might be structured access \autocite{shevlaneStructuredAccessEmerging2022}, where customers and businesses across all (participating) nations could obtain API access. Alternatively, with a sufficiently secure information system, the trained model (or variants thereof) and its weights could be securely transferred to licensed entities in each participating country, whether they be private corporations or public agencies. These licensed entities could in turn offer API access within their countries, or fine-tune the model for particular use cases. Securely transferring the weights poses an extremely challenging problem; at the minimum, it would require extremely strong information security at the licensed entities to prevent theft, as well as protections against misuse of these capabilities. Should the risks appear sufficiently low – perhaps after a model has been surpassed by more capable ones – the model weights could be published publicly.

A CERN for AI could see cooperation between otherwise adversarial countries. One advantage of the CERN model is that countries could build trust by incrementally ratcheting up investments in a ``tit for tat'' manner. The larger the scale of the investments, the less likely it is that one of the participants could be hiding a similarly sized project. Participants could withhold investments to reflect any potential concerns about the safety or security of the project. Thus, a CERN for AI could, if incentives were aligned sufficiently to begin this ratcheting process in earnest, eventually be a stabilizing force in a potential future AI ``arms race.''\footnote{There has been extensive debate on the term ``AI arms race'' \autocite{caveAIRaceStrategic2018, scharreDebunkingAIArms2021, belfieldWhyPolicyMakers2022}.}

However, a CERN for AI could represent one of the most radical expansions of the power of international organizations in human history. Given the mixed track record of international regulation of technology, it is worth being clear-eyed about the large risks associated with such an effort, and the difficulty of success \autocite{thiererExistentialRisksGlobal2023}. In the worst-case scenario, centralizing control of AI in a single organization could increase the risk that the technology is monopolized by an oppressive or illegitimate government \autocite{caplanTotalitarianThreat2008}. More mundanely, there is simply no widespread agreement on what governance of such an organization could look like, or how it could simultaneously satisfy all stakeholders’ demands. The governance structure of a CERN for AI would be an important determinant of how desirable it is, and it is far from clear whether existing proposals provide a satisfactory answer.

\subsection{Enforcement}\label{ssec:enforcement}

Allocation is a blunt tool for public policy. It depends on having reliable ex ante information about which actors or projects are likely to be beneficial or harmful, and the ability to differentially allocate compute toward beneficial users and away from harmful ones.

Reality is more complicated. Users of compute will engage in some combination of beneficial, benign, and harmful computational activities to various degrees. Determined actors will also often find a way to circumvent restrictions on their access to compute \autocite{fistPreventingAIChip2023}.

Regulators will then need to make sure that these users are abiding by rules regarding AI development and usage---or are punished or thwarted if they don’t. We use ``\textit{enforcement}'' to refer to a regulator’s capacity to prevent or respond to violations of rules.

Enforcement naturally complements the visibility and allocation capacities. By exercising their visibility capacity, regulators can more effectively target their monitoring and investigation resources to find rule violations. Regulators could then use traditional enforcement tools, such as civil or criminal penalties, to deter or prevent further violations. Regulators can also use their allocation capacity to block or reduce flows of compute to actors they think are likely to violate rules regarding compute use, or as a penalty for past violations.\footnote{Export controls on compute are arguably (at least partially) an example of this.} Indeed, the restriction of access to computing power (e.g., in the form of AI chips) could be applied to enforce a rule.

However, regulators can leverage compute to enforce rules in other, more novel ``technically-enabled'' ways.\footnote{The revised export controls proposed by the U.S. government include a request for public comments on mechanisms relevant to this context. Specifically, they ask: ``Today’s AC/S IFR seeks public comments on proposed technical solutions that limit items specified under ECCN 3A090 or 4A090 from being used in conjunction with large numbers of other such items in ways that enable training large dual-use AI foundation models with capabilities of concern'' \autocite{usbisImplementationAdditionalExport2022}(p. 104).} By modifying the computing hardware itself (and its associated software), policymakers may be able to effectively limit the workloads that the hardware can perform, thereby outright preventing (some) potentially harmful uses of compute. They could also swiftly---and automatically---respond after the fact if harmful uses occur.

We describe the following illustrative ways of leveraging compute to enforce rules below:
\begin{enumerate}
    \item Enforcing ``compute caps'' via physical limits on chip-to-chip networking
    \item Hardware-based remote enforcement
    \item Preventing risky training runs via multiparty control
    \item Digital norm enforcement
\end{enumerate}

The interventions we discuss here can be implemented primarily in either software or hardware. Hardware implementations are likely more robust to tampering.\footnote{Many of the features that make compute difficult to manufacture also make it difficult to modify once manufactured. Therefore, the hope is that some of the mechanisms discussed here will be difficult and expensive to circumvent even if the hardware is possessed by an untrustworthy actor.} They also ``travel with'' the AI chips themselves, and continue to function throughout the hardware’s lifetime, regardless of where the hardware is and regardless of who owns it.\footnote{They are therefore more robust to failures of allocation, such as allowing bad actors to possess large quantities of compute.} By architecting away the ability to even run certain workloads, they can remove the ability to use the hardware for the prohibited purpose rather than merely disincentivizing it. Software-based implementations are more flexible, but are able to be easily modified---including by malicious actors.

Technically-enabled enforcement could reduce the need for costly physical enforcement (or threats thereof in order to deter certain actions); given the stakes of AI, the magnitude and complexity of enforcement resources required to reliably deter misuse or negligence could otherwise be very large in the future. Furthermore, the possibility of automating enforcement drastically increases the probability that penalties can be successfully applied or that certain harms can be prevented.\footnote{Of course, technically-enabled enforcement may not always be the best way to enforce rules for AI. The regulatory application of this tool requires sensitivity to its context (see e.g. \textcite{mulliganPerfectEnforcementLaw2007}).} And these could be applied selectively: rather than employing broad measures like restricting access to chips, regulators could focus on modulating specific workloads, such as training models above a certain computational budget, or use cases, such as aggregating chips for use in a supercomputer.\footnote{There is no mechanism that differentiates ``good AI'' from ``bad AI.'' Rather, these assurances, and their corresponding mechanisms, are wide-ranging: from influencing the cost of AI model training to delaying deployment, increasing compute costs, or even applying specific constraints like preventing chips from training models on biological data. The desirability of each assurance is eventually informed by the threat model.} We discuss risks from these measures, e.g. to privacy, in \Cref{sec:risks}.

We emphasize that technically-enabled enforcement in this context is highly speculative: the feasibility and robustness of these mechanisms are unproven. These examples are therefore presented more as directions for investigation rather than shovel-ready interventions. We also omit discussion of many of the security and engineering details that would need to be resolved to make these mechanisms effective and robust to attacks. Any technical additions to chips will likely introduce additional security risks; these must be carefully weighed against potential benefits. We list some research directions regarding the security and technical feasibility of these mechanisms in \Cref{sec:appB}.

These drawbacks underline the need for technically-enabled enforcement to be accompanied with traditional methods of enforcement. They cannot operate effectively in isolation and should be complemented by other governance regimes, including methods to verify the integrity of these mechanisms.
\newpage
\textbf{\textit{Enforcing ``compute caps'' by technically limiting chip-to-chip networking}}\\\\
Our first example is a relatively blunt method of leveraging compute to prevent violations of a rule. Training highly capable AI systems currently requires accumulating and orchestrating thousands of AI chips; if these systems are potentially dangerous, then limiting this accumulated computing power could serve to limit the production of potentially dangerous AI systems. How might this be accomplished? Instead of broadly limiting access to AI chips to prevent the development of potentially dangerous AI systems, regulators can implement a more targeted approach.

This strategy would involve restricting the networking capabilities of these high-performance chips to prevent them from linking together to form large, powerful clusters. A mechanism for restricting cluster scalability could involve limiting communication outside of a pre-authorized number of chips. While communication between pre-authorized chips could occur at unrestricted bandwidth, communication with external chips or systems could be drastically limited. This confined communication limits the scalability into the large clusters required for the efficient training of large AI models. Determining the optimal bandwidth limit for external communication is an area that merits further research.

Implementing limits on chip-to-chip networking could relax some of the trade-offs involved with broadly denying access to chips. However, the challenge lies in making these mechanisms as targeted as possible. It is true that current frontier AI training runs are extremely communication-intensive and require record-breaking numbers of AI chips, and yet imposing new limitations could also inadvertently affect other workloads. This suggests that the chip-level interventions required to limit large accumulations of compute should be designed to leave consumer use cases unaffected.\footnote{For example, the consumer gaming experience does not benefit from large numbers of accumulated GPUs.}

\textbf{\textit{Hardware-based remote enforcement}} \\\\
In situations where AI systems pose catastrophic risks, it could be beneficial for regulators to verify that a set of AI chips are operated legitimately or to disable their operation (or a subset of it) if they violate rules. Modified AI chips may be able to support such actions, making it possible to remotely attest to a regulator that they are operating legitimately, and to cease to operate if not. Remote enforcement at the chip level could leverage existing cryptographic technology \autocite{sommerhalderHardwareSecurityModule2023, sabtTrustedExecutionEnvironment2015}. One potential application of this technology is in enabling (ex post) visibility of workloads, but it can also be used for automatically enforcing rules.\footnote{Wherein an AI developer uses chips that store privacy-preserving logs of their workloads, and a regulator verifies after the fact that the developer is adhering to any requirements for their workloads (we discuss this in \Cref{ssec:visibility}).}

Consider export controls on AI chips. Using traditional methods of enforcement incurs high administrative costs and inflates the scope of the controls as they have to focus on who accesses the chips, rather than what they are being used for.\footnote{That is, they are targeted broadly at the level of countries and organizations (users) on the theory that those targets run an unacceptable risk of using compute for harmful purposes. This user-level targeting is by necessity, as it is not currently possible for governments to reliably monitor or control how these chips are being exported. These export controls can have the drawback of limiting beneficial or benign use cases (e.g., scientific research or innovation in societally beneficial domains), even those that might benefit the countries imposing export controls in the first place. 
Additional side effects include increasing incentives for domestic development of semiconductor development by targeted countries, curbing the revenue of semiconductor companies located in democracies, increasing geopolitical tensions, and conveying the impression that researchers from certain backgrounds are being targeted as people (rather than the harmful use cases themselves).} If remote authorization mechanisms are used, these export controls could be ``digitized'' \autocite{reinschDigitizingExportControls2021, bureauofindustryandsecurityCommentFRDoc2020}. Specialized co-processors that sit on the chip could hold a cryptographically signed digital ``certificate,'' and updates to the use-case policy could be delivered remotely via firmware updates. The authorization for the on-chip license could be periodically renewed by the regulator, while the chip producer could administer it.\footnote{In principle, remote enforcement need not be ``baked in'' at the hardware level; one can imagine higher-level software that enforces rules on a data center; indeed, many cloud computing providers operate similar software.} An expired or illegitimate license would cause the chip to not work, or reduce its performance.\footnote{It is not just regulators who would benefit from these mechanisms. For example, chip producers could automatically enforce violations of their own terms of service.}

Remote enforcement mechanisms come with significant downsides, and may only be warranted if the expected harm from AI is extremely high. Notably, such mechanisms could themselves pose significant security \autocite{andersonWhoControlsSwitch2010} and privacy risks, as well as potential for the abuse of power. The inclusion of a mechanism to disable the device remotely could be manipulated by malicious actors or even misaligned autonomous AI systems to disable or otherwise manipulate computing infrastructure. This could lead to substantial financial losses or even pose risks to human safety in certain scenarios. Thus, if this approach is desirable at all, these mechanisms should focus on a specific subset of AI development and scenarios---for example, where rapid enforcement is particularly valuable.

\textbf{\textit{Preventing risky training runs via multiparty control}}
\\\\
Another future-oriented, speculative proposal, which may be justified only in extreme scenarios, involves a strategy to prevent undesirable AI training runs. This would operate by distributing the control over the metaphorical ``start switch'' either among multiple parties or to a governing third party. The power to decide how large amounts of compute are used could be allocated via digital ``votes'' and ``vetoes,'' with the aim of ensuring that the most risky training runs and inference jobs are subject to increased scrutiny.

The implementation of this could parallel the previous example of remote enforcement; multilateral control could be implemented through the use of multisignature cryptographic protocols \autocite{cramerSecureMultipartyComputation2015}. The software and hardware for AI chips could be modified to initiate processing instructions only when the workload is cryptographically signed by all parties. Institutionally, a number of configurations seem worthy of exploration. In a domestic setting, the control rights can be distributed to government regulators, independent auditors, or an international body, who should be incentivized to accurately assess the risk of the training run.

While this may appear drastic relative to the current state of largely unregulated AI research, there is precedent in the case of other high-risk technologies: nuclear weapons use similar mechanisms, called permissive action links (``PALs''). PALs are security systems that require multiple authorized individuals in order to unlock nuclear weapons for possible use. By requiring the involvement of multiple parties, the system reduces the risk of human error or malicious intent, and increases the level of accountability for decisions related to nuclear weapons use.

From one perspective, this mechanism could diffuse power, by making it harder for lone actors to unilaterally take actions with massive externalities \autocite{bostromUnilateralistCurseCase2016}. But from another perspective, it could concentrate enormous power in the hands of every party that has the right to veto potential technical advances. We have seen how well-intentioned efforts to give many stakeholders the ability to veto decisions that could affect them can block various desirable forms of progress (e.g., \textcite{verweyNoPermitsNo2021}), including progress towards the very goals that vetocratic policies aimed to advance \autocite{fukuyamaVetocracyClimateAdaptation2022}. As with all policy measures, the substantive and procedural elements of this policy will determine its desirability.

A separate problem is information security. Vote- and veto-holders must be informed of the relevant features of the training run to make an informed decision. But some details of the training run could be sensitive---either to individuals or commercial actors.\footnote{We discuss these issues further in \Cref{ssec:limitations}.} The information shared with vote- and veto-holders would therefore have to be very carefully scoped. It may also be possible to construct ``zero-knowledge'' proofs of certain claims about proposed training runs that do not disclose sensitive information. More research into this possibility seems valuable (e.g., \textcite{buterinMyTechnooptimism2023}, \Cref{sec:appB}).
\newpage
\textbf{\textit{Digital norm enforcement}}\\\\
In some cases, enforcement via compute can enable more flexible and fine-grained prevention and response. One example involves implementing digital controls over compute resources from infrastructure-as-a-service (IaaS) entities, like cloud computing providers. Instead of outright denying access to chips, regulators can set restrictions on the total amount of compute usage permitted. These restrictions are digitally enforced by the IaaS companies themselves. Access to large-scale compute resources could be made conditional upon users complying with risk-reducing policies. For example, an AI developer (building on the IaaS’s compute) planning a large-scale deployment could be required to submit audit results of their AI model as a precondition for access \autocite{eganOversightFrontierAI2023}. Access could be easily restricted at any time if potential violations were detected.

Ideally, decision-making regarding these conditional accesses should not be left at the discretion of IaaS companies, since they face flawed incentives (such as a profit incentive to overgrant access). An alternative would be to have decision-making governed by regulatory mandates and rely on the technical capabilities of IaaS companies for enforcement. As discussed in \Cref{ssec:limitations}, this approach is akin to how digital services are shut down for legal violations, such as hosting illegal online drug markets.

This method allows for more flexible and context-sensitive regulation than broad brush policies (like denying chips). Regulation could adapt to the rapidly evolving landscape of AI development and deployment while ensuring compliance with established legal and ethical standards.

\section{Risks of Compute Governance and Possible Mitigations}\label{sec:risks}

While governing AI via compute has significant potential as discussed above, pushing compute governance to extremes---especially when used as a tool for visibility and enforcement---bears significant risks that policymakers should carefully evaluate. As we have tried to emphasize above, compute governance is a double-edged sword: it can be used to promote widely shared objectives like safety, but it can also be used to infringe on civil liberties, prop up the powerful, and entrench authoritarian regimes. We discuss examples of such unintended consequences of compute governance below, including: threats to privacy; additional opportunities for leakage of commercially sensitive information; other negative economic impacts; and risks from centralization and concentration of power.

Further, compute governance is a promising tool for AI governance in large part due to empirical factors that could change. We discuss such limitations to the feasibility and efficacy of compute governance. These include: algorithmic and hardware progress; low-compute specialized models with dangerous capabilities; and evasion, circumvention, and decoupling.

To close out this section, we provide several overarching recommendations for guarding against some of these concerns. These include focusing on AI chips that are designed for AI supercomputers (excluding consumer-grade hardware as far as possible), using privacy-preserving practices and technologies, favoring compute-based measures for risks where ex ante measures are justified, periodically revisiting controlled computing technologies, implementing all controls with substantive and procedural safeguards, and using governable compute to protect society against risks from ungovernable compute.

\subsection{Limitations}\label{ssec:limitations}
\textbf{\large Unintended Consequences}\label{ssec:unintended-consequences}

\textbf{\textit{Threats to personal privacy}} \\\\ In modern society, computational activity is core to most aspects of virtually every person’s life. The economic, social, political, cultural, intellectual, recreational, and health spheres are all largely enabled and mediated by computation. Thus, it is possible that any revelation or monitoring of an actors’ computational activities could reveal private and sensitive information.

A number of the compute governance possibilities we explore (e.g., \hyperref[bit:report_training]{required reporting of large-scale training compute usage from cloud providers and AI developers}, \hyperref[bit:chip_registry]{international AI chip registry}, and \hyperref[bit:workload_monitoring]{privacy-preserving workload monitoring.}) involve giving some actor more visibility into specific computational activities. For example, required reporting from cloud providers on customer usage could reveal sensitive information about companies or individuals. This visibility may reveal information about computational activities in which individuals have a legitimate privacy interest,\footnote{However, we note that most of the visibility mechanisms we discuss above are targeted at corporate model developers, not consumers.} or in which companies have a trade secret interest. It is reasonable to worry, then, that increasing visibility into AI-relevant computation could carry significant risks to privacy and civil liberties (e.g., \textcite{thiererExistentialRisksGlobal2023}; \textcite{howardAISafetyAge2023}).

Even in the context of large computing clusters, trade-offs between monitoring and privacy or security arise and cannot be addressed solely through means previously discussed, such as structured access via APIs. For example, cloud computing raises ``tenant'' privacy considerations---where customers seek assurance that their cloud provider is not, for example, stealing their IP---that need to be protected strictly and that pose challenges for AI-related monitoring. Government (especially military) data centers may be particularly sensitive to disclosure, and the semiconductor supply chain is regularly targeted for espionage purposes, which could compromise some efforts discussed here absent significant effort.

\textbf{\textit{Opportunities for leakage of sensitive strategic and commercial information}} \\\\ Many of the compute governance ideas discussed above---especially those in \Cref{ssec:visibility}---involve sharing information about compute and compute usage with policymakers. As discussed, there can be large benefits to this sort of visibility. But where these approaches have poor information security or are overly broad, they could create opportunities for the disclosed information to leak, to the competitive detriment of the regulated companies. Such leaks could also undermine trust and exacerbate racing dynamics, making it more challenging to establish effective policy for the governance of AI.

Frontier AI labs increasingly withhold information about the processes used to create their flagship models, including the amount of compute used to create them.\footnote{For example, compare GPT-2 \autocite{radfordLanguageModelsAre2019} with GPT-4 \autocite{openaiGPT4TechnicalReport2023}.} Revealing this information could, for example, help commercial competitors and geopolitical rivals understand how great of an investment would be needed to replicate the capabilities of an existing model. In some instances, the details sought by regulators may be considered highly confidential within the frontier AI labs themselves, accessible to only a select group of employees. Thus, secrecy helps AI labs preserve their economic competitiveness, and also slows diffusion of capabilities advances to geopolitical rivals. However, as this information is made available to policymakers, additional opportunities for this information to leak arise.

Similarly, cloud compute providers often do not release much information about the location, capacity, and operation of their large data centers. They invest a substantial amount in physical security and cybersecurity \autocite{pilzComputeScaleBroad2023}. Policymaker demands for access to or visibility into the supply chain or operation of these data centers could create additional vectors for attack or compromise of sensitive information.

Poor information security could dramatically increase the costs of compliance for AI companies, leak trade secrets, and accelerate proliferation of potentially dangerous capabilities \autocite{anderljungFrontierAIRegulation2023}. As discussed in \Cref{ssec:guardrails}, compute governance measures must therefore be carefully scoped and implemented with information security in mind.

\textbf{\textit{Negative economic impacts}} \\\\ Research by the U.S. Bureau of Economic Analysis suggests that the digital economy accounts for 10\% of U.S. GDP \autocite{highfillNewRevisedStatistics2022}.\footnote{This number is expected to grow significantly; the revised definitions for GDP due to be adopted by the UN in 2025 will likely set out a consistent and more inclusive method for measuring the digital contribution across countries, and work is underway to define and measure the contribution of AI \autocite{briggsPotentiallyLargeEffects2023}.} The ``permissionless'' nature of most computational activity is a large part of why digital technologies have been such a force for economic growth \autocite{thiererPermissionlessInnovationContinuing2014}. It is therefore reasonable to worry that placing burdens on access to certain compute---the substrate of the digital economy---could impose meaningful economic costs \autocite{thiererExistentialRisksGlobal2023}.

For example, we consider KYC requirements for access to large-scale computation above. A skeptic might worry that even a presently high threshold for KYC checks will ultimately cover a sizable portion of the AI industry as compute usage increases, causing significant frictions to economic activity. We also consider export controls, but the history of export control policy is replete with debates around the trade-offs between strategic benefits from controlling exports to rivals and increasing domestic production, including general skepticism toward the effectiveness of many controls \autocite{mastandunoEconomicContainmentCocom1992}. Some of the more dramatic governance approaches we explore above---such as the CERN for AI and multiparty control of large-scale compute usage---contemplate centralizing or concentrating the development of the most capable, compute-intensive, general AI systems. However, if that is not accompanied by widespread ability to build on and deploy such systems, we may fail to harness the creativity of the market, with accompanying loss of economic growth.

\textbf{\textit{Risks from centralization and concentration of power}} \\\\ Right now, control over computation is fairly widely distributed.\footnote{However, as discussed above, the supply chain for AI chips and large data centers is extremely concentrated. Existing compute providers do not seem to leverage this existing power for political or ideological purposes, though perhaps they will in the future. This dynamic resembles the leverage that social media and other communications platforms could (and often do) exercise over speech on their platform, which is the subject of ongoing controversy (e.g., \textcite{klonickNewGovernorsPeople2017}).} Greater central regulatory or allocative authority over large concentrations of compute will increase centralized control over an increasingly crucial economic and political resource. This carries serious risks \autocite{thiererExistentialRisksGlobal2023, howardAISafetyAge2023}.

Some of the risks from centralized control are technical. Remote enforcement mechanisms like kill switches can introduce security risks and the potential for control or manipulation \autocite{andersonWhoControlsSwitch2010}. Compute visibility mechanisms may create concentrated repositories of information that are attractive to bad actors.

Other risks are political. With increased government control over AI-relevant compute, powerful actors---including corporations---may try to wield the power of the state for their own ends, e.g., attempting regulatory capture. More fundamentally, history shows that centralizing power can carry significant---and even catastrophic---downsides, such as entrenching existing inequalities \autocite{goldenCentralizationBargainingWage2006}, suppressing dissent \autocite{wallachCensorshipSovietBloc1991}, creating poor epistemic standards among governing powers \autocite{andersonEpistemologyDemocracy2006}, and promoting poor economic decision-making \autocite{acemogluWhyNationsFail2012, scottSeeingStateHow1999}.

\textbf{\large Issues of Feasibility and Efficacy}\label{ssec:issues-feasibility-efficacy}

\textbf{\textit{Algorithmic and hardware progress}}\\\\ Compute governance is more effective when, all else equal, (1) it takes a large amount of compute to achieve a certain level of capabilities, (2) the cost per unit of compute is high, and (3) using a large amount of compute requires usage of a large data center.\footnote{This is because larger data centers are (1) easier to detect, (2) more expensive to build, (3) less common, and (4) more likely to be used for larger training runs, given the efficiencies of hosting a training run in a single data center.}

However, certain long-run trends are slowly weakening each of these. Due to algorithmic progress, it takes fewer and fewer computational operations each year to achieve a given level of AI performance \autocite{hernandezMeasuringAlgorithmicEfficiency2020, erdilRevisitingAlgorithmicProgress2022}.\footnote{Compute itself is arguably a significant driver of algorithmic progress \autocite{barnettComputebasedFrameworkThinking2023}, as it enables experimenting with more architectures, scaling up what works, and gaining insights that may be only visible at scale.} Due to Moore’s Law\footnote{Moore’s Law originally referred to the density of transistors on a chip \autocite{mooreCrammingMoreComponents1998}, but has since been used colloquially to refer to the general exponential improvements in the performance of chips (in large part due to increasing transistor density).} and more specialized architectures, a dollar can buy many more operations every year \autocite{hobbhahnTrendsGPUPriceperformance2022, hobbhahnTrendsMachineLearning2023}. Also, major progress in communication-efficient training could allow more decentralized training---i.e., splitting a single training run across multiple data centers---allowing training runs of a constant size to be hosted on multiple smaller data centers \autocite{yuanDecentralizedTrainingFoundation2022}. This makes it harder to identify and distinguish data centers potentially useable for large training runs. It is unclear whether these trends will continue in the long run, and what their limits, if any, are. Thus, each year it becomes more feasible to train models to a given level of performance using less, cheaper, and more decentralized compute, and consequently somewhat less governable.\footnote{Dramatically improved computing hardware would certainly change aspects of AI development, but might not necessarily alter the role or importance of compute governance. Semiconductors have powered computing for decades and will likely continue to do so. Alternative compute architectures seem to face significant challenges: quantum computing is likely still distant and poorly suited for training AI models \autocite{sevillaForecastingTimelinesQuantum2020}. Neuromorphic chips are primarily useful for inference, and likely still require the silicon supply chain in the short term. Optical computing remains mostly speculative. While new hardware may improve efficiency, it would not eliminate the need for computational power to develop AI systems.}

The extent to which this effect undermines compute governance largely depends on the importance of relative versus absolute capabilities. Increases in compute efficiency make it easier and cheaper to access a certain level of capability, but as long as scaling continues to pay dividends, the highest-capability models are likely to be developed by a small number of actors, whose behavior can be regulated via compute \autocite{pilzIncreasedComputeEfficiency2023}. This dynamic could change if the scaling paradigm diminishes in effectiveness \autocite{lohnScalingAICost2023} or if decentralized training becomes feasible.\footnote{While progress in decentralized training may allow more actors to train models of a certain capability, such efforts would likely still be enormously resource-intensive.}

That is to say, over time the amount of compute needed to train a system with a \textit{particular level} of capability (e.g. GPT-4 or Claude 2 level in 2023) will decrease, but the amount of compute needed to train a system with a \textit{frontier level} capability (a hypothetical GPT-5 and GPT-6 or Claude 3 and Claude 4) will increase.

\textbf{\textit{Low-compute specialized models with dangerous capabilities}} \\\\ Specialized models trained on high-quality data require significantly less training compute to reach a high level of performance on particular tasks, compared to today’s most well-known generally capable models. For example, AlphaFold 2 achieved superhuman performance on protein folding prediction using fewer than $10^{23}$ operations---two orders of magnitude less compute than models like GPT-4 \autocite{epochParameterComputeData2022}. If such low-compute models could cause significant harm, compute governance could be ineffective or inadvertently impose on harmless activity. Compute governance seems most appropriate where risk originates from a small number of hugely compute-intensive general models. This fact is also recognized in the 2023 U.S. Executive Order on AI, where reporting requirements are imposed on models trained on biological sequence data using three orders of magnitude less compute than other models---$10^{23}$ operations vs. $10^{26}$ operations \autocite{thewhitehouseExecutiveOrderSafe2023}---in light of such models’ potential for biological weapons design \autocite{sandbrinkArtificialIntelligenceBiological2023}.

Dangerous capabilities can also arise through changes made to AI systems post-training. For example, with just \$200 and one GPU, researchers were able to untrain (via fine-tuning) the safety features of Meta’s Llama 2 Chat (the model's weights were publicly available). This intervention caused the subverted model to respond to requests for harmful information in the vast majority of cases \autocite{lermenLoRAFinetuningEfficiently2023}. This was despite Meta’s investments in safety testing and red teaming \autocite{touvronLlamaOpenFoundation2023}. A broader set of policy approaches will be needed to further investigate and mitigate these risks.

Once trained, high-compute models can be run using less costly computational resources. Some important and (potentially dangerous) AI capabilities may be accessible without high-end compute. For instance, protein folding capabilities can be harnessed with only a handful of inferences \autocite{jumperHighlyAccurateProtein2021}. One can imagine successor models trained on biological data that could potentially use small amounts of inference compute to identify novel pathogens. Moreover, there is growing interest in the downsizing of AI models to be compatible with consumer or edge devices like smartphones or laptops. For example, Stable Diffusion v1.5 (albeit operating slowly) can now run locally on a phone \autocite{vincentQualcommDemosFastest2023}, potentially giving rise to the proliferation of visual ``deepfakes.''

In general, compute governance measures would be unable to reliably ``reach'' the computing hardware sufficient to create or run a small number of instances of such low-compute models. Regulation of such low-compute models will require other policy approaches.

\textbf{\textit{Incentives for diversion, evasion, circumvention, and decoupling}} \\\\ Actors are likely to attempt to circumvent and evade compute governance interventions, especially where their access to AI chips or their privacy is severely affected. Cutting off access to compute, for example---either preemptively or reactively---is a blunt instrument and has many downsides. We are already seeing such dynamics play out as a result of U.S. export controls on AI chips to China \autocite{fistChineseFirmsAre2023}.

In the short term, there are attempts to circumvent these AI chip export controls via chip smuggling, using non-controlled chips, or accessing cloud compute \autocite{fistChineseFirmsAre2023, grunewaldAIChipSmuggling2023}. Attempts by non-state groups to evade controls on other materials, such as explosives, chemicals, biological agents, and radioactive material, are common \autocite{allenImprovedExportControls2022}.

In the medium and long term, however, denying compute could further incentivize other attempts to get around a limit. Squeezing one part of the supply chain puts pressure on other parts. Actors without access to high-end chips are incentivized to find ways to utilize larger quantities of lower-grade chips. Restricting Chinese access to AI chips creates even stronger economic incentives to build a supply chain free of U.S. involvement. Though this would be incredibly challenging, over time, this could potentially create a wholly separate supply chain, reducing strategic interdependence and the ability to govern AI using compute---often referred to as ``decoupling.''

Separately, additional scrutiny on training runs above a certain threshold could further incentivize research into algorithmic breakthroughs. However, those incentives are already very strong since they can increase one’s ``effective compute'' given a certain quantity of actual compute.

\subsection{Guardrails for Compute Governance}\label{ssec:guardrails}

Given these serious downside risks, compute governance efforts should be thoughtfully designed and executed, and include safeguards to protect against abuse. We explore some possible approaches to doing so here. A recurring theme of these heuristics is the need for compute governance measures to be carefully scoped to tackle the largest risks while reducing the impacts on consumers and individuals.

Our five principles are:
\begin{enumerate}
    \item Exclude small-scale AI compute and non-AI compute from governance
    \item Implement privacy-preserving practices and technologies
    \item Focus compute-based controls where ex ante measures are justified
    \item Periodically revisit controlled computing technologies
    \item Implement all controls with substantive and procedural safeguards
\end{enumerate}

This list is not intended to be exhaustive; we think additional research on guardrails for compute governance has very high value.

\textbf{\textit{Exclude small-scale AI compute and non-AI compute from governance}} \\\\ Many of the concerns listed above are most concerning if we assume that compute governance is applied to all forms of compute at all scales. But this is not what we in this report mean by compute governance. Part of the appeal of compute governance is the ability to distinguish reasonably well between compute that is likely to be put to particularly risky uses and compute that is used for overwhelmingly beneficial and benign purposes. In particular, as we have discussed, AI-relevant chips are a small and distinct subset of all computer chips. The large-scale computational resources needed for frontier AI systems are both unattainable for virtually all but the wealthiest consumers and reasonably easy to distinguish from other computations with minimally intrusive inspections.

One way to scope compute governance to avoid some of the downsides to privacy and concentration of power would therefore be to clearly exclude consumer-scale compute and non-AI chips\footnote{Of course, it may make sense to govern other specialized computing hardware for reasons other than AI governance. For example, the U.S. government controls other types of computing hardware, such as radiation-hardened chips (see, e.g., ECCN 9A515, 4A001 \autocite{bureauofindustryandsecurityExportAdministrationRegulations2023}). The U.S. is also considering imposing controls on quantum computing hardware \autocite{williamsBidenAdministrationQuantum2022}. Since our primary concern is AI compute, we do not mean to imply that such controls are inappropriate.} from many of the mechanisms discussed here. The Biden export controls and recent executive order on AI focus on \textit{industrial-scale} compute for AI, targeting only the most advanced AI data center chips, the very largest data centers,\footnote{The computing cluster needs to meet an aggregated computing performance of more than $ 10^{20} $ operations per second, a chip interconnectivity of more than 100 Gbit/s, and be housed in a single data center.} and frontier training runs bigger than any yet run. For example, the executive order directs the U.S. Department of Commerce to establish Know-Your-Customer requirements for the provision to foreigners of enough compute to train a $ 10^{26} $ operations model.\footnote{Provided the model is trained in a data center that needs to be reported to the Department of Commerce.} Buying that amount of compute from a cloud compute provider would currently cost no less than \$100 million at on-demand prices.\footnote{GPT-4: $2\times 10^{25}$ FLOP for training \autocite{epochAITrends2023}; H100 performance: 990 teraFLOP/second (peak FP16 tensor performance without sparsity) \autocite{nvidiaNVIDIAH100Tensor}; Assuming 30\% utilization; $ \sim $ \$5.60 per hour per H100 (AWS p5.48xlarge H100 instance, 3-year reserved price, estimate) \autocite{morganH100GPUInstance2023}; AWS p4d.24xlarge A100 instance, 3-year reserved price \autocite{amazonAmazonEC2P4d2023}. Result: $ \sim $ \$100M.} No individual consumer, or even university lab or start-up, is going to be operating at that level, only large companies.

Moreover, it is important to note that the AI chips and large data centers that are the focus of this report constitute a minute fraction of all computational activity, meaning that governance measures targeted at them should leave the overwhelming majority of chips (and computations thereon) untouched.

\begin{figure}[ht]
    \centerline{\includegraphics[width=1.2\linewidth]{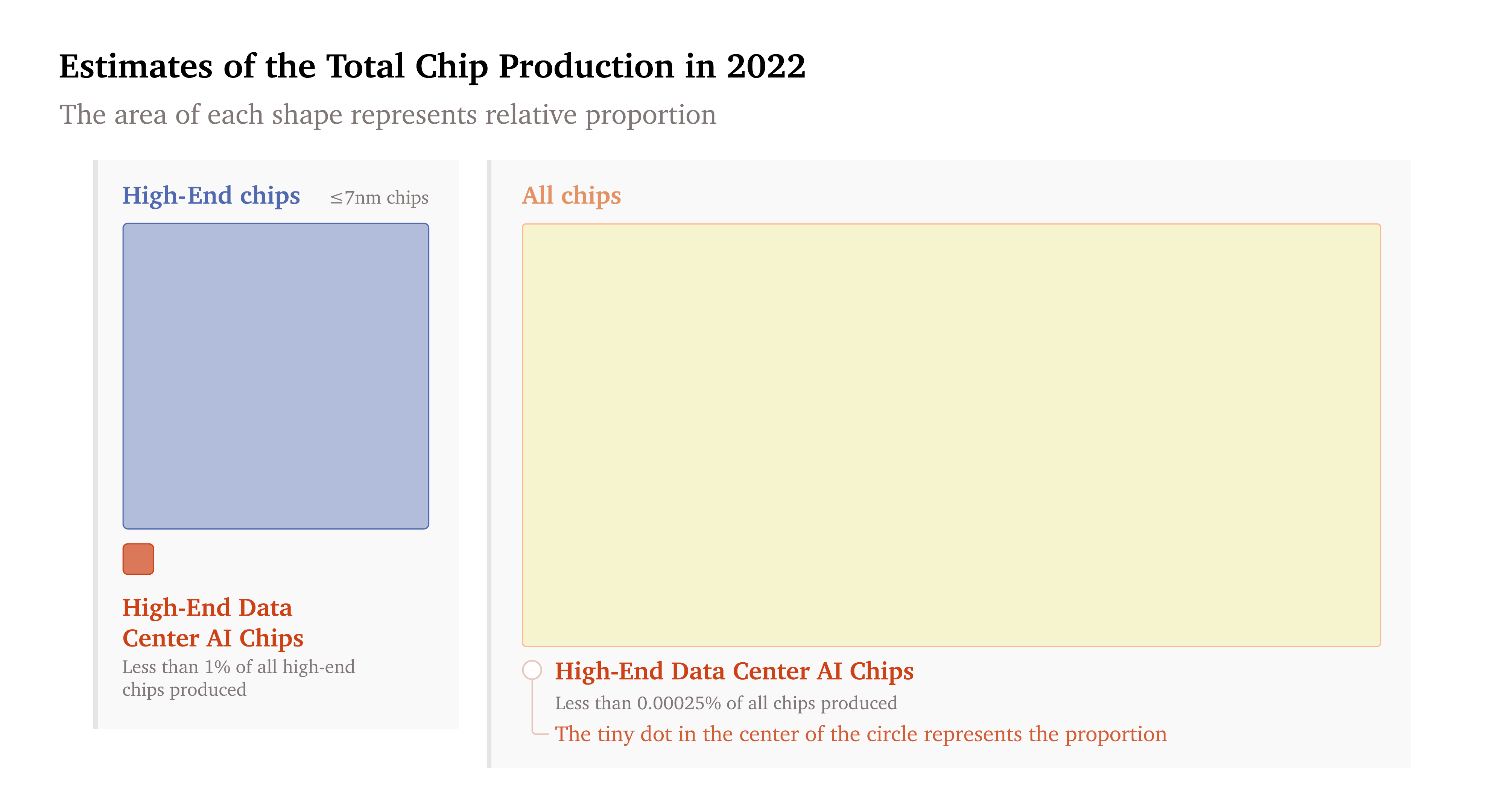}}
    \caption[\textbf{Data-center AI chips are a minor segment of overall and high-end chip production.} For 2022, we estimate that the number of high-end data center AI chips constituted less than 1\% of all high-end ($\leq$7 nm) chips and less than 1 in 400,000 (0.00026\%) of every chip produced.]
    {\textbf{Data-center AI chips are a minor segment of overall and high-end chip production.} For 2022, we estimate that the number of high-end data center AI chips constituted less than 1\% of all high-end ($\leq$7 nm) chips and less than 1 in 400,000 (0.00026\%) of every chip produced.\protect\footnotemark}
    \label{fig:datacenter_fraction}
\end{figure}

\footnotetext{\textcite{heimWhatShareAll2024} outlines the method for these estimates.}

However, this may not always be the case: there is a risk that consumer-scale and AI computation of concern become less separable over time. AI is not inherently limited to data center-grade AI chips, and the landscape of AI hardware will continually evolve in response to technological advancements, regulatory constraints, and the changing needs of AI applications. No foundational facts rule out the technical possibility of training models by linking together many gaming GPUs, either in a dedicated cluster or via massively decentralized training (which is currently technically infeasible). While there would indeed be a performance penalty for doing so, this may not be significant enough to deter a motivated actor. In such situations, governments may need to rely more on tools beyond compute governance to meet their goals.

\newpage
\textbf{\textit{Implement privacy-preserving practices and technologies}} \\\\ Where compute governance touches large-scale computing that contains or could reveal personal information, care must be taken to narrowly tailor the compute governance measures so that they accomplish much of the possible risk-reduction with minimal intrusion on privacy. Take KYC for cloud AI training: applying KYC only to direct purchasers of large amounts of cloud AI compute capacity (as Executive Order 14110 proposes) would impose almost no privacy burdens on consumers. KYC could also feasibly draw on indicators already readily available---such as chip hours, types of chips, and how GPUs are networked---preserving privacy for compute providers and consumers \autocite{eganOversightFrontierAI2023}.

One obvious guardrail that should apply to any compute governance measure that could expose (or create opportunities to leak) sensitive information\footnote{This should be construed broadly, to include personally sensitive information as well as information that is sensitive from a commercial or national security perspective.} (see \Cref{ssec:guardrails}) is to design the measure with information security in mind. A full overview of how to do so is beyond the scope of this paper. However, we would strongly encourage policymakers to consider commonsense measures such as narrowly tailoring the information disclosed to policymakers, using secure channels for communication, and limiting access to sensitive information.

New technologies may also expand the amount of risk-reduction that can be achieved for any given level of intrusion on privacy---or equivalently, reduce the intrusion on privacy needed for any amount of risk reduction \autocite{traskPrivacyTradeoffsStructured2020, bluemkeExploringRelevanceData2023}. For example, new hardware and software technologies could enable regulators to receive limited reliable information about whether computations complied with regulations---perhaps just a single bit of information that indicates compliance---without making any other data available to them. These technologies, if feasible and secure, could dramatically reduce the potential for compute governance to be used for surveillance (and therefore concentration of power) and other privacy infringements.

Privacy-enhancing technologies may also make new sorts of agreements possible. In arms control agreements, state actors often desire verification methods that are both highly reliable---so that they can be assured that their counterparties are not defecting from the agreement to achieve a strategic advantage---and secrecy preserving---so that inspections do not reveal secret information, other than that needed to demonstrate compliance \autocite{oneillVerificationAgeInsecurity2009, coeWhyArmsControl2020}. In the nuclear context, ``information barriers'' have been developed to provide just enough information about warheads to verify compliance with a given agreement, while ensuring appropriate secrecy beyond that (see sources collected at Nuclear Threat Initiative \autocite{ntiInformationProtectionInformation2015}). Some proposals have been developed to navigate such challenges---for example, cryptographic escrow as a technique for addressing North Korea’s security concerns while enabling enforcement of agreements \autocite{philippeCryptographicEscrowTreaty2019}. Drawing on the best of science, engineering, institutional design, and other sources can help alleviate trade-offs where they arise \autocite{traskPrivacyTradeoffsStructured2020}.

\textbf{\textit{Focus compute-based controls where ex ante measures are justified}} \\ \\ Compute governance (especially in its ``allocation'' and ``enforcement'' forms) is often a blunt tool, and generally functions upstream of the risks it aims to govern and the benefits it seeks to promote. Policymakers have often preferred ex post mechanisms that impose some cost (such as a tax, fine, or penalty) for externalities and other dispreferred outcomes after they have occurred (e.g., \textcite{galleTaxCommandNudge2013}).

There are exceptions, however. In particular, certain types of harms justify ex ante efforts at prevention, such as where the harm is so large that no actor would be able to compensate for it ex post. Catastrophic risks and risks to national security often have this nature. Compute controls could therefore be targeted only at risks that are of such quality or magnitude that leaving regulation to ex post mechanisms would fail to adequately address them \autocite{anderljungFrontierAIRegulation2023, koltAlgorithmicBlackSwans2023}. For more detailed discussion, see \Cref{ssec:development_vs_depoyment}.

\textbf{\textit{Frequently revisit controlled computing technologies and thresholds}}\\\\ Regulatory thresholds (like a training compute threshold of $10^{26}$ operations) or list-based controls on technologies, such as those used in export controls, can become outdated fairly quickly. This applies in both directions: changing circumstances might mean that controls are either too loose---e.g., because a new technology has not yet been controlled, or an old technology has become newly riskier---or too strict---e.g., because a controlled item is freely attainable on the open market \autocite{mastandunoEconomicContainmentCocom1992}. In the fast-moving domain of AI, more significant changes to policy may be needed more frequently than in other domains. Compute regulators should therefore ensure that their governance mechanisms are regularly reviewed at least every year, assessing their particulars---e.g., lists of controlled technologies, particular thresholds used, methods for detecting violations---as well as whether they are achieving their intended goals.\footnote{As a possible model, the Federal Select Agents Program statutorily requires the administering agency to review controlled agents at least biennially (7 U.S.C. § 8401).}

\textbf{\textit{Ensure strong substantive and procedural safeguards}} \\\\ As we acknowledged above, compute writ large is a societally important technology with many beneficial and benign use cases. In the future, compute’s importance is likely to increase, and so the stakes of preventing mismanagement of this important resource are likely to increase.

Any implemented compute control measures should therefore include both substantive and procedural safeguards, at the statutory level if possible.\footnote{Of course, this too must be balanced with the need for some flexibility given rapidly changing technical circumstances.} Substantively, such controls could prevent downsides from compute governance by, for example, limiting the types of controls that can be implemented, the type of information that regulators can request, and the entities subject to such regulations. Domestically, procedural safeguards could include such measures as notice and comment rulemaking, whistleblower protections, internal inspectors general and advocates for consumers within the regulator, opportunities for judicial review, advisory boards, and public reports on activities.

\section{Conclusion}\label{sec:conclusion}

Compute has properties that are unique among the various inputs to AI capabilities, and it is particularly important for governance of compute-intensive frontier AI models. Prominent AI governance proposals and practices in the past few years reflect this realization. With this paper, we hope to provide a better theoretical understanding of the promises and limitations of compute governance as a vehicle for AI governance, and spur more creative thinking on the future of compute governance.

A few themes of this paper are worth reiterating. Of the inputs to AI, compute is the most regulable, due to its \textit{detectability}, \textit{excludability}, \textit{quantifiability}, and \textit{supply chain concentration}. Where inputs-based governance of AI is warranted, therefore, compute provides a good lever for such regulation.

We identify three core governance capacities that compute can enhance, and provide examples of each: (1) increasing regulatory \textit{visibility} into AI capabilities and use, (2) \textit{allocating} resources toward safe and beneficial uses of AI, and (3) \textit{enforcing} prohibitions against irresponsible or malicious development or use of AI. However, we emphasize the many potential limitations and downsides to some approaches to compute governance, especially with regard to centralization of control over an increasingly important technology. We therefore conclude by providing heuristics that, if followed, should help compute governance measures to be carefully scoped to tackle the largest risks while reducing the impacts on consumers and individuals.

A number of the ideas in this paper are exploratory or tentative. In particular, many of the policy mechanisms described in \Cref{sec:enhance} are sketches of possible directions for compute governance, not fully fledged policy proposals. We hope that further work will determine whether and how these mechanisms can be designed and implemented in accordance with the limiting principles set forth in \Cref{sec:risks}. In \Cref{sec:appB}, we list additional open questions in compute governance.

Hardware and software progress will over time erode the effectiveness of many compute governance mechanisms, as these secular trends drive down the hardware cost of achieving a particular level of AI capabilities. In \Cref{sec:risks} we propose limiting compute governance mechanisms to AI chips. If this advice is heeded, many AI capabilities---including risky ones---will become increasingly achievable using ``ungovernable'' compute. To mitigate these risks, society will have to use more powerful, governable compute timely and wisely, to develop defenses against emerging risks posed by ungovernable compute.

\newpage
\section*{Acknowledgments}\label{sec:acknowledgments}

Thanks to Alex Savard, Allan Dafoe, Andrew Lohn, Andrew Trask, Carrick Flynn, Chris Phenicie, David Robinson, Gretchen Krueger, Jaan Tallin, Jade Leung, Katarina Slama, Larissa Schiavo, Lewis Ho, Lucy Lim, Magnus Løiten, Matthijs Maas, Mauricio Baker, Michael Lampe, Paul Scharre, Rosie Campbell, Sam Manning, Sean O hEigeartaigh, Tim Fist, Tom Davidson, Tom Westgarth, and Yonadav Shavit for feedback on earlier versions of this paper, and Eden Beck for editorial revision. Thank you to Alex Savard for graphic design help. Miles dedicates this paper to the memory of his father, Jan Brundage.

GPT-4 and Claude were used to suggest ideas and provide feedback during the writing process.

\newpage
\addappheadtotoc
\appendix
\section{The Compute-Uranium Analogy}\label{sec:appA}

\begin{figure}[ht]
    \centerline{\includegraphics[width=1.3\linewidth]{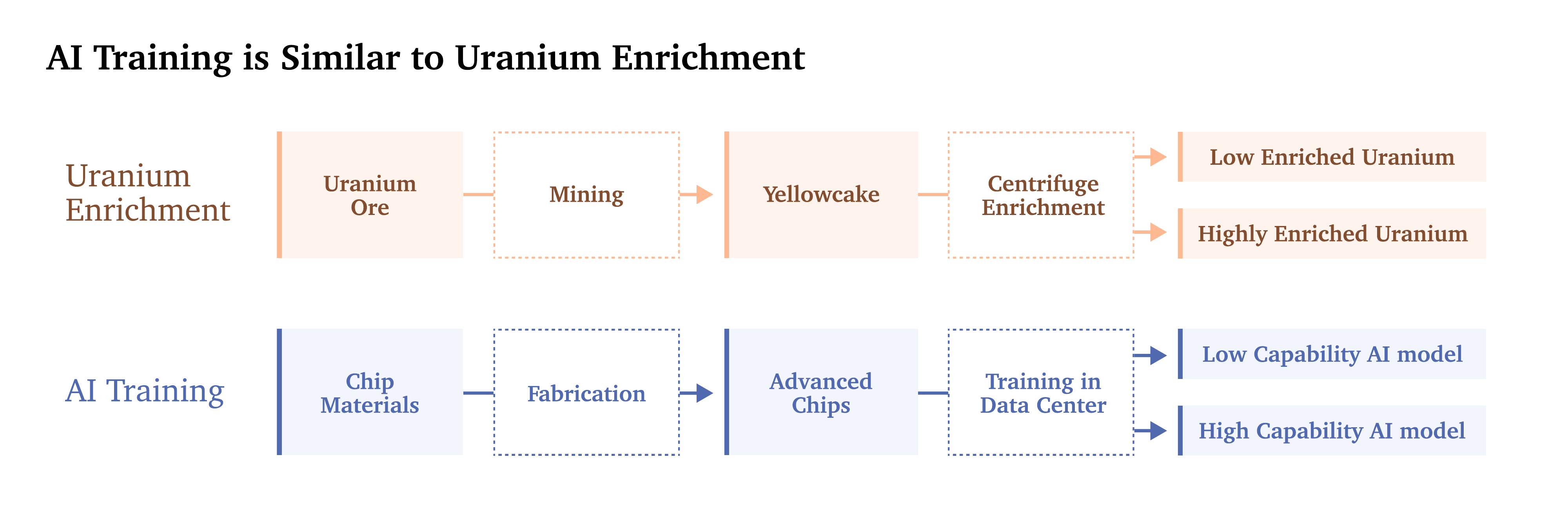}}
    \caption{\textbf{The analogy between uranium enrichment and AI training.} For both AI (chips) and nuclear energy (uranium), there is a key input that is difficult to produce and potentially regulable.}
    \label{fig:uranium_compute}
\end{figure}

There is a suggestive analogy between a key physical input to two powerful technologies: compute and data centers in the case of AI, and uranium and enrichment facilities in the case of nuclear energy or weapons.\footnote{Nuclear weapons can also be made with plutonium, though similar considerations apply.} Uranium mining and enrichment and compute fabrication and training both lead to outputs that can be used for both safe and harmful purposes, require significant capital investments, and can be differentiated by quantitative measures of quality (e.g., operations per watt or the level of enrichment). One way of envisaging this analogy is shown in \Cref{fig:uranium_compute}.

Uranium ore goes through a process of mining to produce yellowcake, which then goes through a process of enrichment to produce either low or highly-enriched uranium. One can draw an analogy with compute: materials go through a process of fabrication to produce chips, which then are used in a process of training to produce a model (below or above some level of capability). Each process is lengthy, difficult, expensive, and potentially amenable to monitoring.

This analogy, while imperfect,\footnote{Like other analogies, the comparison between training and uranium enrichment has its limitations. In particular, while both are dual-use, it is possible to infer a narrower set of potential uses for enriched uranium, while high-capability models can be applied to a wider variety of use cases.\\ The analogy is inexact as the chips are the physical location where the process occurs, rather than the material that is processed---which is data using particular algorithms. So an individual AI chip processing data, say, can be compared to an individual centrifuge enriching uranium, while a data center can be compared to an enrichment plant.} is encouraging. Risks associated with nuclear enrichment have been (more or less) managed for decades. A wide variety of technical and political measures are used to control the production, flow, and use of nuclear materials. Many of these can be thought of as ``accounting'': keeping and checking careful records of who is creating nuclear material, where it goes, how it’s used, and how it’s disposed of. There are national and international regimes to track and monitor mines, yellowcake, enrichment plants, and enriched uranium. These measures include export controls, inspections by the International Atomic Energy Administration (IAEA), remote monitoring, unique identifiers such as serial numbers, and regulations around the use and disposal of nuclear materials. Enrichment capacity has been a key focus of nuclear nonproliferation regimes, as it is the key determinant of ``breakout time'': the minimum time for a state to produce enough weapons-grade enriched uranium fuel for a single nuclear weapon. Taken together, these measures have contributed to a low rate of proliferation and the prevention of a nuclear conflict since 1945.

Analogously, we may also want to consider regimes to track and monitor chip fabrication, the resulting advanced AI chips, and their final destination in data centers. Much like how unique IDs and tamper-proofing are employed to track uranium, complemented by monitoring through inspections and intelligence sources like satellite footage, we could envision similar methodologies being theoretically applicable to the advanced AI chip supply chain \autocite{bakerNuclearArmsControl2023}. Compute may even have some advantages over nuclear. For example, there are between 6 and 59 manufacturers for each of the dual-use goods under the Nuclear Suppliers Group’s purview \autocite{doyleNuclearSafeguardsSecurity2019}. By contrast, some steps in the compute supply chain have only a single company. 

Our use of this analogy should not be interpreted to mean that we overlook its limitations. Difficult political battles were required to achieve today’s modern institutions, and we recognize that nonproliferation governance continues to be a contested space, as demonstrated in the last nuclear nonproliferation treaty review conference \autocite{unNonproliferationTreatyReview2022, potterScenesHowNot2023}. As \textcite{stewartWhyIAEAModel2023} writes, modern nuclear nonproliferation governance evolved over decades, through international crises rather than preemptively, resulting in a governance patchwork instead of a universally coherent standard. And we would not be able to apply the same technical methods used for monitoring uranium to compute. The nonradioactivity of compute makes it more difficult to detect at ports and other border crossings than nuclear material. Many efforts related to tracking nuclear material, including international inspections, depend on the ability to correlate radioactivity emissions with the precise chemical properties of uranium (as well as plutonium).

There is another significant limitation associated with the nuclear analogy; namely, while the emphasis on hardware excludability advances nonproliferation aims, there are no obvious parallels in nuclear proliferation to the release of model weights. Recent historical cases demonstrate that access to information about nuclear weapons design---non-rivalrous, easily replicable and transferable---is often an insufficient condition for nuclear proliferation \autocite{kempNonproliferationEmperorHas2014, ouagrham-gormleyBarriersBioweaponsChallenges2014}. We can reasonably conclude that a bad actor with access to scientific information would not seriously undermine the existence of the nuclear nonproliferation regime. The release of model weights, on the other hand, poses a significant threat to nonproliferation compute regimes, because their public availability would allow an individual with a moderate amount of machine learning expertise to bypass the large compute requirements needed for training a model.\footnote{Because the computing power necessary to run a model is much less than the computing power needed to train a model.} This is an argument not only for strong cybersecurity controls, but for avoiding a repeat of the historical setbacks suffered by nuclear nonproliferation regimes and adopting preemptive governance mechanisms before the wide availability of model weights potentially undermines safeguards.

Despite these limitations with the analogy, it is striking that society has also safely produced fairly large quantities of nuclear power \autocite{ritchieNuclearEnergy2023} and that there have been zero instances of nuclear terrorism in the nearly 80 years since the advent of nuclear weapons technology. The political and technical regimes that govern nuclear technology likely deserve some credit for this situation. While this system is certainly imperfect---rogue states like North Korea have still managed to build up their nuclear capacity in part via illegal proliferation networks \autocite{chestnutIllicitActivityProliferation2007, reissRedhanded2005}---it is nevertheless a proof of concept for an institutional design that governs a highly sought after, dual-use technology at global scale.

\section{Research Directions}\label{sec:appB}

\textbf{\textit{Policy implications of increased compute efficiency}} \\\\
Given ongoing algorithmic and hardware progress, an AI capability that is initially only available to a small number of well-resourced developers will slowly diffuse to increasingly compute-limited actors over time \autocite{pilzIncreasedComputeEfficiency2023}.

This makes it challenging to construct enduring compute-based policy. Instead, compute governance interventions may be about influencing not whether but \textit{when} certain capabilities are made available, to whom, and for what purpose. This view suggests that society will need to use the time bought by certain compute governance interventions to prepare for the widespread diffusion of advanced AI capabilities. To what extent is this picture correct? If so, what can policymakers do to increase society’s resilience to the diffusion of increasingly capable AI systems?

Relatedly, what does the offense-defense balance look like for different AI capabilities? One could potentially use the compute required to develop different capabilities as one input to the offense-defense balance. Will the proliferating systems favor the offense or defense? Will it be possible to use compute resources to counter harm, such as by using highly performing systems or deploying defensive applications on a large scale?

\textbf{\textit{Trustworthy verification of compute capabilities and usage}} \\\\
Implementing compute governance requires the ability to verify claims about actors’ compute capabilities and usage \autocite{brundageTrustworthyAIDevelopment2020}. However, validating these claims often involves accessing sensitive information that actors wish to keep private. On-site inspections intended to verify the number of chips an actor has access to might inadvertently reveal other sensitive information, and monitoring computational workloads is likely unacceptably intrusive in the international context with current techniques. This presents a challenge for arms control, especially when trust is low and competition is high \autocite{coeWhyArmsControl2020}. How severe is this trade-off? What sorts of hardware, software, and institutional techniques might help alleviate it?

\textbf{\textit{Regulatory flight as a result of compute governance measures}}\\\\
Certain compute governance interventions can induce regulatory flight, where activities are moved to less heavily regulated jurisdictions. For example, recent U.S. chip export controls incentivize the creation of an advanced chip supply chain without U.S. inputs. Similarly, customers may prefer compute providers with the least ability to monitor their compute capabilities and usage. What compute governance interventions are most likely to see their effectiveness undermined by regulatory flight? How can the chance of such flight be reduced? For instance, mechanisms for detecting black market AI chips in smaller countries may be important, since this compute could potentially be targeted by malicious actors who exploit differences in regulation.

\textbf{\textit{Countries caught in geopolitical competition}}\\\\
The compute supply chain is increasingly shaped by geopolitical competition between great powers. How will other countries respond? To what extent can they avoid getting involved in the conflict, or will they be forced to ally with one side? What effects would this have on semiconductor supply chains?

\textbf{\textit{Incentivizing responsible compute provision}}\\\\
How should responsible compute provision practices be incentivized and enforced by policymakers? Certain practices may require enforcement from the government via regulation or export controls. Other practices could be incentivized via liability regimes, where compute providers are held partly liable for lax attempts at identifying and thwarting misuse. Others may be implemented voluntarily by the compute provider industry.

\textbf{\textit{Limits of scaling}} \\\\ 
What are the fundamental and practical limits to compute scaling? The past decade’s growth in compute usage for notable AI systems has been driven by increases in spending as well as reductions in the cost of compute. If these trends continue, training a state-of-the-art AI system would cost approximately 2.2\% of U.S. GDP in 2032, similar to the annual cost of the Apollo program \autocite{heimThisCanGo2023, lohnAICompute2022}. Such spending would only be possible if the returns to scaling compute are immense. Further, many have argued that reductions in compute efficiency are likely to slow down, as Moore’s Law starts to hit its limits \autocite{shalfFutureComputingMoore2020}.

\textbf{\textit{Piloting compute governance levers}} \\\\
To have confidence in deploying some of the levers described above at large scale, pilot studies may be needed to test their viability (such as the compute measurement pilot suggested by Brundage \autocite{brundageTrustworthyAIDevelopment2020}), much as the Joint Verification Experiment demonstrated the feasibility of seismographic detection of nuclear testing \autocite{usgovWhiteHouseStatement1988, sykesComparisonSeismicHydrodynamic1989}.

\textbf{\textit{Security-privacy trade-offs}}\\\\
Researchers should also more carefully analyze the trade-offs related to security and privacy, and assess what sorts of hardware, software, and institutional techniques might help alleviate such trade-offs. Further piloting, analysis, and debate are needed in order to more fully understand how compute can and can’t enable effective AI governance.

\clearpage

\printbibliography

\end{document}